\documentclass[11pt,a4paper]{article}
\usepackage{amsmath, amsfonts,amssymb,mathtools,nicefrac,nccmath,cases}
\usepackage{booktabs,multirow}
 
\usepackage{microtype}

\usepackage[usenames,dvipsnames,table]{xcolor}
\usepackage{colortbl}
\usepackage{tikz}
\usetikzlibrary{shapes,arrows}
\usetikzlibrary{calc,shapes.callouts,shapes.arrows,bending,intersections}
\usetikzlibrary{shapes.geometric}
\usetikzlibrary{arrows.meta,arrows}
\usetikzlibrary{decorations.pathmorphing}
\usetikzlibrary{decorations.markings}
\usetikzlibrary{shapes.multipart}
\usetikzlibrary{tikzmark}
\usepackage{caption}
\usepackage{subcaption}
\usepackage{bm}
\usepackage[nosort]{cite}
\usepackage{colortbl}
\usepackage{amsthm}
\usepackage{soul}
\usepackage{dsfont}

\usepackage{float}
\usepackage{tabularx}
\usepackage{multirow}
\usepackage{hhline}

\usepackage{xcolor}
\usepackage[linktocpage=true]{hyperref}
\definecolor{nicecolor}{rgb}{0.1, 0.3, 0.4}
\hypersetup{colorlinks=true,linkcolor=nicecolor,citecolor=nicecolor,urlcolor=nicecolor}

\def\hybrid{\topmargin -20pt    \oddsidemargin 0pt
  \headheight 0pt \headsep 0pt
  \textwidth 6.5in        
  \textheight 9in         
  \textwidth 6.25in       
  \textheight 9 in       
  \marginparwidth .875in
  \parskip 5pt plus 1pt 
  \jot = 1.5ex
}
\hybrid
\numberwithin{equation}{section}
\numberwithin{table}{section}
\usepackage{physics}

\usepackage{float}
\usepackage{tabularx}
\newcolumntype{D}{>{\centering\arraybackslash}X}
\newcolumntype{L}{>{$}l<{$}}
\newcolumntype{R}{>{$}r<{$}}
\newcolumntype{C}{>{$}c<{$}}
\usepackage{IEEEtrantools}
\usepackage{makecell}

\newcommand{\beq}{\begin{equation}}  \newcommand{\eeq}{\end{equation}}
\newcommand{\bal}{\begin{aligned}}   \newcommand{\eal}{\end{aligned}}
\newcommand{\bea}{\begin{eqnarray}}  \newcommand{\eea}{\end{eqnarray}}
\def\beqa{\begin{eqnarray}}
\def\eeqa{\end{eqnarray}}

\newcommand{\bmat}{\left(\begin{array}}
\newcommand{\emat}{\end{array}\right)}

\newcommand{\bbC}{\mathbb{C}}
\newcommand{\bbR}{\mathbb{R}}

\newcommand{\cO}{\mathcal{O}}

\newcommand{\cE}{\mathcal{E}}

\newcommand{\cS}{\mathcal{S}}

\newcommand{\cN}{\mathcal{N}}

\newcommand{\cH}{\mathcal{H}}

\newcommand{\cM}{\mathcal M}

\usepackage{cancel}

\newcommand{\be}{\begin{equation}}
\newcommand{\ee}{\end{equation}}

\newcommand{\im}{\mathbf{i}}
\newcommand{\slt}{\mathfrak{sl}(2)}

\newcommand{\bbZ}{\mathbb{Z}}
\newcommand{\subs}{\subset}

\newcommand{\Gr}[2]{\mathrm{Gr}^{#1}_{#2}}
\newcommand{\conj}[1]{\ensuremath{\overline{#1}\vphantom{#1}}} 

\newcommand{\Ranexp}{\ensuremath{\bbR_{\mathrm{an, exp}}}}
\newcommand{\bbN}{\ensuremath{\mathbb{N}}}
\newcommand{\us}{{\ensuremath{\underline{s}}}}
\newcommand{\um}{{\ensuremath{\underline{m}}}}
\newcommand{\zbar}{\ensuremath{\bar{z}}}

\newcommand{\mysetminusD}{\hbox{\tikz{\draw[line width=0.6pt,line cap=round] (3pt,0) -- (0,6pt);}}}
\newcommand{\mysetminusT}{\mysetminusD}
\newcommand{\mysetminusS}{\hbox{\tikz{\draw[line width=0.45pt,line cap=round] (2pt,0) -- (0,4pt);}}}
\newcommand{\mysetminusSS}{\hbox{\tikz{\draw[line width=0.4pt,line cap=round] (1.5pt,0) -- (0,3pt);}}}
\newcommand{\mysetminus}{\mathbin{\mathchoice{\mysetminusD}{\mysetminusT}{\mysetminusS}{\mysetminusSS}}}

\newcommand{\ut}{\ensuremath {\underline{t}}}
\newcommand{\uz}{\ensuremath {\underline{z}}}

\newcommand{\bbQ}{\ensuremath {\mathbb{Q}}}
\newcommand{\ui}{\ensuremath {{\underline{i}}}}
\newcommand{\uj}{\ensuremath {{\underline{j}}}}
\newcommand{\ux}{\ensuremath {{\underline{x}}}}
\newcommand{\uy}{\ensuremath {{\underline{y}}}}
\newcommand{\uk}{{\ensuremath {\underline{k}}}}
\newcommand{\hi}{\ensuremath {\hat{\imath}}}
\newcommand{\hj}{\ensuremath {\hat{\jmath}}}
\newcommand{\vtil}{\ensuremath {\tilde{v}}}
\newcommand{\wtil}{\ensuremath {\tilde{w}}}
\newcommand{\util}{\ensuremath {\tilde{u}}}

\renewcommand{\ut}{\ensuremath {\underline{z}}}

\definecolor{Gray}{gray}{0.95}

\newtheorem{lemma}{Lemma}
\newtheorem{theorem}{Theorem}
\newtheorem{claim}{Claim}
\newtheorem{fact}{Fact}

\newcommand{\hH}{\ensuremath{\hat{H}}}
\newcommand{\hQ}{\ensuremath{\hat{Q}}}
\newcommand{\hC}{\ensuremath{\hat{C}}}
\newcommand{\hh}{\ensuremath{\hat{h}}}
\newcommand{\ha}{\ensuremath{\hat{a}}}
\newcommand{\hb}{\ensuremath{\hat{b}}}
\newcommand{\hc}{\ensuremath{\hat{c}}}
\newcommand{\hd}{\ensuremath{\hat{d}}}
\newcommand{\tH}{\ensuremath{\tilde{H}}}
\newcommand{\tQ}{\ensuremath{\tilde{Q}}}
\newcommand{\tC}{\ensuremath{\tilde{C}}}
\newcommand{\tilh}{\ensuremath{\tilde{h}}}

\definecolor{colorloc1}{RGB}{10,125,173} 
\definecolor{colorloc2}{RGB}{0,48,143} 

\usepackage[framemethod=default]{mdframed}
\newmdenv[skipabove=10pt,
skipbelow=7pt,
rightline=false,
leftline=true,
topline=false,
bottomline=false,
linecolor=colorloc2,
backgroundcolor=colorloc1!5,
innerleftmargin=4pt,
innerrightmargin=0pt,
innertopmargin=0pt,
leftmargin=2pt,
rightmargin=0pt,
linewidth=2pt,
innerbottommargin=0pt,
frametitlebackgroundcolor=colorloc2]{lbBox}

\usepackage[framemethod=default]{mdframed}
\newmdenv[skipabove=10pt,
skipbelow=7pt,
rightline=false,
leftline=false,
topline=false,
bottomline=false,
linecolor=colorloc2,
backgroundcolor=colorloc1!5,
innerleftmargin=4pt,
innerrightmargin=0pt,
innertopmargin=0pt,
leftmargin=2pt,
rightmargin=0pt,
linewidth=2pt,
innerbottommargin=0pt,
frametitlebackgroundcolor=colorloc2!100]{tbBox}
\newenvironment{importantboxtitle}[1]{\begin{tbBox}[frametitle={\textcolor{white}{ \textbf{\sffamily{#1}} }},nobreak=true] \vspace{1 mm}
		 } {\vspace{1.5 mm}\end{tbBox}}

\begin{document}


  \baselineskip=14pt
  \parskip 5pt plus 1pt

	\vspace*{-1.5cm}
  
  \vspace{4cm}
  \begin{center}        

    {\huge Tameness, Strings, and the Distance Conjecture}   
  \end{center}
  
  \vspace{0.5cm}
  \begin{center}        
    {\large  Thomas W.~Grimm, Stefano Lanza, Chongchuo Li}
  \end{center}
  
  \vspace{0.15cm}
  \begin{center}  
    \emph{Institute for Theoretical Physics,
      Utrecht University}\\
    \emph{Princetonplein 5, 3584 CE Utrecht, The Netherlands}
    \\[.3cm]
  \end{center}
  
  \vspace{2cm}
  
  
\begin{abstract}
The Distance Conjecture states that an infinite tower of modes becomes exponentially light when approaching an infinite distance point in field space. We argue that the inherent path-dependence of this statement can be addressed when combining the Distance Conjecture with the recent Tameness Conjecture. The latter asserts that effective theories are described by tame geometry and implements strong finiteness constraints on coupling functions and field spaces. By exploiting these tameness constraints we argue that the region near the infinite distance point admits a decomposition into finitely many sectors in which path-independent statements for the associated towers of states can be established. We then introduce a more constrained class of tame functions with at most polynomial asymptotic growth and argue that they suffice to describe the known string theory effective actions. Remarkably, the multi-field dependence of such functions can be reconstructed by one-dimensional linear test paths in each sector near the boundary. In four-dimensional effective theories, these test paths are traced out as a discrete set of cosmic string solutions. This indicates that such cosmic string solutions can serve as powerful tool to study the near-boundary field space region of any four-dimensional effective field theory. To illustrate these general observations we discuss the central role of tameness and cosmic string solutions in Calabi-Yau compactifications of Type IIB string theory.

\end{abstract}

\thispagestyle{empty}
\clearpage
  
\setcounter{page}{1}
  
  
\newpage

\tableofcontents

\newpage

\section{Introduction}
\label{sec:intro}

It is widely believed that only a limited subset of quantum field theories can be regarded as effective field theories (EFTs) arising from a UV-complete quantum gravity theory. 
Singling out the criteria that EFTs have to satisfy in order to admit a UV quantum gravity completion is the core of the \emph{Swampland program} (see, for example, \cite{Palti:2019pca,vanBeest:2021lhn} for reviews on the program). In recent years several of these criteria have been proposed and collected 
within the so-called `Swampland conjectures'. The Swampland conjectures put constraints on many crucial features of the EFTs, such as the gauge sector, 
the allowed symmetries, or the spectrum of objects that the EFT can describe. 
Such conjectures are typically formulated independently, but several relations have subsequently led to the formation of an interconnected web. 
Eventually one might thus hope that the Swampland conjectures can be reduced to a set of basic principles that 
the quantum nature of gravity imposes on the effective descriptions. The aim of this work is to further combine some of the Swampland conjectures
and thereby clarify their statements. 
In particular, we will investigate how the constraints set by the Distance Conjecture \cite{Ooguri:2006in} can be more generically addressed by employing the 
recently formulated Tameness Conjecture \cite{Grimm:2021vpn}.

The Distance Conjecture \cite{Ooguri:2006in} poses constraints on the explorable field space of any effective field theory. Consider an effective field theory with a set of scalar fields, known as 
\textit{moduli}, that 
are not subjected to any scalar potential. 
The field space of these moduli can be highly nontrivial and, in particular, can admit boundaries at which the field space metric degenerates. 
In turn, we can distinguish the boundary points by using the length of the minimal geodesic distance that is required to reach them, which can either be finite or infinite from a regular field space point. The family of infinite distance points are central in the Distance Conjecture, which asserts that in any consistent EFT of quantum gravity, 
an infinite tower of exponentially light states becomes relevant near the infinite distance point. In other words, infinite distance limits in an EFT should be viewed as 
an artifact of the effective description and cannot be reached within the same EFT due to the emergence of additional light states.

The obstructions that the Distance Conjecture suggests are path independent: no matter how the infinite distance locus is reached, the effective description is 
eventually rendered invalid close to an infinite distance point.
However, it is known from many examples  \cite{Grimm:2018ohb,Lee:2018urn,Lee:2018spm,Lee:2019tst,Marchesano:2019ifh,Font:2019cxq,Lee:2019xtm,Grimm:2019wtx,Lee:2019wij,Grimm:2019bey,Grimm:2019ixq,Baume:2019sry,Gendler:2020dfp,Lanza:2020qmt,Klaewer:2020lfg,Ashmore:2021qdf,Brodie:2021ain} that the details on how this happens precisely generically depend on the path and that the 
physical interpretation can change from one path to another.  
For example, along different paths leading to the same infinite distance point, the leading tower of light states invalidating the EFT might well be different. 
We are thus facing a number of questions that are crucial in order to understand the physics emerging in the near-boundary region: 
(1) Given an infinite tower of states invalidating the EFT along a given path, along which other paths does this tower remain relevant? 
(2) How many families of infinite tower of states are relevant when considering all paths? (3) Is there a special set of paths that allows us to 
probe the physics near a given boundary?  
In particular, we might wonder if there is only a \textit{finite} number of towers realizing the Distance Conjecture that can be systematically 
explored to characterize the boundary.
As is clear from the outset, these pivotal questions can be answered only by knowing universal features that the EFTs display close to any field space boundary.

The recent Tameness Conjecture \cite{Grimm:2021vpn} suggests a novel way to constrain field spaces, parameter spaces, and coupling functions of any EFT compatible with quantum gravity.
While motivated originally by the finiteness conditions on EFTs arising in string theory, it proposes to use a general mathematical framework, known as tame or o-minimal geometry, 
within which any effective field theory should be formulated and studied. In particular, the Tameness Conjecture notes that couplings should not be arbitrary functions of the moduli fields or parameters that enter the EFT, but rather should be sufficiently \emph{tame} functions. As will become clearer in Section~\ref{sec:tamenessconj}, this means that the couplings have to be \emph{definable} 
in a so-called o-minimal structure. Such o-minimal structures were originally introduced in the context of model theory, which is part of mathematical logic, but later found application in various areas of mathematical research. In particular, we will be interested in their relation with topology (see \cite{dries_1998} for an introductory reference on the subject).  Remarkably, assuming that the couplings are definable in an o-minimal structure imposes sets of functional constraints and allows us to address path-dependency questions arising in the Distance Conjecture.

\begin{figure}[thb]
		\centering
		\includegraphics[width=10cm]{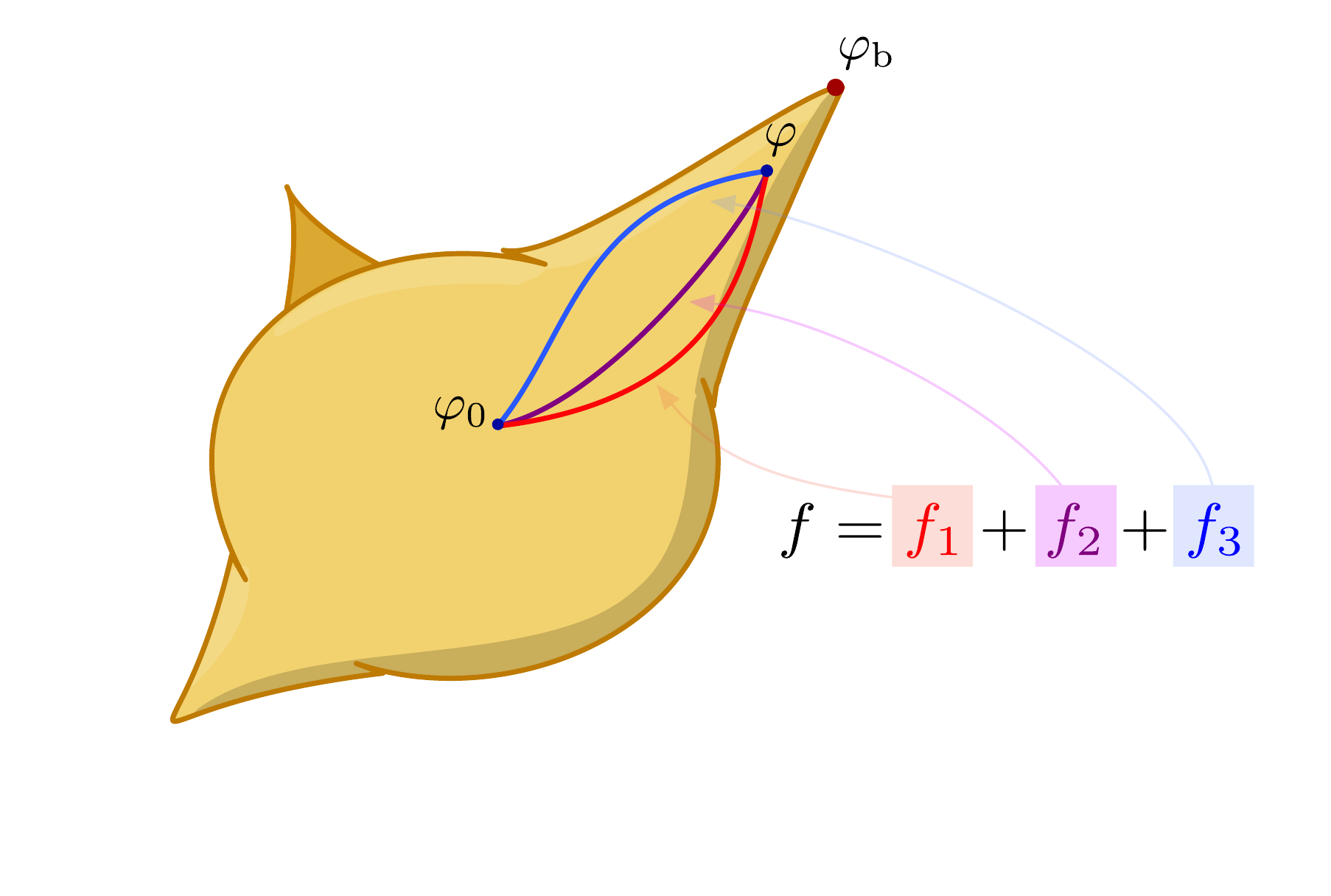}\vspace{-2.5em}
		\caption{A function $f$ may exhibit different leading asymptotic behaviors along different paths approaching the infinite field distance boundary. Here is depicted a function $f$ that displays the different leading behaviors $f_1$, $f_2$ and $f_3$ along, respectively, the red, purple and blue paths connecting a regular point $\varphi_0$ of the moduli space to the near-boundary point $\varphi$. \label{Fig:Infinite_Dist_Gen_RM_RP}}
	\end{figure}

While the tameness of EFT couplings is a general property, it becomes particularly powerful when considering their functional form in 
the near-boundary region of the moduli space. In fact, the definability in an o-minimal structure allows one to infer general properties about the growth or fall-off of the EFT couplings towards the field space boundary. As a first step, we need to specify the o-minimal structure in which the EFT couplings are defined. In \cite{Grimm:2021vpn} it was proposed that 
the o-minimal structure that is relevant for stringy EFTs is $\mathbb{R}_{\rm an,exp}$. This structure appears in many geometric applications and 
allows for defining the exponential function and all restricted analytic functions. We will introduce $\mathbb{R}_{\rm an,exp}$ in more detail below. 
In this work we will further restrict the structure in which the EFT couplings are defined, in order to give more precise statements about their growth towards the field space boundary. In fact, for concrete applications in stringy EFTs, not all functions definable in the o-minimal structure $\mathbb{R}_{\rm an,exp}$ seem to arise.  
To constrain the allowed set of functions further we will be following \cite{BKT} and introduce special families of tame functions
in $\mathbb{R}_{\rm an,exp}$, that we name \emph{monomially tamed} and \emph{polynomially tamed}. Roughly speaking, given a subregion $\mathcal{U}$ that touches the boundary of moduli space, monomially tamed functions are those which display a leading monomial behavior in the moduli fields along any path in $\mathcal{U}$ towards the field space boundary. Instead, a polynomially tamed functions in $\mathcal{U}$ are finite sums of such monomials for which one cannot single out a leading monomial behavior in $\mathcal{U}$; rather, they exhibit different leading behaviors according to the chosen path that leads to the field space boundary, as depicted in Figure~\ref{Fig:Infinite_Dist_Gen_RM_RP}. 
We will propose that in most of the concrete stringy EFTs, EFT couplings can be ascribed to these two families of functions. EFTs that stem from Type IIB ten-dimensional string theory compactified on
a Calabi-Yau threefold, eventually supplemented by an orientifold projection, provide evidence for this claim. For instance, as we will show in Section~\ref{sec:IIB}, the couplings involving the vector multiplet sector within 4D $\mathcal{N}=2$ Type IIB EFTs are fully determined by Hodge inner products. As demonstrated in \cite{BKT} that Hodge inner products are polynomially tamed 
near the boundaries of the complex structure moduli space. Consequently, the field space metric, the gauge coupling functions, the masses and physical charges of D3-particles are also polynomially or monomially tamed.

The aforementioned restricted notion of tameness of the EFT couplings has a remarkable implication: important information about the physics emerging towards the field space boundaries can be understood by only examining  what happens along certain special paths that stretch towards the boundary. To single out such paths we note that this reduction to one-dimensional slices was 
central in \cite{BKT} where it was shown that the relevant curves are linear paths with rational slopes. Following this work, we will indeed show that, 
if the EFT couplings are mutually bounded along these test paths, such bounds can be extended in a wider region of the moduli space. This feature will allow us to give a recipe to concretely test the Distance Conjecture: it is enough that the Distance Conjecture is obeyed along these special curves, and it holds in a wider region of the moduli space path independently.

Thus, such curves serve as \emph{test paths} and can be employed to diagnose pathologies that the effective description may develop. Such  linear test paths constitute a very restricted family of all the possible paths that may drive the fields towards the boundary. Can we get a phenomenological understanding of why such test paths are `special'? The curves that in \cite{BKT} have been employed to test the behavior of monomially and polynomially tamed functions can be viewed as the backreactions induced by cosmic string solutions. As shown in the seminal work \cite{cstring} and recently explored in \cite{Lanza:2020qmt,Lanza:2021udy}, along BPS cosmic string solutions the moduli fields develop a linear backreaction when regions too close to the singularity are probed. Therefore, it is enough to show that a massless tower of states emerges along all the allowed string backreactions, and we know that such a pathology is spread throughout a wider region of the moduli space. These observations fit well with the Distant Axionic String Conjecture proposed in \cite{Lanza:2020qmt,Lanza:2021udy}. This conjecture postulates that, in four-dimensional EFTs, any infinite distance limit is characterized by the appearance of a tensionless axion string, namely a string that is magnetically coupled to an axion. In \cite{Lanza:2021udy}, such a claim was checked in a large class of minimally supersymmetric examples by showing that, along the BPS backreactions of axion strings, the emergence of an infinite tower of states and consequent lowering of the EFT cutoff are always accompanied by an axion string that becomes tensionless. We will here show that these statements are enough to prove that axion strings do indeed emerge along any path that drives the moduli towards infinite distance limits.

This work is articulated as follows. In Section~\ref{sec:DC} we review the Distance Conjecture and the Tameness Conjecture, and we illustrate how the latter can help addressing some features of the former. Indeed, therein, special families of tame functions are introduced, namely the monomially and polynomially tamed functions. We will then show that, assuming that EFT couplings are described by such special tame functions, the Distance Conjecture can be rephrased in a path-independent fashion.  In Section~\ref{sec:linear_paths} we show that, provided that EFT couplings are either monomially or polynomially tamed, crucial information about the near-boundary physics can be obtained by examining how the EFT couplings behave on certain special curves. We will further show how, at the EFT level, such curves can be regarded as the backreaction of cosmic strings, which then serve as candidate objects to test the EFT couplings. In Section~\ref{sec:IIB} we provide evidence for the statements made in the previous sections. Specifically, we prove that monomially and polynomially tamed functions dictate the couplings entering the vector multiplet sector of the EFTs obtained after compactifying Type IIB string theory on Calabi-Yau three-folds. The appendices contain important mathematical results that are used throughout the main text. Indeed, Appendix~\ref{sec:rm_rp} collects some properties that the monomially and polynomially tamed functions exhibit. In Appendix~\ref{sec:proof_lemma4.5}, following~\cite{BKT}, we show how polynomially tamed functions can be bounded by monomially tamed functions by studying their behavior on curves. Finally, in Appendix~\ref{sec:proof_lemma4.7} we prove that the Hodge inner products, that dictate the couplings of the Type IIB EFTs studied in Section~\ref{sec:IIB}, grow as polynomially or monomially tamed functions.

\section{The Distance Conjecture and the Tameness Conjecture}
\label{sec:DC}

In this section we first introduce the Distance Conjecture \cite{Ooguri:2006in}. As we will see, the Distance Conjecture predicts how any effective description gets broken when approaching an infinite distance point in field 
space due to the emergence of an infinite tower of states that become light at a certain rate. However, this statement does not make precise 
assertions about either the specific path that leads to the infinite distance point or the number different towers becoming massless towards infinite distance. 
Here we will introduce a framework, the one of `tame geometry', within which both issues can be addressed in full generality.  Developing on the recently formulated Tameness Conjecture \cite{Grimm:2021vpn}, we will illustrate that, if the couplings distinguishing the effective field theory are sufficiently `tamed' (in a sense that will become clear in Section~\ref{sec:tamenessconj}), one can formulate the Distance Conjecture in a path independent fashion.

\subsection{Distance Conjecture and path dependence}
\label{sec:DC_pd}

Let us begin by recalling the original statement of the Distance Conjecture \cite{Ooguri:2006in}. Consider a $D$-dimensional 
effective theory with a number of real scalar fields $\varphi^i$ spanning a field space $\cM$. The effective field theory under consideration is 
assumed to include Einstein gravity coupled to some scalar fields $\varphi^i$, $i =1,\ldots,n$ via an action of the form
\beq
\label{DC_Sgen}
S^{(D)} = M_{\rm P}^{D-2} \int {\rm d}^Dx \sqrt{-g} \Big( \frac{1}{2} R - \frac12 G_{ij} \partial_\mu \varphi^i \partial^\mu \varphi^j + \ldots \Big) \ ,
\eeq
where $R$ denotes the $D$-dimensional Ricci scalar, $G_{ij}$ is the field space metric on $\cM$ and the dots indicate additional couplings that potentially also include the scalars $\varphi^i$.
Let us denote the shortest geodesic distance between two points $\varphi,\varphi_0 \in \cM$ by $d(\varphi,\varphi_0)$. 
The statement of the Distance Conjecture \cite{Ooguri:2006in} can be split into two parts:
\begin{itemize} 
	\item[(1)] Let $\cM$ be a moduli space of dimension at least one, i.e.~assume that there are scalar fields $\varphi^i$ that are not subjected to a scalar potential. For any point $\varphi \in \cM$ and any positive number $C$ there exists another point $\varphi_0 \in \cM$ such that $d(\varphi,\varphi_0) >C$. This implies that the space $\cM$ cannot be 
	compact and that it admits at least one boundary point $\varphi_{\rm b} \in \partial \cM$ which is at infinite distance from any point of $\cM$. 
	\item[(2)] When approaching an infinite distance point $ \varphi_{\rm b} \in \partial \cM$ with $d(\varphi,\varphi_{\rm b}) \rightarrow \infty$, there exists an infinite tower of states that becomes exponentially light. More precisely, 
	consider a $\varphi \in \cM$ for $d(\varphi,\varphi_0)$ sufficiently large, the masses of the states at $\varphi$ compared with the masses at $\varphi_0$ behave as \footnote{Notice that the the masses $M_n(\varphi)$ are assumed to be computed for canonically normalized fields. 
	}
	\beq \label{MP=MQ}
	M_n(\varphi) \sim M_n(\varphi_0) e^{-\lambda d(\varphi,\varphi_0)}\ ,
	\eeq
	with $\lambda$ an unspecified real parameter. In refined versions of this conjecture it is claimed that $\lambda$ is $\mathcal{O}(1)$ \cite{Klaewer:2016kiy,Baume:2016psm}.
\end{itemize}

\begin{figure}[thb]
	\centering
	\includegraphics[width=10cm]{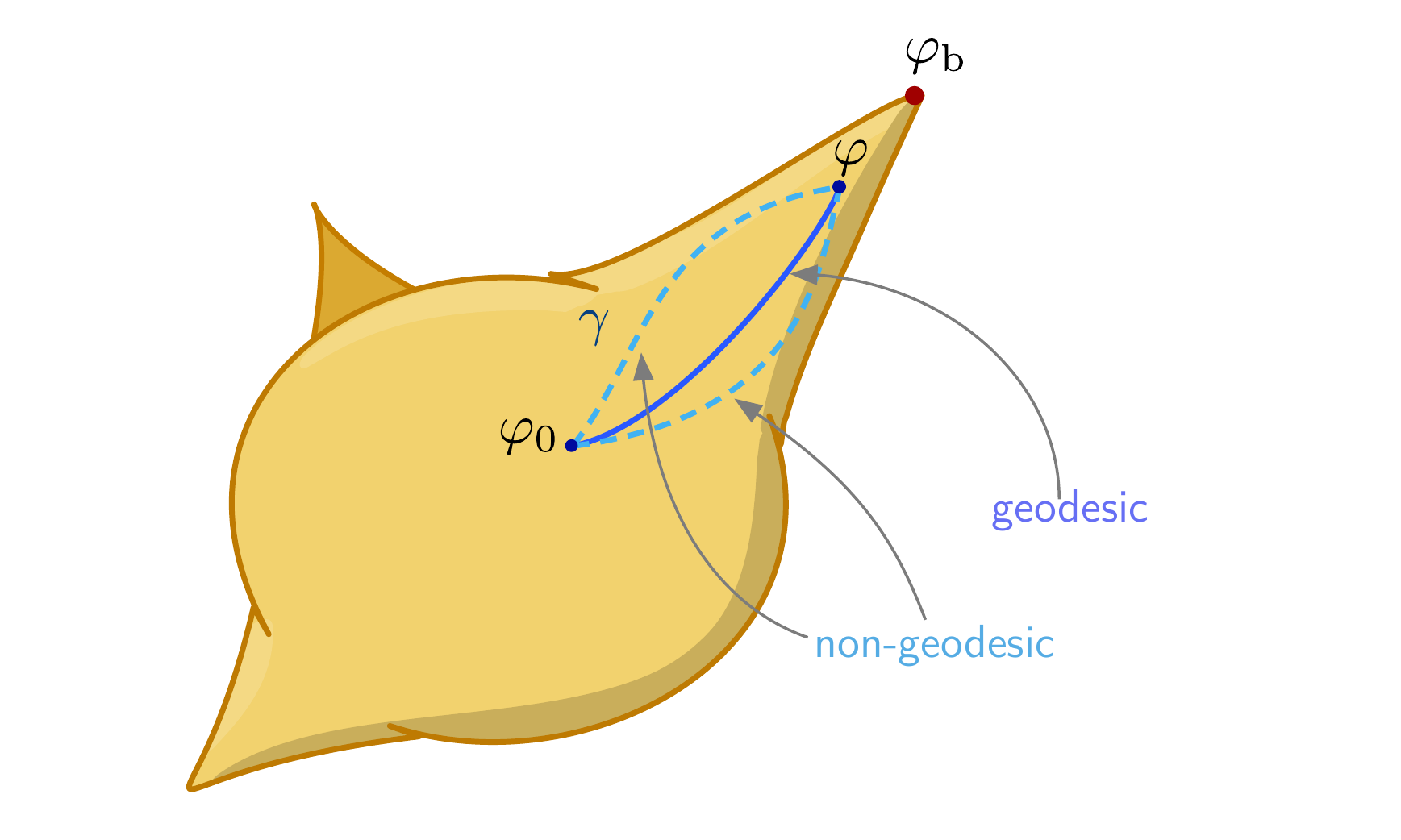}
	\caption{Starting from a point $\varphi_0$, a point $\varphi$ close to the boundary $\varphi_{\rm b}$ can be reached via a path $\gamma$ that be either a geodesic or non-geodesic. \label{Fig:Infinite_Dist}}
\end{figure}

Let us stress that the first part of the conjecture solely restricts the geometry of $\cM$, while the second part 
delivers a more precise statement about what happens to the effective theory
when approaching an infinite distance point $\varphi_{\rm b}$. According to the conjecture, any effective theory with fixed energy scale $\Lambda$
eventually breaks down near $\varphi_{\rm b}$, since at this point an infinite tower of massless degrees of freedom ought to be included.
This can be also stated by noting that there is a quantum
gravity cut-off  $\Lambda_{\rm QG}$ associated to the infinite tower of states that becomes 
exponentially small compared with $\Lambda$ when approaching $\varphi_{\rm b}$. Effective field theories are thus only 
valid for finite scalar field excursions measured by the shortest geodesic distance. 
Instead, the second part of the conjecture asserts that the masses of the states constituting an infinite tower need to fall-off exponentially in the geodesic distance. However, no information is provided about the path along which 
one approaches the boundary point $\varphi_{\rm b}$. Starting at some 
point $\varphi_0$ in the field space $\cM$ we can consider any path $\gamma: [0,1) \rightarrow \cM$ with one 
end being at $\gamma(0)=\varphi_0$ and the other end reaching towards the point $\varphi_{\rm b}$. The considered points $\varphi$ 
appearing in \eqref{MP=MQ} are, by construction, on the path $\gamma([0,1))$ sufficiently close to $\varphi_{\rm b}$ such that $d(\varphi,\varphi_0)$ is large,
as depicted in Figure~\ref{Fig:Infinite_Dist}. 
The path $\gamma$ can be very complicated and there is no a priori assertion that it has to be a geodesic. Nevertheless 
the distance $d(\varphi,\varphi_0)$ appearing in the exponential fall-off in \eqref{MP=MQ} is asserted to be the geodesic distance. 

Evidence for the Distance Conjecture has been collected by studying various effective theories arising from 
string theory (see, for example, \cite{Grimm:2018ohb,Lee:2018urn,Lee:2018spm,Lee:2019tst,Marchesano:2019ifh,Font:2019cxq,Lee:2019xtm,Grimm:2019wtx,Lee:2019wij,Grimm:2019bey,Grimm:2019ixq,Baume:2019sry,Gendler:2020dfp,Lanza:2020qmt,Klaewer:2020lfg,Ashmore:2021qdf,Brodie:2021ain}). In particular, in the study of the vector multiplet sector of 
$\cN=2$ effective theories arising from Type IIA and Type IIB on Calabi-Yau threefolds, the existence of 
a whole network of infinite distance points has been established in \cite{Grimm:2018ohb,Grimm:2018cpv} and a candidate tower of states 
has been identified for many of the associated limits. More precisely, it has been argued in 
\cite{Grimm:2018ohb,Grimm:2018cpv} that in many limits a tower of states with charges 
$q_n = T^n q_0$ can be constructed by acting with an appropriately chosen monodromy symmetry $T$ on some seed charge $ q_0$. 
It was a crucial aspect of  \cite{Grimm:2018ohb,Grimm:2018cpv} to show that the tower constructed in this way was actually satisfying the desired behavior for 
the Distance Conjecture in whole sectors of near the infinite distance point. This apparent feature is not predicted by the Distance 
Conjecture and we will explain in the remainder of this section what is the underlying reason for this local universality of the towers. 
It will turn out that it can be related to the tameness of the masses of the states and hence is fundamentally 
linked with the Tameness Conjecture \cite{Grimm:2021vpn} reviewed in subsection \ref{sec:tamenessconj}.

Any explicit check of the Distance Conjecture can be generically split into two steps: (1) 
identifying an infinite tower of states that become massless towards the infinite distance point and (2)
showing that the masses of said states fall-off as suggested by \eqref{MP=MQ}. Note that it can be 
notoriously difficult to perform these steps along any path $\varphi(s) \equiv \gamma(s)$ in $\cM$, especially when 
considering higher-dimensional field spaces $\cM$. In order to test the conjecture we need to ensure 
that along each path we retain the behavior 
\beq \label{Mass+ineq}
\frac{M_n(\varphi)}{M_n(\varphi_0)} \sim e^{-\lambda d(\varphi,\varphi_0)} \ ,\qquad \qquad  M_1(\varphi) <  M_2(\varphi) < M_3(\varphi) < \ldots ,
\eeq
where $M_n, n=1,2,...$ denote the masses of the states in the infinite tower. 
The Distance Conjecture hereby does not give an answer to the following questions:
\begin{itemize}
	\item Does one need to find a different tower for each individual path? How far can one deform a path and still use the same tower to satisfy the conjecture? 
	\item How many different towers of states are becoming massless at an infinite distance point?
\end{itemize}
It is clear that any restriction on the considered paths might lead to an incomplete picture and prevent us from 
answering these questions. As it stands, the Distance Conjecture might require us to construct a gigantic set of towers of 
states in a path-dependent way. As we will argue in Section~\ref{sec:Tame_DC} the above questions 
can be answered when additionally enforcing the Tameness Conjecture. This will open the possibility 
to only study a special set of paths and then use the resulting insights to infer information about all paths that leads to the infinite distance points. 

\begin{figure}[thb]
	\centering
	\includegraphics[width=\textwidth]{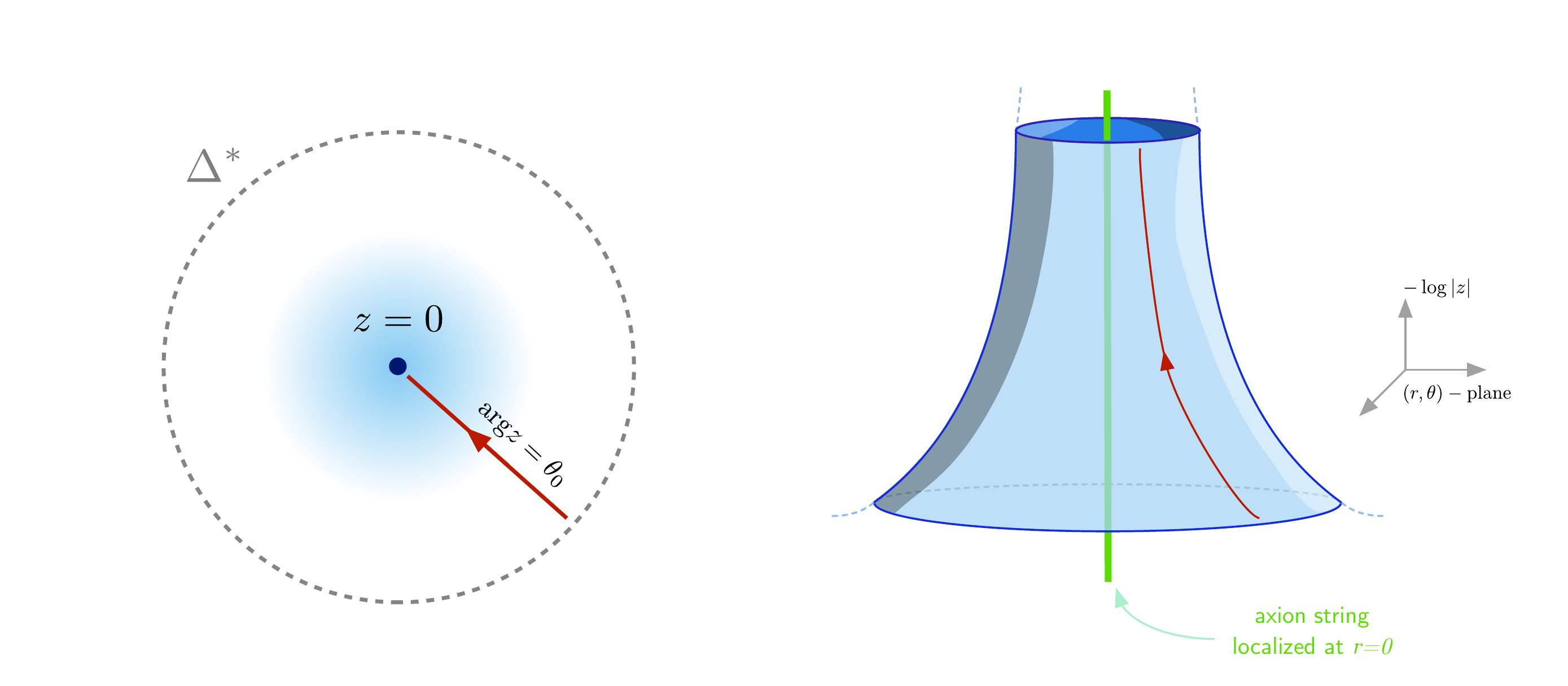}
	\caption{On the left is depicted a path with fixed arg$(z)=\theta_0$ towards $z=0$ in the punctured unit disk $\Delta^*=\{0<|z|<1 \}$. Such a path is in correspondence with the backreaction of an axion string, that is depicted on the right: the axion string, whose core is depicted in green, induces a backreaction on $z$, depicted in blue, in the $(r,\theta)$-plane transverse to the string; along the radial coordinate $r$, ${\rm Im}\,z \to 0$, as happens along the red path at fixed arg$(z)$. \label{Fig:String_HalfPlane}}
\end{figure}

There are various motivations that would lead one to only consider a special set of paths. Firstly, for practical purposes one 
could restrict to only geodesics when moving to the infinite distance point. Then can directly replace $d(\varphi_0,\varphi)$ in \eqref{Mass+ineq} 
with the length of the considered curve. 
Secondly, on a more fundamental level, the proper identification of the light states can be more apparent along 
a set of `special paths'. In particular, in the recent study of the Distant Axionic String Conjecture \cite{Lanza:2021udy} a concrete 
suggestion for the tower of states was made when considering certain linear paths as we will review in more detail in Section~\ref{sec:linear_paths}. 
In a nutshell, the conjecture suggests that each infinite distance limit can be understood by examining 
strings in the effective theory that couple to the axion-like scalar emerging in the limit. Namely, one investigates 
the backreaction of a string on the scalar fields of the effective theory. Along the radial coordinate transverse to the string, some of these fields acquire a nontrivial profile,  and they are driven to boundary in the near-core region of the string. This renders such axion strings powerful tools to investigate the physics emerging towards any infinite-distance field space boundary. 
From a field theoretical viewpoint, as stressed in \cite{Lanza:2020qmt}, the string backreaction can be regarded as an RG flow of the coupling
when changing the energy scale. 
This interpretation allows one to map the spacetime backreaction onto 
the fields of the theory as a path within the moduli space. 
Concretely, this implies the following. 
Let us consider a two-dimensional field space $\cM$ with complex coordinate $z$ and assume that the 
infinite distance point is at $z=0$ on $\partial \cM$. Locally, we can model the near boundary region 
by considering the punctured disk, i.e.~the unit disk $|z|<1$ with the center $z=0$ removed. As showed in  \cite{Lanza:2021udy} and further summarized in Section~\ref{sec:axion_strings}, axion strings generate a backreaction that can be mapped to the lines with arg$(z)=\theta_0$ in the punctured disk as depicted in Figure~\ref{Fig:String_HalfPlane}. For each line a tower of states arises from the oscillation modes of the string. While not necessarily the lightest tower, 
it gives a candidate set of states that can be used in the Distance Conjecture. It is then important to 
show that these states remain relevant away from these special paths. It will be 
one of the goals of Section~\ref{sec:linear_paths} to show that this exactly happens if an 
appropriate tameness condition is imposed on the couplings of the effective theory.

\begin{figure}[thb]
	\centering
	\includegraphics[width=\textwidth]{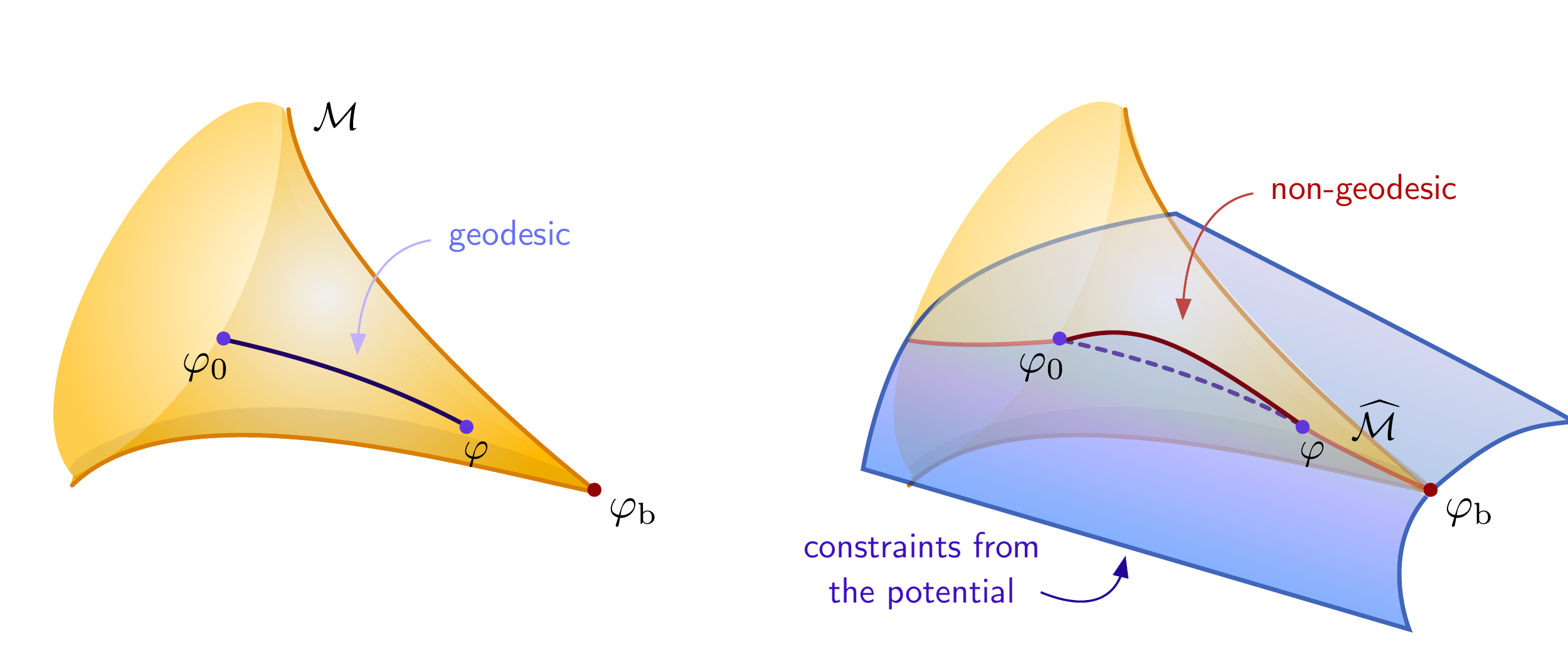}
	\caption{On the left is depicted the near-boundary region of a two-dimensional moduli space $\mathcal{M}$. On the right, the constraints introduced by the scalar potential, depicted as the blue surface, reduce the moduli space to the one-dimensional space $\widehat{\mathcal{M}}$, depicted as the red line, obtained as the intersection of the moduli space $\mathcal{M}$ and the potential constraint surface. Consider two points $\varphi_0$ and $\varphi$ in $\widehat{\mathcal{M}}$: while they can be linked via a geodesic path in $\mathcal{M}$ (depicted in blue on the left picture, and with a blue, dotted line on the right), in $\widehat{\mathcal{M}}$ they can  be linked only along the one-dimensional space $\widehat{\mathcal{M}}$, along the dark-red path, and such a path may be non-geodesic in $\mathcal{M}$. \label{Fig:M_cutoff}}
\end{figure}

Let us close this section by noticing that the original form of the conjecture assumes that $\cM$ 
is a moduli space with no scalar potentials for the $\varphi^i$. It was subsequently suggested in \cite{Klaewer:2016kiy} to apply this conjecture to 
theories with scalar potentials and much recent research \cite{Baume:2016psm,Valenzuela:2016yny,Blumenhagen:2017cxt,Grimm:2019ixq,Grimm:2020ouv,Calderon-Infante:2020dhm} has focused on clarifying this possibility. Such an application immediately leads to a puzzle, since then the notion of field space $\cM$ depends on the energy scale $\Lambda$ of the effective 
theory. Lowering the energy scale might require us to integrate out massive scalars, thereby reducing the original field space $\cM$
to a  subspace $\widehat \cM \subset \cM$. The value $d_{\widehat \cM}(\varphi,\varphi_0)$ of the shortest 
geodesic distance between $\varphi,\varphi_0 \in \widehat \cM$ will in general differ from $d_{\cM}(\varphi,\varphi_0)$ measured in $\cM$ and we always have 
\beq
d_{\cM}(\varphi,\varphi_0) \leq d_{\widehat \cM} (\varphi,\varphi_0) \ , \qquad \varphi,\varphi_0\in \widehat \cM \subset \cM\ ,
\eeq
(see Figure~\ref{Fig:M_cutoff} for a pictorial representation). 
This inequality implies that even if we assume that 
the Distance Conjecture is satisfied in $\cM$, it might well be the case that it is violated in $\widehat \cM$. To see this, imagine 
that we have found a tower of states with masses $M_n$ that become exponentially light with the geodesic distance $d_\cM$ in $\cM$ with a rate specified by \eqref{Mass+ineq}. Now, performing the computation 
in $\widehat \cM$ the fall-off of the tower should be determined by $d_{\widehat \cM}$. However, generically the masses $M_n$ of the above tower does not generically scale as $M_n \sim e^{-\lambda_n d_{\widehat \cM}}$, for $e^{-\lambda_n d_{\widehat \cM}}$ falls off faster than $e^{-\lambda_n d_{\cM}}$. In fact, for the Distance Conjecture to be satisfied in $\widehat \cM$
we have to find a new tower of states that becomes massless faster and there is no reason for this tower to coincide with the same tower of state that realizes the Distance Conjecture in $\cM$. In other words, assuming the Distance Conjecture in $\cM$ does not imply the Distance Conjecture in $\widehat \cM$, if this 
smaller field space is obtained by a general potential.

Therefore, it may then appear that the Distance Conjecture loses its predictability when the effective description displays a nontrivial potential for the moduli. However, one can address the issue from another perspective: if a scalar potential is present, the moduli fields $\varphi^i$ are constrained to follow given paths in the wider space $\cM$, which are here flat directions of the potential. It is then crucial to have an understanding of whether the Distance Conjecture holds also for such non-geodesic paths. This issue was explored on different grounds also in \cite{Calderon-Infante:2020dhm}, where it was proposed that the Distance Conjecture can be viewed as a constraint on the scalar potential: namely, the flat directions that the potential allows need to be such that the Distance Conjecture is realized along those as well. In what follows, we will see that the general tame structure of any consistent EFT allows us to investigate the realization of the Distance Conjecture also along such non-geodesic paths.

\subsection{Tameness Conjecture} \label{sec:tamenessconj}

Previously we have explained that the Distance Conjecture alone does not 
allow us to develop a very detailed picture of what happens near an infinite distance 
point. However, in the following we argue that the situation changes if we invoke another 
recent conjecture,  the Tameness Conjecture, that implements finiteness constraints into the structure of the effective theory. 
In essence, this conjecture states that all effective theories that 
can be consistently coupled to quantum gravity can be defined using `tame geometry', which is a currently very active 
field of mathematics linking geometry and logic. 
More precisely, we recap stating that effective theories that 
can be consistently coupled to quantum gravity admit the following properties:
\begin{itemize}
\item[(1)] All effective theories valid below a fixed finite energy cut-off scale 
are labelled by a definable parameter space and must have
scalar field spaces and coupling functions that are definable in an o-minimal structure.
\item[(2)] The relevant o-minimal structure is $\bbR_{\rm an,exp}$.
\end{itemize}  
Here we note that part (2) gives a strengthening of the conjecture, since it specifies an o-minimal 
structure denoted by $\bbR_{\rm an,exp}$. Note that the conjecture is very far reaching, since it asserts the 
existence of a general parameter space and constrains its properties. As we will explain momentarily, it thereby 
excludes that this parameter space can contain infinite discrete components such as a lattice. It was stressed 
in \cite{Grimm:2021vpn} that this finiteness condition is deeply linked to the coupling to gravity. 
Furthermore, the conjecture restricts valid field spaces and every coupling function varying over the parameter space 
and field space. In particular, we will apply this condition to the masses and distances appearing in the Distance Conjecture.

\begin{figure}[thb]
	\centering
	\includegraphics[width=7cm]{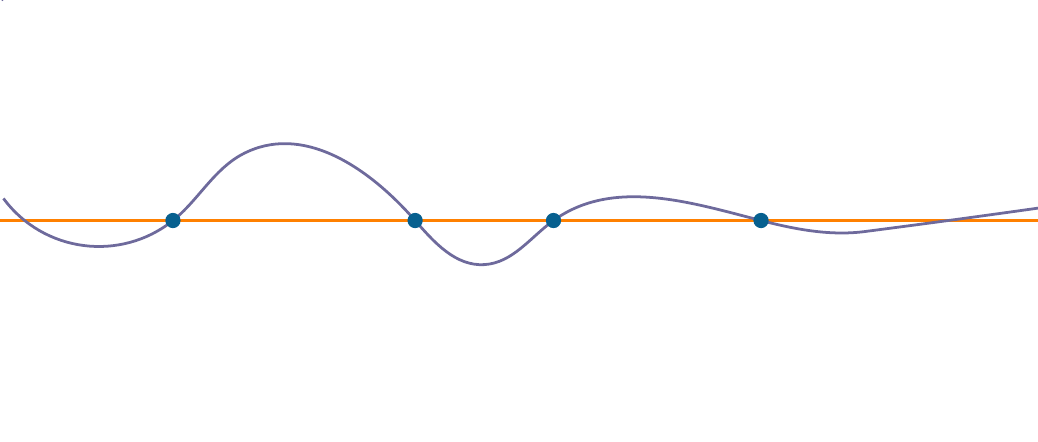} \qquad \includegraphics[width=7cm]{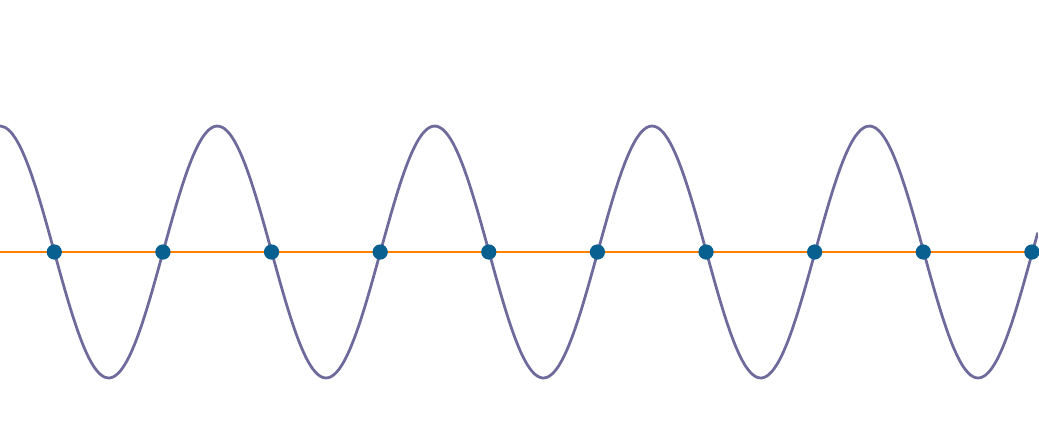}
	\caption{On the left, an example of tamed function with a finite set of zeros. On the right, an example of non-tamed function, with an infinite number of zeros. \label{Fig:Tam_examples}}
\end{figure}

To explain the conjecture, we first have to define what we mean by an 
\textit{o-minimal structure} $\cS$. The basic idea hereby is to introduce sets of subsets of $\bbR^n$ denoted by $\cS_n$, for all $n=1,2,...$, that 
form a structure and then add a tameness constraint making it into an o-minimal structure. The collection of all $\cS_n$, $n=1,2,...$ are called 
$\cS$-\textit{definable sets}, or definable sets if the o-minimal structure has been specified before.
The conditions on the sets $\cS_n$ are as follows: (1) $\cS_n$ contains the zero-set of any polynomial in $n$ variables; (2) 
$\cS_n$ is closed under finite intersections, finite unions, and complements; (3) the Cartesian product of a set in $\cS_n$ and a set in $\cS_m$ is in $\cS_{m+n}$; 
and (4) linear projections $\bbR^{n+1} \rightarrow \bbR^n$ applied to a set in $\cS_{n+1}$ give a set in $\cS_n$.   
Finally, the tameness condition is 
then stated as: 
\begin{itemize}
 \item The $\cS$-definable sets in $\bbR$ are the finite unions of points and intervals.\footnote{Note that these intervals can be closed or open and be of finite or infinite length.} 
\end{itemize}
It is a remarkable fact that this tameness property solely imposed on the subsets of the real line $\bbR$ 
constrains the space of allowed sets in all $\cS_n$ so significantly that strong finiteness properties can be inferred. 
This is rooted in the projection property, which implies that any projection to a real line of a higher-dimensional 
definable set should lead only finite unions of points and intervals. We depict a definable and a non-definable set in 
$\bbR^2$ in Fig.~\ref{Fig:Tam_examples}. It is then natural to introduce $\cS$-\textit{definable maps} $\varphi: \bbR^n \rightarrow \bbR^m$ by requiring that their graph, which is 
a subset of $\bbR^{n+m}$, is definable in $\cS_{n+m}$.  
Using $\cS$-definable sets and definable functions one can then define an $\cS$-definable topological space and a definable manifold by requiring 
that it admits definable atlas with appropriate definable transition functions \cite{dries_1998}. In fact, one can use this as a starting point 
for introducing many other geometric structures by adding the definability criterium as an additional constraint. The resulting geometry framework 
is provides that the aforementioned \textit{tame geometry}.

It is a remarkable fact that there exist multiple examples of o-minimal structures that extend the simplest 
structure, denoted by $\bbR_{\rm alg}$, generated by polynomial equations only. $\bbR_{\rm alg}$ is hereby 
obtained by collecting all zero-sets $P(x_1,...,x_n)=0$ in $\bbR^n$ and completing this set by including their 
unions, complements, intersections, and projections. The resulting set of semi-algebraic sets is the smallest 
o-minimal structure. The strategy to find non-minimal examples is to carefully extend 
the set of functions that is used to define the definable sets. There are two o-minimal structures 
that will be of importance in this work and we will introduce in some detail in the following. 

\noindent
\textbf{O-minimal structure $\pmb{\bbR}_{\textbf{an}}$}: This o-minimal structure is obtained by extending $\bbR_{\rm alg}$ by also considering zero-sets of so-called restricted analytic 
functions. More precisely, we also include subsets of $\bbR^n$ defined by the equations $P(x_1,...,x_n,f_1,...,f_n)=0$, where $P$ is a polynomial and $f_i(x_1,...,x_n)$ are restricted analytic functions.
 Before defining restricted analytic functions, let us recall that an analytic function defined on a domain is a function that coincides with its own Taylor series on that domain. Analytic functions are necessarily smooth, but the converse is not true.
Roughly speaking, restricted analytic functions are restrictions of analytic functions to smaller domains. More precisely,  such functions are all restrictions $f|_{B(R)}$ of 
functions $f$ that are analytic on a ball $B(R_0)$ of finite radius $R_0$ to a ball $B(R)$ of strictly smaller
radius $R < R_0$. Applying this definition to the punctured disk $\Delta^*$ introduced in Fig.~\ref{Fig:String_HalfPlane}, we realize that a restricted analytic function $f$
is more constrained than an analytic function when examining its behavior near the puncture, i.e.~the point $z=0$. By definition, such an $f$ needs to come from a function that is also 
analytic at the puncture and hence implies that $f$ cannot `go wild' when approaching the puncture. As an example of an analytic but not restricted analytic function consider $f(z) = 1/z$ over the punctured disk $\Delta^*$. This function is analytic over $\Delta^*$, but its singularity at $z = 0$ forbids it being restricted analytic over $\Delta^*$. Note, however, that $1/z$ is still definable in $\bbR_{\rm an}$, since its graph can be given by an algebraic equation.

\noindent
\textbf{O-minimal structure} $\pmb{\bbR}_{\textbf{an,exp}}$: This o-minimal structure will be central for the discussion of the upcoming sections. It is obtained by further extending the former ${\bbR}_{\rm an}$ so as to include real exponentials. Specifically, ${\bbR}_{\rm{an,exp}}$ can be understood as the subsets of $\bbR^n$ described by the equation $P (x_1, \ldots , x_n, f_1, \ldots , f_m, e^{x_1}, \ldots, e^{x_n}) = 0$, where, as above, $P$ is a polynomial and $f_i(x_1, ..., x_n)$ are restricted analytic functions.

It is worth stressing that the choice of the domain is crucial in order to correctly tell whether the function is definable in a given o-minimal structure. The complex exponential $e^z: \mathbb{C} \to \mathbb{C}$ is not definable in general. In fact, rewriting $e^z = e^{r+ 2\pi {\bf i}\phi} = e^r (\cos (2\pi \phi) + {\bf i} \sin (2\pi \phi))$, $e^z$ has an infinite, discrete set of zeros for $r, \phi \in \mathbb{R}$. Only if we further reduce the domain of $\phi$, $e^z$ is definable. In most of the applications below, $\phi$ is an axion, whose fundamental domain is bounded, say $0 + \varepsilon < \phi < 1 - \varepsilon$, with $\varepsilon \ll 1$. Under such an assumption, $e^z$ is definable in ${\bbR}_{\rm{an,exp}}$.

Let us now explicitly see how the Tameness Conjecture reflects on the couplings that effective field theories are allowed to display. To this end, let us consider a generic effective field theory, focusing on the bosonic spectrum only, valid up to the energy cutoff $\Lambda_{\text{\tiny{EFT}}}$ and coupled to Einstein gravity. We assume that the EFT is endowed with:
\begin{itemize}
	\item a set of real parameters $\lambda^\kappa$, $\kappa = 1\, \ldots, k$, spanning a subset of $\mathbb{R}^k$; in concrete EFT models obtained from compactifying a higher-dimensional string theory, such parameters may stem either from compactification data, or from the integration of fields more massive than $\Lambda_{\text{\tiny{EFT}}}$;
	\item a set of real scalar fields $\varphi^{a}$, $a = 1, \ldots, m$, locally parametrizing a (pseudo-)moduli space $\mathcal{M}_\lambda$, fibered over the parameter space; thus, the local field space metric $G_{ab}(\varphi,\lambda)$ generically depends not only on the fields $\varphi^a$, but also on the parameters $\lambda^\kappa$;
	\item a set of abelian $p$-form gauge fields $A^\mathcal{I}_{p_{\mathcal{I}}}$, $\mathcal{I} = 1, \ldots, N$, with field strengths $F^\mathcal{I}_{p_{\mathcal{I}}+1} = {\rm d}A^\mathcal{I}_{p_{\mathcal{I}}}$; the associated gauge kinetic function $f_{\mathcal{I}\mathcal{J}}(\varphi,\lambda)$ is allowed to depend on both the moduli fields $\varphi^a$ and the parameters $\lambda^\kappa$.
\end{itemize}
A prototypical, generic $D$-dimensional action built out of these ingredients would include the following terms\footnote{Notice that, in our conventions, a $p$-form gauge field $A_p$ has zero mass dimensions.}
\begin{equation}
	\label{Defin_Action}
	\begin{aligned}
	S^{(D)} &= \int \Big( \frac{1}2 M^{D-2}_{\rm P} R *1 - \frac12 M^{D-2}_{\rm P} G_{ab}(\varphi,\lambda) {\rm d} \varphi^a \wedge * {\rm d} \varphi^b 
	\\
	&\qquad\qquad - M^{D-2(p_\mathcal{I}+1)}_{\rm P} f_{\mathcal{I}\mathcal{J}}(\varphi,\lambda) F^\mathcal{I}_{p_{\mathcal{I}}+1} \wedge * F^\mathcal{J}_{p_{\mathcal{J}}+1} - V(\varphi,\lambda) + \ldots \Big)\,.
\end{aligned}
\end{equation}
Here, $V(\varphi,\lambda)$ denotes the scalar potential which the moduli fields $\varphi^a$ are subjected to, and eventually higher-derivatives terms can be included to the action. Moreover, the action \eqref{Defin_Action} is not required to be supersymmetric. Then, the Tameness Conjecture is a restriction on all the couplings entering \eqref{Defin_Action} -- i.e. the field space metric $G_{ab}(\varphi,\lambda)$, the gauge kinetic matrix $f_{\mathcal{I}\mathcal{J}}(\varphi,\lambda)$, the scalar potential $V(\varphi,\lambda)$, etc. 

Concretely, let us consider a generic coupling $g$. Here we neglect additional structures dressing the coupling, namely we assume the coupling $g$ to be a scalar coupling, without any index. Then, the coupling can be regarded as map $g(\varphi,\lambda)$, with domain in $\mathbb{R}^{m+k}$ and values in $\mathbb{R}$. As stated above, requiring that the coupling is definable in a given o-minimal structure implies that the graph of the function $g(\varphi,\lambda)$, namely the set of points $(\varphi,\lambda, g(\varphi,\lambda))$, has to be definable in $\mathcal{S}_{m+k+1}$. According to the strong version of the Tameness Conjecture, the o-minimal structure in which the EFT couplings ought to be defined is $\mathbb{R}_{\rm{an,exp}}$. Therefore, according to the explanation above, the coupling $g(\varphi,\lambda)$ can be understood as originating from a locus built as follows. We first introduce a set of an arbitrary number of \emph{auxiliary} variables, $x_q$, $q=1,\ldots,l$, which do not enter the EFT either as couplings or fields, and we assume that, at a first stage, the coupling $g$ is a function of these auxiliary variables as well, $g(\varphi,\lambda,x)$. Then, the coupling $g$ originates from the locus:
\begin{equation}
	\label{Defin_Loci}
	\begin{aligned}
		\exists\; x_1, \ldots, x_l: \qquad &P_i(\varphi,\lambda,x,g, f_1, \ldots, f_m, e^\varphi, e^\lambda, e^x, e^g)=0\,,
		\\
		&Q_j(\varphi,\lambda,x,g, f_1, \ldots, f_m, e^\varphi, e^\lambda, e^x, e^g)>0\,,
	\end{aligned}
\end{equation}
where $P_i$, $Q_j$ are polynomials and $f_i(\varphi,\lambda,x,g)$ are restricted analytic functions in the variables $(\varphi^a,\lambda^\kappa, x_q,g(\varphi,\lambda,x))$. The dependence of the coupling $g(\varphi,\lambda,x)$ on the auxiliary variables $x_q$, may be necessary in order to identify the locus \eqref{Defin_Loci}. However, by solving the defining equations of the locus \eqref{Defin_Loci} for $x_q$, one can eliminate the explicit dependence of $g$ on $x_i$. We stress that the definition of the locus \eqref{Defin_Loci} is, in principle, sensitive to change of parametrizations of the fields $\varphi^a$ and the parameters $\lambda^\kappa$. For instance, redefining the fields as $\varphi^a = e^{t^a}$, the locus \eqref{Defin_Loci} is not described by polynomials in $t^a$. Thus, we shall assume that there exists an appropriate parameterization $(\varphi,\lambda)$ of a patch within $\cM_\lambda$ such that the coupling $g$ can be computed as the locus \eqref{Defin_Loci}. Moreover, it is worth noticing that the finiteness of the constraints and the form of \eqref{Defin_Loci} needed to specify a coupling definable in $\mathbb{R}_{\rm an,exp}$ is a deep mathematical result that is related to model completeness \cite{10.2307/2152916}. 

Let us motivate the appearance of $e^g$ and the auxiliary variable $x$ in \eqref{Defin_Loci} via two simple examples. First, take the logarithmic function $g(\varphi) = \log(\varphi)$. It is straightforward to check that the graph of the logarithm is $\Ranexp$-definable, which can also be written as the vanishing locus of the polynomial $P(\varphi, e^g) = \varphi - e^g$ with the help of $e^g$. To see the necessity of the auxiliary variables, take the double exponential function $g(\varphi) = \exp(\exp(\varphi))$. In order to show the definability of the double exponential, denote $\Gamma(\exp)$ the graph of the exponential function, and note that the graph of the double exponential function is simply a coordinate projection of the definable set $(\bbR \times \Gamma(\exp)) \cap (\Gamma(\exp) \times \bbR)$, hence the double exponential is $\Ranexp$-definable. This fact is expressed in \eqref{Defin_Loci} by introducing an auxiliary variable $x$, and considering the locus of the polynomials $P_1(g, e^x) = g - e^x, P_2(x, e^\varphi) = x - e^\varphi$. It is worth noting that in our first example, the (global) logarithmic function does not belong to the class of restricted analytic or exponential functions that define the $\Ranexp$-structure. In fact, it can be shown\cite[Corollary~4.7]{MR1289495} that every $\Ranexp$-definable function can be written piece-wisely as polynomials of compositions of restricted analytic, exponential\footnote{It is worth mentioning that by \cite[(4.9)]{MR1289495}, the exponential function can be replaced by compositions of the reciprocal, and $n$-th roots for all positive integer $n$.}, and logarithmic functions.

Examples of couplings that are definable $\mathbb{R}_{\rm an,exp}$ include polynomials in either the $\varphi^a$ and the parameters $\lambda^\kappa$. However, also exponential functions of the form $e^{p(\varphi,\lambda)}$, where $p(\varphi,\lambda)$ is a generic polynomial in $(\varphi^a,\lambda^\kappa)$, are definable in $\mathbb{R}_{\rm an,exp}$. Instead, the Tameness Conjecture hinders the appearance of \emph{any} periodic coupling defined in $\mathbb{R}^n$. Other notable examples of functions definable in ${\bbR}_{\rm{an,exp}}$ include the period integrals of Calabi-Yau manifolds. This important result, proved in \cite{BKT}, allows for delivering nontrivial evidences of the aforementioned strong version of the Tameness Conjecture, as we will see in section~\ref{sec:IIB}. Therein we will investigate the four-dimensional EFTs obtained after compactifying Type IIB string theory on a Calabi-Yau manifold, or an orientifold projection thereof. In the former case, the couplings of the vector multiplet sectors are fully determined by the period integrals, thus residing in ${\bbR}_{\rm{an,exp}}$ \cite{Grimm:2021vpn}; in the latter case, the couplings of a chiral multiplet sector are definable in ${\bbR}_{\rm{an,exp}}$, for they are also specified by the periods only.

\subsection{A special class of tame functions}
\label{sec:Tame_bound}

As we have seen, the Tameness Conjecture constrains the form of the couplings that appear in any effective field theory that is consistent with quantum gravity. However, the constraints imposed appear to be rather mild, and the loci \eqref{Defin_Loci} lead to a great variety of couplings.
However, in concrete stringy EFTs we typically face couplings of a more restricted form than the ones delivered by the loci \eqref{Defin_Loci}. 

In order to better characterize the couplings, we first need to specify the region of $\cM_\lambda$ in which the EFT is defined. The effective description under investigation is assumed to be well-controlled: for instance, one can assume that the EFT is defined in regimes of small string coupling, or large internal volume. Typically, the corners in which the EFT is well-defined are near-boundary regions of $\cM_\lambda$, for some given values of the parameters $\lambda^\kappa$. We will denote $\cE$ such an asymptotic field space region in (the cover of) $\cM$, and we assume that it can be parametrized by a set of $m$ real local coordinates $\varphi^a$. Within $\cE$ the real fields $\varphi^a$ span different domains, and it is convenient to split them accordingly. Thus, we introduce 
real local coordinates $s^i>1$ and $\phi^\alpha$, such that the field space boundary is 
at $s^1 , ..., s^n \rightarrow \infty$. The residual $m-n$ fields $\phi^\alpha$ are assumed to stay bounded $|\phi^\alpha| < \delta$.  In other words, we identify the asymptotic region as described by the following
\begin{equation} \label{def-cE}
	\mathcal{E} = \{ |\phi^\alpha| < \delta,\ \  s^1 ,  s^2 , \ldots , s^n > 1 \}\, .
\end{equation}

Loosely speaking, the fields $s^i$ spanning a non-compact domain may be assumed to be those which tell information about the validity of the EFT: as the boundary $s^i \to \infty$ is approached, the corrections that would modify the effective description become more and more negligible. On the other hand, the fields $\phi^\alpha$ lying in a compact domain are not expected to make the EFT depart from its regime of validity for any value $|\phi^\alpha| < \delta$. To get a feeling for this parametrization, let us mention a few well-known examples. The simplest case 
is a circle compactification in which we have one real field $s$ parametrizing the radius of the circle. The decompactification 
limit $s\rightarrow \infty$ is in the open interval $\cE=\{s>1\}$. A more involved example is provided by the K\"ahler moduli 
space of some Calabi-Yau manifold. In this case $s^i,\phi^\alpha$ can be identified with the set of K\"ahler moduli arising 
by expanding the K\"ahler form as $J = s^i \omega_i + \phi^\alpha \omega_\alpha$. If one considers the complexified
K\"ahler form with the NS-NS B-field, the new scalars will be part of the $\phi^\alpha$.
While these examples might be 
instructive, we note that the following discussion is general.

Now consider an EFT defined in the domain \eqref{def-cE}, and consider any coupling $g(s,\phi,\lambda)$ that the EFT is endowed with. The dependence of the coupling $g(s,\phi,\lambda)$ on the fields $(s,\phi)$ is expected to come from two different contributions, namely
\begin{equation}
	\label{Tame_g}
	g(s,\phi, \lambda) = g_{\text{pert}}(s,\phi,\lambda) + g_{\text{non-pert}}(s, e^{-s}, \phi, \lambda)\,.
\end{equation}
Here $g_{\text{pert}}(s,\phi,\lambda)$ encapsulates the perturbative contributions to $g(s,\phi, \lambda)$ and, in the chosen parametrization, $g_{\text{pert}}(s,\phi,\lambda)$ is expected to behave as a rational function in the fields $s^i$ for large $s^i$. Instead, $g_{\text{non-pert}}(s, e^{-s}, \phi, \lambda)$ collects all the non-perturbative corrections to the coupling and, as such, $g_{\text{non-pert}}$ is exponentially suppressed in the non-compact fields $s^i$. Couplings such as \eqref{Tame_g} are endowed with a crucial property: \emph{given the subregion of the moduli space \eqref{def-cE} and fixed parameters $\lambda^\kappa$, any coupling with the structure \eqref{Tame_g} can be upper bounded by a monomial in the fields $s^i$}. These special class of tame functions, which are definable in $\mathbb{R}_{\rm an,exp}$, are indeed enough to fully characterize the couplings of most of the known stringy EFTs. Following \cite{BKT}, these special tame functions can be split into two families, that we call \emph{monomially tamed} and \emph{polynomially tamed} functions.\footnote{Notice that in \cite{BKT} these families of tame functions are referred to as \emph{roughly monomial} and \emph{roughly polynomial}, respectively.}
Here, we just limit ourselves to their (loose) definition and main properties, and we refer to Appendix~\ref{sec:rm_rp} for additional details:
\begin{itemize}
	\item \textbf{Monomially tamed functions:} We consider a function $g: \mathcal{U} \rightarrow \bbR$, with $\mathcal{U} \subseteq \mathcal{E}$ a given subset of the asymptotic region \eqref{def-cE}. We 
	require that on $\mathcal{U}$ the coupling $g$ can be written as
	\begin{equation} \label{g-expansion}
		g(s,\phi) = \sum_{ \mathbf{m}}\rho_{\mathbf{m}}(e^{-s^i},\phi^\alpha)   (s^1)^{m_1} \cdots (s^n)^{m_n}\, , 
	\end{equation}
	where $\rho_{\mathbf{m}}$ are coefficient functions that we require to be restricted analytic on $\mathcal{U}$ and
	are labelled by a multi-index $\mathbf{m}=(m_1,...,m_n)$ with integer $m_i$. This expansion 
	is restricted further by requiring that $g$ behaves as
	\beq \label{g-isroughlymono}
	g(s) \sim (s^1)^{k_1} \cdots (s^n)^{k_n} \qquad \text{on} \quad \mathcal{U}\ , 
	\eeq
	for some integers $k_1,\ldots,k_n$.
	Here and in the following the symbol $\sim$ should be read as a boundedness statement, i.e.~we write $f \sim h$ is there exists $C_1,C_2$, such that $C_1 h < f < C_2 h$, with $C_1$, $C_2$ real numbers having the same sign.
	In particular, the condition \eqref{g-isroughlymono} implies that near the infinite distance point there is leading monomial term in the expansion \eqref{g-expansion} for every 
	path $s^i \rightarrow \infty$. Thus, clearly, \eqref{g-isroughlymono} gives a strong constraint on the functional form of $g$: in fact, the restricted analytic functions $\rho_{\mathbf{m}}(e^{-s^i},\phi^\kappa)$ that appear in the expansion \eqref{g-expansion} need to guarantee that the double bound $ C_1 (s^1)^{k_1} \cdots (s^n)^{k_n} < g < C_2 (s^1)^{k_1} \cdots (s^n)^{k_n}$, for some $C_1$, $C_2$ has to hold throughout the region $\mathcal{U}$.
	
	It is worth remarking that one could extend the definition of monomially tamed function to a tame function obeying the more general
	\beq \label{g-isroughlymonob}
	g(s) \sim (s^1)^{\alpha_1} \cdots (s^n)^{\alpha_n} \qquad \text{on} \quad \mathcal{U}\ , 
	\eeq
	with $\alpha_1,\ldots,\alpha_n \in \mathbb{R}$. However, \eqref{g-isroughlymonob} may be reduced to \eqref{g-expansion} by redefining $s^i \to (s^i)^\frac{k_i}{\alpha_i}$.
	
	\item \textbf{Polynomially tamed functions:} Similarly to a monomially tamed function, a polynomially tamed function $f: \mathcal{U} \rightarrow \bbR$ exhibits the finite expansion
	\begin{equation} \label{f-expansion}
		f(s,\phi) = \sum_{ \mathbf{m}}\rho_{\mathbf{m}}(e^{-s^i},\phi^\alpha)   (s^1)^{m_1} \cdots (s^n)^{m_n}\, , 
	\end{equation}
	with $\rho_{\mathbf{m}}(e^{-s^i},\phi^\alpha)$ restricted analytic. However, in contrast with the monomially tamed functions, \eqref{g-isroughlymono} does not hold generically in $\mathcal{U}$. Thus, being less constrained, polynomially tamed functions are more general than the monomially tamed ones. Indeed, due to the properties of the restricted analytic functions (see Appendix~\ref{app:characterization_m_and_p_tamed_functions}), one can at most upper bound a polynomially tamed function by a monomial as
	\begin{equation} \label{f-constr}
		f(s,\phi)  \prec  (s^1)^{N_1} \cdots (s^n)^{N_n}\, , 
	\end{equation}
	for some set of integers $N_1, \ldots, N_n$.
	Here we have introduced the symbol `$\prec$', that has to be understood as follows: given two positive definite functions $f$ and $g$, $f \prec g$ if there exists a real, positive constant $C$ such that $f < C g$ for every point in $\mathcal{U}$. It is however worth stressing that the bound \eqref{f-constr} might be too rough, and in Section~\ref{sec:roughlylinear} we will provide a recipe in order to refine \eqref{f-constr}.
\end{itemize}

Therefore, the most evident difference between monomially and polynomially tamed functions is in their growth properties within the subset $\mathcal{U}$. On the one hand, owing to \eqref{g-isroughlymono}, monomially tamed functions display a definite growth in $\mathcal{U}$, fixed by the set of integers $k_1,\ldots,k_n$. Said differently, the leading growth of monomially tamed functions on \emph{every} path leading to the field space boundary is univocally fixed by the single set of integers $k_1,\ldots,k_n$. On the other hand, a polynomially tamed function does not exhibit a clear leading term in the whole $\mathcal{U}$. Only at most in some subsets of $\mathcal{U}$, one can single out a leading monomial term in the polynomially tamed expansions, and there are cases where such a leading monomial term does not exist. Thus, as a result, the growth of a polynomially tamed function is generically path dependent within $\mathcal{U}$. Moreover, it is worth stressing that whether a function is monomially tamed may depend on the choice of the domain $\mathcal{U}$: as an example, considering a smaller subset $\hat{\mathcal{U}} \subset \mathcal{U}$, a polynomially tamed function might become monomially tamed, as
\begin{equation}
	\label{f-rphatu}
	f(s,\phi) \sim  (s^1)^{N_1} \cdots (s^n)^{N_n} \qquad \text{in $\hat{\mathcal{U}} \subset \mathcal{U}$}.
\end{equation}
Thus, in such a case, the polynomially tamed function exhibit a definite behavior in the smaller subset $\hat{\mathcal{U}}$. It is worth stressing however, that a subset $\hat{\mathcal{U}} \subset \mathcal{U}$ such that a polynomially tamed function reduces to a monomially tamed one may not exist. In fact, if $f \prec  (s^1)^{N_1} \cdots (s^n)^{N_n}$ holds strictly everywhere in $\mathcal{U}$, its leading growth cannot be monomial. Therefore, whenever a leading growth can be identified within a subset $\hat{\mathcal{U}} \subset \mathcal{U}$, it has to be of the form
\begin{equation}
	\label{f-rphatub}
	f(s,\phi) \sim  \rho(e^{-s^i},\phi^\alpha)  (s^1)^{N_1} \cdots (s^n)^{N_n} \qquad \text{in }\ \hat{\mathcal{U}} \subset \mathcal{U}\,,
\end{equation}
with $\rho(e^{-s^i},\phi^\alpha)$ a restricted analytic function. An example of one-variable polynomially tamed function that is not monomially tamed is $f(s) = e^{-s} s$ defined over $\cE = \{s>1\}$: while $f \prec s$ in $\cE$, $f \sim s$ does not hold.

\begin{figure}[thb]
	\centering
	\includegraphics[width=12cm]{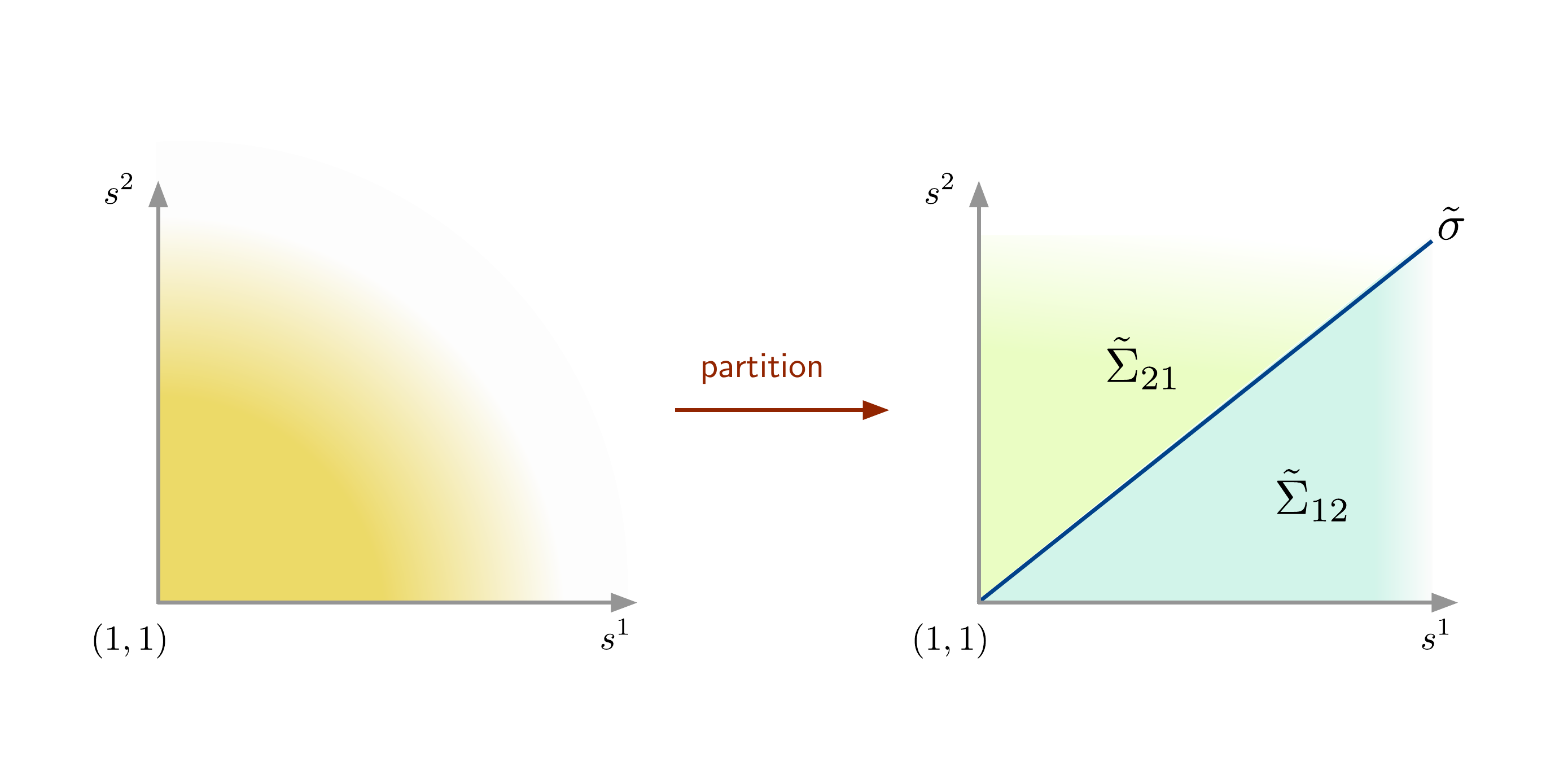}
	\caption{An example of moduli space $\mathcal{M}$ parametrized by two fields $s^1$ and $s^2$. The region close to the boundary point $\phi_{\rm b}$, reached as $s^1, s^2 \to \infty$, can be covered by two sets, $\tilde\Sigma_{12} = \{s^1 > s^2 >1 \}$ and $\tilde\Sigma_{21} = \{s^2 > s^1 >1 \}$, and the locus $\tilde{\sigma} =\{s^1 = s^2>1\}$. 
	\label{Fig:Infinite_Dist_Sectors}}
\end{figure}

Indeed, the behavior of any given polynomially tamed function might be a guiding principle in order to better characterize the region $\mathcal{U}$. Consider a general polynomially tamed function $f$. Here, for the sake of generality, we shall assume that the polynomially tamed function is defined over the full asymptotic region $\mathcal{E}$ defined in \eqref{def-cE}. Our aim is to \emph{partition} the asymptotic region in smaller subsets where the function $f$ exhibits simple behaviors. First, recall that a partition of a set $\mathcal{E}$ consists of a collection of disjoint subsets $\mathcal{U}_{\mathsf{A}} \subs \cE$, with $\mathsf{A} = 1, \ldots, N$, which covers $\mathcal{E}$:
\begin{equation}
	\label{def-partU}
	\bigcup\limits_{\mathsf{A}}\, \mathcal{U}_{\mathsf{A}} = \mathcal{E}\,, \quad\textrm{and}\quad \mathcal{U}_{\mathsf{A}} \cap \mathcal{U}_{\mathsf{B}} = \varnothing\, , \textrm{ if } {\mathsf{A}}\neq {\mathsf{B}}\,.
\end{equation}
As a simple example of partition of the asymptotic region, one can cover the subsets of $\cE$ as follows. We define the subsets
\begin{equation} \label{def-tildeSigmaI}
	\tilde\Sigma_I = \{ |\phi^\alpha| < \delta,\ \  s^1 >  s^2 >\cdots > s^n > 1 \}\, ,
\end{equation}
where the index $I$ labels all permutations of the $s^i$ in the hierarchy and we have only displayed the 
simplest permutation with $I = (1, 2, \ldots, n)$. Picking a region \eqref{def-tildeSigmaI} singles out a specific ordering for the sizes of the fields 
$s^i$ (see Figure~\ref{Fig:Infinite_Dist_Sectors} for a pictorial, two-dimensional representation representation). Then, a partition of $\cE$ is realized as the union of all the sets \eqref{def-tildeSigmaI} and the $(2^{n - 1} - 1)$ loci $\{ s^1 = s^2 > s^3 > \cdots > s^n > 1 \}, \ldots, \{s^1 = s^2 = \cdots = s^n > 1\}$.
Generally, how the partition of $\cE$ is carried is arbitrary. However, we choose to partition $\cE$ in compliance with the \emph{monotonicity theorem}\footnote{Strictly speaking, we should be using the regular cell decomposition in \cite[Exercise~(2.19)]{dries_1998}. We abuse the terminology and call it monotonicity theorem to emphasize the monotonic property of the function $f$ on each of the subset $\mathcal{U}_{\mathsf{A}}$.} (see, for instance, \cite[Chapter~3]{dries_1998}). The theorem states the following: for each definable function $f : \cE \to \mathbb{R}$, there is a partition of $\cE$ in finite subsets $\mathcal{U}_A$ such that the restriction $f |_{\mathcal{U}_{\mathsf{A}}}$, for each $s^i$, is either strictly increasing, strictly decreasing or constant.\footnote{Recall that a function $f$ is strictly increasing (decreasing) for a given field direction $s^i$ if, for any given $s^i_{(1)} < s^i_{(2)}$, $f(s^i_{(1)}) < f(s^i_{(2)})$ ($f(s^i_{(1)}) > f(s^i_{(2)})$).} 

While given a polynomially tamed function one can always find a partition such that, on each of its subsets, the function behaves as predicted by the monotonicity theorem, it is however important to stress that this does not solve the path dependence issue that we raised above. In fact, albeit on each subset $\mathcal{U}_{\mathsf{A}}$, a polynomially tamed function has a definite growth behavior, it does not necessarily display a leading term. We will then assume that, beyond the partition above, a given polynomially tamed function $f$ may induce a finer partition $\{\hat{\mathcal{U}}_{\mathsf{A}}\}$ such that, on each of these subsets $\hat{\mathcal{U}}_{\mathsf{A}}$, $f$ has a definite leading behavior:
\begin{equation}
	\label{f-rphatuc}
	f(s,\phi) \sim  \rho(e^{-s^i},\phi^\alpha)  (s^1)^{N_1} \cdots (s^n)^{N_n} \quad \text{in }\ \hat{\mathcal{U}}_{\mathsf{A}} \subset \mathcal{U}\,.
\end{equation}
Clearly, if $\rho(e^{-s^i},\phi^\alpha)  \sim 1$ on $\hat{\mathcal{U}}_{\mathsf{A}}$, the restriction $f|_{\hat{\mathcal{U}}_{\mathsf{A}}}$ is monomially tamed. The leading behavior \eqref{f-rphatuc} can be one of the terms appearing in the general expansion \eqref{f-expansion}; as such, also this finer partition is expected to be composed by a finite number of subsets.

\subsection{Taming the Distance Conjecture}
\label{sec:Tame_DC}

In light of the newly introduced terminology and characterization of the asymptotic region, let us now come back to the original questions that we posed about the Distance Conjecture. Similarly to 
the strategy in the Tameness Conjecture, our basic idea is to 
single out a set of functions that can occur in the Distance Conjecture and then give us a more complete picture 
of the physics in the infinite distance limit. In particular, in this section, we will assume that the Distance Conjecture holds in the near-boundary region of the field space, and we will investigate how  the EFT obstructions predicted by the Distance Conjecture may occur in an EFT that is characterized by tame couplings. 
As discussed in the previous section, it is natural to reduce the class of functions drawn from the o-minimal structure $\mathbb{R}_{\rm an,exp}$ in the study of concrete stringy EFTs. Indeed, in relation to the Distance conjecture we will now make concrete assertions about the functional dependence of the geodesic distance $d(s,\phi)$ (here and below we suppress the initial point $s_0,\phi_0$) and the masses $M_n(s,\phi)$, that make use of the monomially and polynomially tamed functions only. We will split the discussion in the three main issues that concur in the realization of the Distance Conjecture. Note that we will initially not restrict to any supersymmetric setting 
and therefore the following discussion might look rather complicated at first. 

\noindent\textbf{Reducing path-dependency of the infinite tower of states.}  The first ingredient that we need in order to realize the Distance Conjecture is an infinite tower of states, with masses $M_n$, that become massless as the infinite distance singularity is approached. Since the masses $M_n$ are physical couplings we require that they are tame. In particular, employing the arguments of Section \ref{sec:Tame_bound}, we require that
\beq
\label{DC_hyp1}
M_n(s,\phi)\ \text{is polynomially tamed in $\cE$}\ .
\eeq
Indeed, in Section~\ref{sec:IIB} we will show that the hypothesis \eqref{DC_hyp1} holds for a large class of stringy EFTs. In the following we will limit ourselves to the analysis of the consequences of such an assertion.  
As stressed in the previous section, polynomially tamed functions do not display a definite leading term throughout $\cE$. Therefore, the masses $M_n$ might exhibit a different fall-off or growth according to the chosen path. However, one can partition the set $\mathcal{E}$ into smaller subsets where $M_n(s,\phi)$ has a definite growth. As a preliminary step, let us notice that, if $M_n(s,\phi)$ is a proper candidate tower that realizes the Distance Conjecture in a subset $\mathcal{U} \subset \cE$, it is necessary that $M_n(s,\phi)$ is strictly decreasing on $\mathcal{U}$. Thus, such a set $\mathcal{U}$ might be a single subset of the partition induced by the monotonicity theorem, or a union of different subsets thereof. Additionally, combining the requirement that $M_n(s,\phi)$ is polynomially tamed and that the states become massless near the boundary, we demand
\begin{equation}
\label{DC_Mn_U}
	M_n(s,\phi) \prec  (s^1)^{N_1} \cdots (s^n)^{N_n} \qquad \text{on}\quad\mathcal{U}\,,
\end{equation}
with $N_1, \ldots, N_n \in \mathbb{Z}_{< 0}$. Thus, the condition \eqref{DC_Mn_U} guarantees that the masses $M_n(s,\phi)$ obey $M_n (s,\phi)\to 0$ along \emph{any} path in $\mathcal{U}$ that approaches the infinite distance singularity. 

\begin{figure}[thb]
	\centering
	\includegraphics[width=7cm]{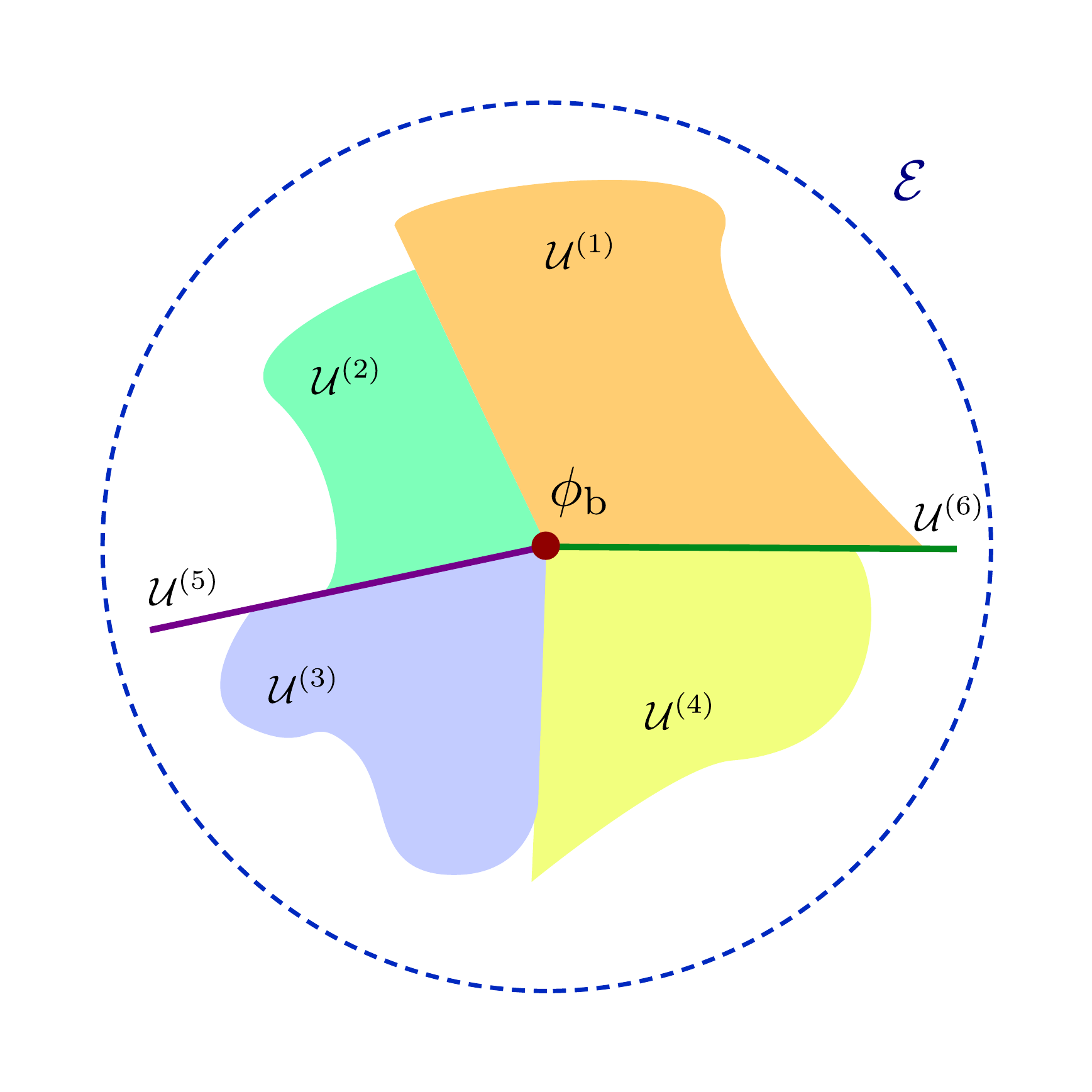}
	\caption{Schematic depiction of how the field space region around the boundary $\phi_{\rm b}$ in such a way that, in any subset $\mathcal{U}^{(a)}$, there exists an infinite tower of states with masses $M_n^{(a)}$ obeying \eqref{DC_Mn_Ua}. Notice that the dimensions of the subsets $\mathcal{U}^{(a)}$ do not have to be the same.
	\label{Fig:Partition_Mn}}
\end{figure}

But what happens in the asymptotic regions outside $\mathcal{U}$? If outside $\mathcal{U}$ the masses $M_n$ do not obey \eqref{DC_Mn_U}, they may not become massless, and thus the tower $M_n$ cannot serve as a candidate for realizing the Distance Conjecture in $\cE \mysetminus \mathcal{U}$. However, different infinite towers of states may be available, with masses $M_n^{(a)}$, where the index $a$ labels the tower. The towers are chosen in such a way that the masses of the constituting states becoming massless in some near-boundary region are strictly decreasing and
\begin{equation}
	\label{DC_Mn_Ua}
	M_n^{(a)}(s,\phi) \prec  (s^1)^{N^{(a)}_1} \cdots (s^n)^{N^{(a)}_n} \qquad \text{on}\quad \mathcal{U}^{(a)} \subset \cE\,,
\end{equation}
where $N^{(a)}_1, \ldots, N^{(a)}_n \in \mathbb{Z}_{< 0}$. The emergence of any infinite tower of states as the infinite distance singularity is approached requires that with such sets $\mathcal{U}^{(a)}$ we can cover the whole asymptotic region, namely $\bigcup\limits_{a} \mathcal{U}^{(a)} = \cE$ (see figure~\ref{Fig:Partition_Mn} for a representation).

\noindent\textbf{Matching the behavior of masses and the geodesic distance.} The Distance Conjecture asserts the specific asymptotic behavior \eqref{Mass+ineq} for the masses of the infinite states that break the effective description. However, in general, it is hard to estimate the behavior of the geodesic distance $d(s,\phi)$. Still, if the Distance Conjecture has to hold, then the behavior of the distance $d(s,\phi)$ has to be such that the behavior of $e^{-\lambda d(s,\phi)}$ is comparable with one of the masses $M_n$ of any tower of states. Therefore, it is enough to minimally assume that 
\beq
\label{DC_hyp2}
e^{-\lambda d(s,\phi)}\ \text{is polynomially tamed in $\cE$}\ .
\eeq
It is interesting to note that this is a tameness condition on the exponential of the distance rather than the distance itself. In this way the exponential function does not necessarily need to be definable and the statement of the Distance Conjecture \eqref{MP=MQ} can be formulated even within $\bbR_{\rm alg}$, e.g.~in highly supersymmetric setting when instanton corrections are absent \cite{Grimm:2021vpn,toappear_withMick}. However, if we insist that the distance itself is tame, we conclude that the statement of the Distance Conjecture \eqref{MP=MQ} requires to use, at least, the o-minimal structure $\mathbb{R}_{\text{exp}}$ in which the exponential function is definable. In the following we will 
stay general and work with $\mathbb{R}_{\text{an,exp}}$ in which, as in $\mathbb{R}_{\text{exp}}$, the 
definability of the exponential of $d(s,\phi)$ implies the 
definability of $d(s,\phi)$.

Now, in order to satisfy \eqref{Mass+ineq} throughout the asymptotic region $\cE$ we proceed as follows. Since $e^{-\lambda d(s,\phi)}$ is polynomially tamed, we can partition $\cE$ in regions $\{\mathcal{U}_{\mathsf{A}}\}$ where the $e^{-\lambda d(s,\phi)}$ is upper bounded by a monomial and has a distinguished leading behavior. Since our analysis concerns infinite distance singularities, in any of the sets $\mathcal{U}_{\mathsf{A}}$ that contains the boundary $\phi_{\rm b}$, to which we will refer to as `boundary' subsets, then
\begin{equation}
\label{DC_dist_phib}
	e^{-\lambda d(s,\phi)}\ \prec  (s^1)^{N^{\mathsf{A}}_1} \cdots (s^n)^{N^{\mathsf{A}}_n} \qquad \text{on any boundary}\quad\mathcal{U}_{\mathsf{A}}\,,
\end{equation}
with $N^{\mathsf{A}}_1, \ldots, N^{\mathsf{A}}_n \in \mathbb{Z}_{< 0}$. Then, on any of these boundary subsets, $e^{-\lambda d(s,\phi)}$ is either monomially tamed, or strictly upper bounded by a monomially tamed function. 

Then, if the Distance Conjecture is satisfied, there must exist some towers of states, subjected to an appropriate partition of $\cE$, such that for each of the subsets $e^{-\lambda d(s,\phi)} \sim M_n^{(a)}(s,\phi)$. For concreteness, take a single subset $\mathcal{U}_{\mathsf{A}}$. On $\mathcal{U}_{\mathsf{A}}$ we introduce an additional, finer partition that is now induced by the behavior of the masses of the states constituting the infinite towers. Namely, we construct a partition $\{\mathcal{U}^{(a,k)}_{\mathsf{A}}\}$ of $\mathcal{U}_{\mathsf{A}}$ such that, on each of these subsets, the tower $M_n^{(a)}$ displays a definite leading behavior
\begin{equation}
\label{DC_Mn_phib}
	M_n^{(a)}(s,\phi) \sim \rho_k (e^{-s},\phi) (s^1)^{n^{(a,k)}_1} \cdots (s^n)^{n^{(a,k)}_n} \qquad \text{on}\quad \mathcal{U}^{(a,k)}_{\mathsf{A}}
\end{equation}
with $n^{(a,k)}_i \in \mathbb{Z}_{<0}$. Then, on each $\mathcal{U}^{(a,k)}_{\mathsf{A}}$ the Distance Conjecture requires that $M_n^{(a)}(s,\phi) \sim e^{-\lambda d(s,\phi)}$. This, in turn, guarantees that, on each set $\mathcal{U}^{(a,k)}_{\mathsf{A}}$ the Distance Conjecture is realized path-independently.

\noindent\textbf{Finiteness of the infinite towers of states.} As we have illustrated, the realization of the Distance Conjecture in the full asymptotic region $\cE$ might require the presence of different towers of states. But how many towers are required? In general, the answer depends on how the partition of $\cE$ is carried over, by applying first the monotonicity theorem, and then refining such a partition by singling out the subsets where the masses have a distinguished leading behavior. However, the underlying tame structure of the effective field theory couplings guarantees that the subsets of $\cE$ that realize the partition are never infinite in number. We can then formulate the following statement:

\begin{importantboxtitle}{Finiteness of the infinite towers of states}
	In order for the Distance Conjecture to be realized within a tame EFT, only a finite number of infinite tower of states is sufficient.
\end{importantboxtitle}

\noindent\textbf{The Distance Conjecture and partial moduli stabilization.} The tameness of the EFT couplings allows for better addressing the issue raised at the end of Section~\ref{sec:tamenessconj} about the realization of the Distance Conjecture within an EFT with sliding cutoff. First, let us mention that the scalar potential, regarded as an EFT coupling, is also conjectured to be tame according to the Tameness Conjecture. As a relevant class of examples, in \cite{Bakker:2021uqw} it was proved that any flux-induced F-theory scalar potential is tame. As stressed in \cite{Grimm:2021vpn}, an important consequence of the tameness of the scalar potential resides in the finiteness on the number of vacua that it can deliver. Here we are interested in the special case where a partial moduli stabilization occurs. Concretely, let us consider an EFT endowed with the (pseudo-)moduli $\{\varphi^a\}$, $a =1,\ldots,m$, spanning the space $\mathcal{M}$, and assume that a scalar potential $V(\varphi,\lambda)$ fixes some of the field directions such that the residual moduli space $\widehat{\mathcal{M}}_\lambda$ is spanned by the fields $\{{\hat\varphi}^{\hat a}\}$, ${\hat a} =1,\ldots,{\hat m}$, ${\hat m} < m$. Now, consider any coupling $g(\varphi,\lambda)$ whose graph is definable in $\mathcal{M}_\lambda$. However, the graph of the coupling $g(\hat{\varphi},\lambda)$ is also definable is $\widehat{\mathcal{M}}_\lambda$: in fact, this is guaranteed by the properties of the definable sets outlined in Section~\ref{sec:tamenessconj} and by regarding the graph of $g(\hat{\varphi},\lambda)$ as an appropriate projection of the one of $g(\varphi,\lambda)$.

Let us now specialize to the case of interest, where the considered coupling $g(\varphi,\lambda)$ is one of those entering the Distance Conjecture, namely either $e^{-\lambda d}$ or the masses $M_n$ of the states constituting the infinite towers. We assume that the Distance Conjecture is realized over the moduli space $\mathcal{M}$ and, in particular, that there exists a finite partition of the near-boundary region of any infinite distance singularity where the Distance Conjecture is realized in a path independent fashion as explained above, with the masses constituting the infinite tower of states falling off as $M_n^{(a)} \sim e^{-\lambda d_{\mathcal{M}}}$. Consider now the subset $\widehat{\mathcal{M}}$ of $\mathcal{M}$ obtained after a partial moduli stabilization. Since the masses are tame on $\mathcal{M}$, they are also tame on $\widehat{\mathcal{M}}$. In particular, the partition of the near-boundary region $\cE$ in \eqref{DC_Mn_Ua} also covers the subset $\widehat{\mathcal{M}}$. Consequently, the same towers that are candidates for realizing the Distance Conjecture in $\mathcal{M}$ may also serve as candidates in order to realize the Distance Conjecture on $\widehat{\mathcal{M}}$. Additionally, the number of infinite tower of masses that are necessary to realize the Distance Conjecture is finite in number, when employing arguments similar to the ones used for $\cM$.
However, in general, assuming solely the definability of the couplings in $\bbR_{\rm an,exp}$ seems not to be enough to guarantee that the fall-off of the masses obeys \eqref{Mass+ineq} in terms of the geodesic distance on $\widehat{\mathcal{M}}$. In order to realize \eqref{Mass+ineq} on $\widehat{\mathcal{M}}$ a different partition of the near-boundary region from the one induced by \eqref{DC_dist_phib}-\eqref{DC_Mn_phib} might be needed and it is tempting to speculate that the Distance Conjecture can inferred for $\widehat \cM$ if the 
potential is moreover polynomially tamed.

In order to give evidence for these statements, let us first recall the simplest prototype example for the Distance Conjecture and 
consider a Kaluza-Klein compactification of a $D$-dimensional theory on a circle with radius $s$. In the effective $(D-1)$-dimensional theory the radius is a 
scalar field with kinetic term 
\beq
S^{(D-1)} = M_{\rm P}^{D-3} \int {\rm d}^Dx \sqrt{-g} \left(- \frac{1}{s^2} \partial_\mu s \partial^\mu s + \ldots \right)\ .
\eeq
The limit $s \rightarrow \infty$ is an infinite distance limit with a logarithmic growth in the geodesic length, $d(s,s_0) = \int^{s}_{s_0} \frac{ds'}{s'} = \text{log}\big(\frac{s}{s_0}\big)$. 
Note that the metric $1/s^2$ is definable in $\bbR_{\rm an,exp}$ in accordance with the Tameness Conjecture. The masses of the Kaluza-Klein states arising 
in the limit $s \rightarrow \infty$ are $M_n(s) = \frac{n}{s} \sim n\, e^{-d(s,s_0)}$ in accordance with the Distance Conjecture. In fact, $M_n(s)$  is also definable 
in $\bbR_{\rm an,exp}$ and both the metric and the masses have a simple polynomial growth 
\beq
G_{ss} \sim \frac{1}{s^2}\ , \qquad M_n \sim e^{-d(s,s_0)} \sim  \frac{1}{s}\ . 
\eeq

This asymptotic polynomial behavior is common to all examples in which the Distance Conjecture 
has been tested so far. However, it does not resolve the path-dependence issues that 
we have raised in Section~\ref{sec:DC_pd}. To see that, let us extend the setting to multiple variables.  Concretely, let us consider a four-dimensional model with three moduli $s^i$, $i =1,2,3$, described by the following action
\beq
S^{(4)} = M_{\rm P}^{2} \int {\rm d}^4x \sqrt{-g} \left(- \sum\limits_{i=1}^3 \frac{1}{(s^i)^2} \partial_\mu s^i \partial^\mu s^i + \cdots \right)\ .
\eeq
Such an action stems, for instance, from the compactification of ten-dimensional string theory on a six-dimensional toroidal orbifold $T^6/\Gamma$, with $\Gamma$ a discrete group; specifically, in Type IIB EFTs, the fields $s^i$ parametrize the imaginary part of the complex structure moduli, or in Type IIA EFTs the fields $s^i$ parametrize the volume each the volume of an internal $T^2$. The geodesic distance between the field space points $(s^i_0)$ and $(s^i)$ is
\begin{equation}
	\label{TameDC_Ex2_d}
	d =  \sqrt{\sum\limits_{i=1}^3 \left(\log \frac{s^i}{s^i_0}\right)^2}\,.
\end{equation}
Thus, again, infinite distance points are reached when any of the fields $s^i \to \infty$. A candidate infinite tower of states realizing the Distance Conjecture is given by the tower of the Kaluza-Klein modes. The lightest among the Kaluza-Klein states have the following behavior in terms of the fields $s^i$:
\begin{equation}
	\label{TameDC_Ex2_Mn}
	M_n^i \sim \frac{M_{\rm P}}{\sqrt{s^i}}\,.
\end{equation}
Clearly, which tower of state is relevant depends on the specific choice of the path that leads to infinite distance. For instance, along paths in which only a single field $s^i \to \infty$, the distance conjecture is realized by three different tower of states as $M_n^i \sim e^{-\frac12 d}$. More generally, one can consider three different `strict' asymptotic regimes
\begin{equation}
	(1)\; s^1 \gg s^2 , s^3 \,, \qquad (2)\; s^2 \gg s^1 , s^3 \qquad (3)\; s^3 \gg s^1 , s^2\,,
\end{equation}
such that, asymptotically in each sector, $M_n^i \sim e^{-\frac12 d}$. Moreover, provided a redefinition of fields $s^i \to (s^i)^2$, the common leading behavior of both $M_n^i$ and $e^{-\frac12 d}$ is monomially tamed. This simple example shows us a general feature that is crucial to consider when testing the Distance Conjecture: we need to specify the asymptotic regime that we are investigating. Such regimes can be captured, for instance, by the sets $\tilde\Sigma_I$ defined in \eqref{def-tildeSigmaI}, or smaller subsets thereof. Once an appropriate subset is chosen, on the one hand,  $e^{-d(s,s_0)}$ may acquire a simple asymptotic behavior; on the other hand, the tower of states that could realize the Distance Conjecture organize \emph{hierarchically}, and one can then single out the lightest tower, namely the one that leads to the breaking of the EFT. 

In the above selected examples it is clear how one can choose asymptotic regimes in order to realize the Distance Conjecture. However, in general, it is hard to determine whether, and to what extent the Distance Conjecture is realized by any given tower of states: the towers can change if the path that leads to infinite distance is chosen differently and, in principle, $e^{-\lambda d(s,s_0)}$ may exhibit a complicated fall-off that is hard to match with the fall-off of the masses of the candidate infinite tower of states. In the following section we will explore what is the minimal information to tell whether the relation \eqref{Mass+ineq} can be realized within a partition of the asymptotic region $\cE$ in a path independent way.


\section{Test strings and tame functions}
\label{sec:linear_paths}

In this section we propose that one can probe the behavior of four-dimensional EFT couplings via strings. Indeed, strings backreact on the scalar fields, and the backreaction solely depends on the charge of the string. Such a backreaction offers a \emph{test path} on which to probe the behaviors of EFT couplings. We will show that the behavior of monomially and polynomially tamed functions on families of these test paths -- or, on the allowed string backreactions -- delivers the minimal information required to characterize the behavior of the function throughout $\Sigma_I$.

Specifically, as we shall see, the leading behavior of any monomially tamed function is fully determined by how the monomially tamed function grows or falls off on such test paths. On the other hand, polynomially tamed functions have a more complicated structure, and cannot be solely determined by examining their behaviors on string backreactions. However, we will deliver a recipe to \emph{bound} polynomially tamed functions by how they behave on string backreactions.

In the following, to begin with, we will recall how to construct cosmic string solutions in four-dimensional EFTs, and we will later promote such solutions as test paths to examine the behavior of monomially and polynomially tamed functions.

\subsection{Cosmic and axion strings}
\label{sec:axion_strings}

\begin{figure}[thb]
	\centering
	\includegraphics[width=\textwidth]{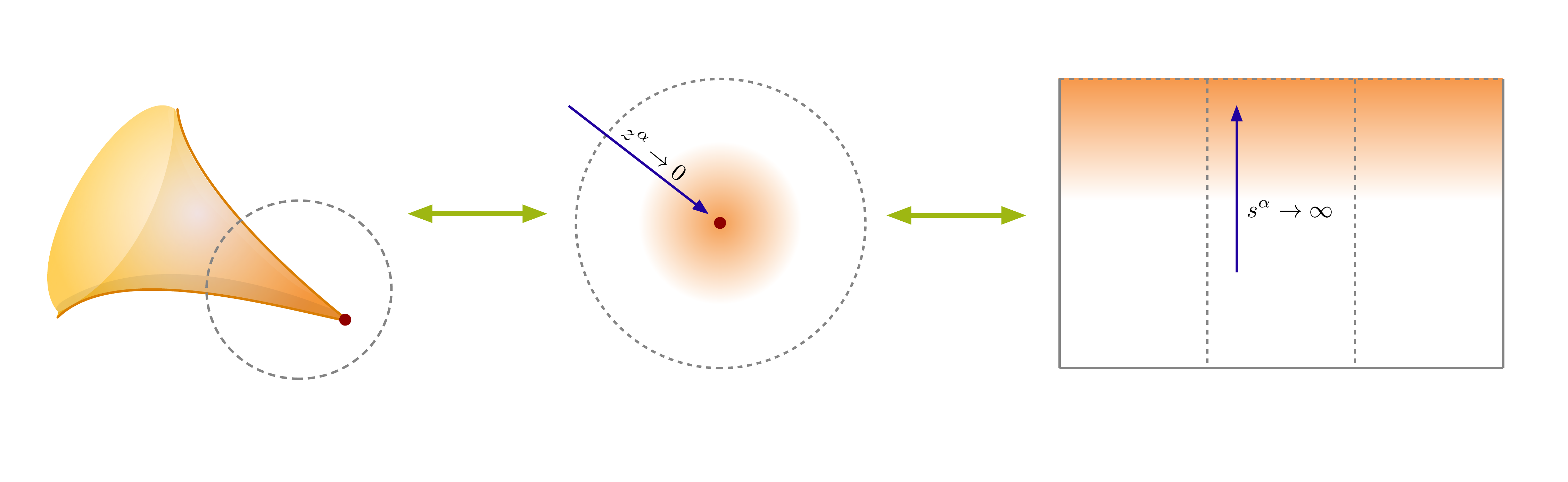}
	\caption{The near-boundary region around the boundary point, depicted on the left, can be described by the polydisk $0<|z^\alpha|<1$, with the boundary located at $z^\alpha = 0$; in turn, via \eqref{CosmStr_varphir}, this can be mapped to the upper-half-plane, where the boundary is reached as $s^\alpha \to \infty$. \label{Fig:Disk_HalfPlane}}
\end{figure}

Here, we will review the cosmic string solutions first studied in \cite{cstring,Dabholkar:1990yf} and later generalized and applied to axion strings in \cite{Lanza:2020qmt,Lanza:2021udy}. To begin with, assume that a local patch within the moduli space $\mathcal{M}$ is parametrized by $N$ complex coordinates $z^\mathcal{A}$.\footnote{It is worth stressing that, at this stage, $z^\mathcal{A}$ is \emph{any} modulus of the EFT.} Singularities can be locally described as the loci
\begin{equation}
  \label{CosmStr_zalpha}
  z^\alpha = 0 \,,\qquad \alpha = 1, \ldots, n\, ,
\end{equation}
for some $n \leq N$. It is then convenient to split the coordinates $z^\mathcal{A}$ as $z^\mathcal{A} = (z^\alpha, \zeta^{\kappa})$, with $\kappa = 1,\ldots, N-n$. The domain that will be of interest for us, within which we assume that the EFT is well defined, is
\begin{equation}
  \label{CosmStr_domain}
  \mathcal{E} = (\Delta^*)^n \times \Delta^{N-n}\, ,
\end{equation}
where we have denoted $\Delta^{N-n} = \{|\zeta^{\kappa}| < 1\}$, a product of disks, and  $(\Delta^*)^n = \{0 < |z^\alpha| < 1\}$, a product of punctured disks. However, it is more convenient to redefine
\begin{equation}
  \label{CosmStr_varphir}
  \varphi^\alpha = \frac{1}{2 \pi \im} \log z^\alpha\,,
\end{equation}
so that the singular locus in \eqref{CosmStr_zalpha} is reached as $\varphi^\alpha \to \im \infty$. Notice that \eqref{CosmStr_varphir} relates $(\Delta^*)^n$ to the upper half plane $\mathcal{H}^n$ as:
\begin{equation}
    \label{Linear_Dom}
       \{0 < |z^\alpha| < 1\} = (\Delta^*)^n \;\ni\; z^\alpha  \quad  \longleftrightarrow \quad     \varphi^\alpha \;\in\; \mathcal{H}^n = \{\Im \varphi^\alpha > 0\}\,
\end{equation}
as depicted in Fig.~\ref{Fig:Disk_HalfPlane}.

We also split the complex coordinates $\varphi^\alpha$ as
\begin{equation}
  \label{CosmStr_as}
  \varphi^\alpha = a^\alpha + \im s^\alpha\,,
\end{equation}
in terms of the real coordinates $a^\alpha$, $s^\alpha$, with $s^\alpha > 0$. As will become clear soon, the real fields $a^\alpha$ can be regarded as \emph{axions}. We will assume that the axions span a compact domain, and that they are identified as $a^\alpha \simeq a^\alpha + 1$. Their partners $s^\alpha > 0$, that build the complex coordinates $\varphi^\alpha$ alongside $a^\alpha$, will be referred to as \emph{saxions}. 

Let us now introduce the class of EFTs that will be under scrutiny. We will focus on either $\mathcal{N}=2$ or $\mathcal{N}=1$ supersymmetric effective field theories. We will assume that the moduli $\zeta^\kappa$ are fixed at a specific point $\zeta^{\kappa}_0$ within $\Delta^{N-n}$, and we will regard them as non-dynamical, `spectator' fields. Then, within the domain $\mathcal{E}$, the sole dynamical complex fields $\varphi^\alpha$, for which we will write down an effective field theory. The fields $\varphi^\alpha$ parametrize a local patch of K\"ahler manifold, and we will denote the associated K\"ahler potential as $K$. The effective action describing the coupling the complex fields $\varphi^\alpha$ to gravity is:
\begin{equation}
\label{CosmStr_S}
S=M^2_{\rm P}\int \left(\frac{1}{2}R*1-K_{\alpha\bar \beta}\,{\rm  d}\varphi^\alpha\wedge*{\rm d}\bar\varphi^{\bar \beta} \right)\, .
\end{equation}

The contributions appearing in \eqref{CosmStr_S} are common to both the $\mathcal{N}=2$ action (see the following \eqref{N2_action}) and the $\mathcal{N}=1$ action (see \eqref{N1_action}). In the former case, the contributions in \eqref{CosmStr_S} can be obtained from the bosonic components of $\mathcal{N}=2$ actions describing the interactions of $n$ vector multiplets and turning off the gauge fields $A^I$. In the case of $\mathcal{N}=1$ actions,  \eqref{CosmStr_S} correspond to the bosonic components of supergravity actions describing the interaction of $n$ chiral multiples in the absence of a scalar potential. In this latter case, setting the scalar potential $V (\varphi^\alpha) = 0$ can be either achieved by turning off background fluxes and neglecting additional corrections to the scalar potential or by stabilizing the `spectator' moduli $\zeta^\kappa$ in such a way that $V (\zeta_0^\kappa) = 0$.

Cosmic string solutions are solitonic solutions of the equations of motion that preserve the two-dimensional Poincar\'e invariance along two spacetime directions. Namely, let us split the spacetime coordinates as $(t,x,\xi, \bar \xi)$, where $\xi$ and $\bar\xi$ are complex coordinates spanning the space directions orthogonal to $(t,x)$. Indeed, introducing the polar coordinates $(r,\theta)$, one can relate the latter to $\xi$ as $\xi=re^{i\theta}$. Then, a cosmic string solution is a solution to the equations of motion stemming from \eqref{CosmStr_S} by imposing the following metric ansatz: 
\begin{equation}
\label{CosmStr_ds2}
{\rm d} s^2= -{\rm d} t^2+{\rm d} x^2+ e^{2D}{\rm d} \xi {\rm d}\bar \xi\, ,
\end{equation}
where $D$ is a warp factor. For simplicity, we will further assume that both scalar fields $\varphi^\alpha$ and the warp factor $D$ depend only on the coordinates $(\xi,\bar\xi)$.

It can be shown that the equations of motion for the fields $\varphi^\alpha$ leads to \cite{Lanza:2021udy}
\begin{equation}
\label{CosmStr_eomphi}
K_{\bar \alpha \beta}\partial \bar\partial \varphi^\beta+K_{\bar \alpha \beta \gamma}\partial \varphi^\beta\wedge {\bar\partial}\varphi^{ \gamma}=0\,,
\end{equation}
where we have introduced $\partial \equiv {\rm d} \xi \wedge \partial_\xi$ and its complex conjugate $\bar\partial \equiv {\rm d} \bar\xi \wedge  \partial_{\bar\xi}$. As shown in \cite{Dabholkar:1990yf,Lanza:2021udy}, the simplest BPS solutions to \eqref{CosmStr_eomphi} are either holomorphic profiles obeying
\begin{equation}
\label{CosmStr_phihol}
\bar\partial \varphi^\alpha = 0
\end{equation}
or anti-holomorphic profiles satisfying $\partial \varphi^\alpha = 0$, along which half of the bulk supersymmetry is preserved. Here, we will pick the holomorphic profiles in \eqref{CosmStr_phihol} as solutions to \eqref{CosmStr_eomphi}. For completeness, let us mention that the Einstein equations deliver the following relation between the warp factor $D$ in \eqref{CosmStr_ds2} and the K\"ahler potential: 
\begin{equation}
\label{CosmStr_warp}
e^{2D}=|f(z)|^2 e^{-K}\, ,
\end{equation}
with $f(z)$ an arbitrary, non-vanishing holomorphic function.

\if{}
In general, solitonic solutions for the action \eqref{CosmStr_S} correspond to infinite energy. Thus, rather than considering the energy associated to the solution \eqref{CosmStr_holphi}-\eqref{CosmStr_warp}, we shall consider the energy per unit area
\begin{equation}
\label{CosmStr_energy}
\mathcal{E} = M_{\rm P}^{2} \int K_{\alpha \bar \beta}\left(\partial\varphi^\alpha \wedge *_2\bar\partial\bar\varphi^{\bar \beta}+\bar\partial\varphi^\alpha\wedge *_2\partial\bar\varphi^{\bar \beta}\right)=M_{\rm P}^{2}\int J_{\mathcal{M}}+ 2M_{\rm P}^{2}\int K_{\alpha\bar \beta}\,\bar\partial\varphi^\alpha\wedge *_2\partial\bar\varphi^{\bar \beta}\, ,
\end{equation}
where $*_2$ is the Hodge-star operator in the two-dimensional space $(\xi,\bar\xi)$, and $J_{\mathcal{M}} = i K_{\alpha\bar\beta} {\rm d}\phi^\alpha \wedge {\rm d}\bar{\phi}^{\bar\beta}$. Indeed, according to the specific choice of holomorphic profile for $\varphi^\alpha$, one can obtain solitonic solutions with finite energy per unit area. Moreover, due to the positive definiteness of the last term in \eqref{CosmStr_energy}, the energy per unit area obeys the BPS bound
\begin{equation}
\label{CosmStr_BPSbound}
\mathcal{E}  \geq M_{\rm P}^{2}\int J_{\mathcal{M}}\, ,
\end{equation}
which is clearly saturated by the holomorphic profiles \eqref{CosmStr_phihol}.
\fi{}

The cosmic string solution \eqref{CosmStr_phihol} is agnostic about the specific profile of the $\varphi^\alpha(\xi)$ in terms of the holomorphic coordinate $\xi$. Here, we are interested in cosmic string solution exhibiting the following monodromy transformation 
\begin{equation}
\label{CosmStr_mon}
\varphi^\alpha \rightarrow \varphi^\alpha  +e^\alpha\, , \qquad \qquad e^\alpha \in \mathbb{Z}\, ,
\end{equation}
when encircling a loop around $r = 0$ (i.e. $\xi =0$). The integers $e^\alpha$ are those that distinguish the monodromy transformation and -- as will become soon clear -- can regarded as the \emph{elementary charges} of the cosmic string solution. We also stress that, in order for a solitonic solution with the property \eqref{CosmStr_mon} to be valid, the EFT has to be invariant under the monodromies \eqref{CosmStr_mon} \cite{Polchinski:2005bg,Banks:2010zn}. Specifically, in the action \eqref{CosmStr_S}, the K\"ahler metric has to be invariant under the monodromies \eqref{CosmStr_mon}.

The holomorphic solution that realizes \eqref{CosmStr_mon} is
\begin{equation}
\label{CosmStr_solvarphi}
\varphi^\alpha=\varphi^\alpha_0+\frac{1}{2\pi \im}e^\alpha \log \left(\frac{\xi}{\xi_0}\right)\, ,
\end{equation}
for some constant $\varphi^\alpha_0$ and $\xi_0$.\footnote{One may wonder if the solution \eqref{CosmStr_solvarphi} is a well defined BPS solitonic solution, with finite energy density. Indeed, it can be checked that the cosmic string solution \eqref{CosmStr_solvarphi} delivers configurations of finite energy density, saturating the BPS bound \cite{cstring,Dabholkar:1990yf,Lanza:2021udy}.}
Indeed, splitting the complex fields $\varphi^\alpha$ as in \eqref{CosmStr_as} into the real fields $a^\alpha$, $s^\alpha$, the solution \eqref{CosmStr_solvarphi} can be recast as
\begin{subequations}
  \label{CosmStr_solvarphib}
  \begin{align}
  \label{CosmStr_sols}
  s^\alpha &=  s_{0}^\alpha -\frac1{2\pi}\log \left(\frac{r}{r_0}\right) e^\alpha\, ,\\
  \label{CosmStr_sola}
  a^\alpha &= \frac{\theta}{2\pi}\,e^\alpha+ a^\alpha_0\, .
  \end{align}
\end{subequations}
This clearly exhibits that the real fields $a^\alpha$ experience the monodromy $a^\alpha \rightarrow a^\alpha + e^\alpha$ after turning around any loop centered at $r = 0$. On the other hand, the fields $s^\alpha$ depend on the radial distance $r = |\xi|$; importantly, as $r \to 0$, $s^\alpha \to \infty$. 

Thus, according to the choice of the elementary charges $e^\alpha$, different saxions are driven towards large vevs as $s^\alpha \to e^\alpha \times \infty$. In order to organize the string solutions, as in \cite{Lanza:2021udy}, we further distinguish \emph{elementary} and \emph{non-elementary flows}. Consider a basis of BPS charges $\{e^\alpha\}$. An elementary flow is a solitonic solution of the kind \eqref{CosmStr_solvarphib} generated by a cosmic string with charge coinciding with a single basis element $e^{\hat\alpha}$ . A non-elementary flow is a BPS solitonic solution generated by an axion string whose electric charge is a linear combination of the basis elements $\{e^\alpha\}$.

The solution \eqref{CosmStr_solvarphi} is rather general: it relies only on the action \eqref{CosmStr_S}, with the assumption that the axion fields ought to experience the monodromy \eqref{CosmStr_mon} once we turn around a loop centered at $\xi = 0$. In some cases, however, we can further elaborate about the phenomenological meaning of the solution \eqref{CosmStr_solvarphi}. As is clear from \eqref{CosmStr_solvarphi}, the solution exhibits a singularity at $\xi = 0$, and assumes that the singularity is not resolved within the effective field theory.\footnote{Notice that we are looking for solutions for which the singularity at $\xi =0$ \emph{cannot} be resolved within the EFT. Non-singular solutions at $\xi =0$ can be built, but they require the presence of additional gauge fields subjected to Higgsing effect. We refer to \cite{Reece:2018zvv,Vilenkin:2000jqa} for details and concrete examples.} Then, the singularity can be understood as accommodating a codimension-two spacetime defect, which we will identify as an \emph{axion string}. In order to see this, preliminarily, let us recall that the cosmic string solution \eqref{CosmStr_solvarphib} exhibits a nontrivial winding for the axion. A rather simple, but strong assumption that guarantees the invariance of the action \eqref{CosmStr_S} under the monodromy \eqref{CosmStr_sola} is to assume that the K\"ahler potential does \emph{not} depend on the real fields $a^\alpha$, that is $K(\varphi, \bar\varphi) = K(s)$, implying that also the K\"ahler metric appearing \eqref{CosmStr_S} is solely saxion-dependent. In other words, \eqref{CosmStr_S} enjoys the exact continuous shift symmetry $a^\alpha \to a^\alpha + c^\alpha$, with $c^\alpha \in \mathbb{R}$, rendering $a^\alpha$ proper `axions', namely zero-form gauge fields.\footnote{Note that the assumption that $K(\varphi, \bar\varphi) = K(s)$ implies that \eqref{CosmStr_S} exhibits a zero-form global symmetry, leading to a three-form conserved current. On the other hand, invariance under the monodromies \eqref{CosmStr_mon} requires invariance only under a discrete gauge group. See, for instance, \cite{Heidenreich:2020pkc} for details on the subject.}

It was shown in \cite{Lindstrom:1983rt,Lanza:2019xxg,Lanza:2019nfa} that it is then convenient to rephrase the action \eqref{CosmStr_S} in terms of dual variables so defined:
\begin{equation}
  \label{AxStr_soldual}
  \ell_\alpha =  - \frac12 \frac{\partial K}{\partial s^\alpha}\,, \qquad 
  H_{3\, \alpha} ={\rm d} B_{2\, \alpha}-M^2_{\rm P}\, \mathcal{G}_{\alpha\beta}*{\rm d}a^\beta\, ,
\end{equation}
with
\begin{equation}
\label{AxStr_G}
\mathcal{G}_{\alpha\beta} \equiv \frac12 \frac{\partial^2 K}{\partial s^\alpha \partial s^\beta}\,.
\end{equation}
Here, the \emph{dual saxions} $\ell_\alpha$ in \eqref{AxStr_soldual} replace the saxions $s^\alpha$ as the real scalar fields enjoying a non-trivial radial flow. Instead, in \eqref{AxStr_soldual}, the axions $a^\alpha$ are traded with the gauge two-forms $B_{2\,\alpha}$ via a standard electro-magnetic duality. It can be shown that in this dual framework, the action \eqref{CosmStr_S} reads
\begin{equation}
\label{AxStr_Sdual}
S_{\rm dual}=\int \left(\frac{M^2_{\rm P}}{2}R*1-\frac{M^2_{\rm P}}{2} \mathcal{G}^{\alpha\beta}\,{\rm  d}\ell_\alpha\wedge*{\rm d}\ell_{\beta} -\frac{1}{2 M^2_{\rm P}} \mathcal{G}^{\alpha\beta}\,H_{3\,\alpha}\wedge*H_{3\,\beta} \right)\, ,
\end{equation}
with $\mathcal{G}^{\alpha\beta}$ the inverse of \eqref{AxStr_G}. Focusing on the bosonic sector only, the dualization of the saxionic and axionic fields in \eqref{AxStr_soldual} is general. However, in supersymmetric theories the complex scalar fields $\varphi^\alpha$ are accommodated in appropriate multiplets alongside with their fermionic partners. In $\mathcal{N}=1$ supergravity such as those examined in Section \ref{sec:hier_N1}, $\varphi^\alpha$ reside in chiral multiplets $\Phi^\alpha$. It can be shown that the dualization \eqref{AxStr_soldual} can be performed at levels of multiplets, with the chiral fields $\Phi^\alpha$ traded with linear multiplets $L_\alpha$, which accommodate the dual saxions $\ell_\alpha$ and the gauge two-forms $B_{2\,\alpha}$ in their bosonic components. We refer to \cite{Lindstrom:1983rt,Grimm:2004uq,Lanza:2019xxg,Lanza:2019nfa} for further details.

Since the dual action \eqref{AxStr_Sdual} manifestly contains the gauge two-forms $B_{2\,\alpha}$, we can include the fundamental objects electrically coupled to the gauge two-forms. These are strings, effectively described by the following action:
\begin{equation}
\label{AxStr_Sstr}
S_{\rm string}= -\int_\mathcal{S} {\rm d}^2\xi\sqrt{-h}\, \mathcal{T}_{\rm str} +e^\alpha \int_\mathcal{S} B_{2\, \alpha}\,.
\end{equation}
Here we have introduced the coordinates $\xi^{\hat\imath}$, $\hat\imath = 1,2$, that parametrize the worldsheet $\mathcal{S}$ of the strings spanning the time direction and one space direction, and $h_{{\hat\imath}{\hat\jmath}}$ the induced metric on the string. Furthermore, $\mathcal{T}_{\rm str}$ denotes the (field-dependent) string tension, and $e^\alpha$ the electric charges of the string. In order for the effective description not to be broken by the inclusion of the strings, we need to require
\begin{equation}
\label{AxStr_Tcutoff}
  \mathcal{T}_{\rm str}  \geq \Lambda^2_{\text{\tiny{EFT}}}\,,
\end{equation}
with $\Lambda_{\text{\tiny{EFT}}}$ the cutoff of the effective field theory. The condition \eqref{AxStr_Tcutoff} guarantees that the strings are described as \emph{fundamental} objects, whose core is not resolved within the EFT,  allowing us to neglect all the stringy oscillatory modes. For this reason, we will call the strings described by the action \eqref{AxStr_Sstr}, equipped with \eqref{AxStr_Tcutoff} \emph{fundamental axion strings}.

We will be specifically interested in BPS strings. In $\mathcal{N}=1$ theories it can be shown that for  fundamental axion strings maximally preserving two supercharges over their worldvolume the string tension $\mathcal{T}_{\rm str}$ has to be linear in dual saxions $\ell_\alpha$ \cite{Bandos:2019lps,Bandos:2019qok,Lanza:2019nfa,Lanza:2019xxg}:
\begin{equation}
\label{AxStr_Tstr}
  \mathcal{T}_{\rm str}  \equiv \mathcal{T}_{\bf e} = M_{\rm P}^2 e^\alpha \ell_\alpha\,,
\end{equation}
supported by the BPS condition
\begin{equation}
  \label{AxStr_BPScond}
  e^\alpha \ell_\alpha > 0 \,.
\end{equation}
We further recall that, by exploiting the dualization \eqref{AxStr_soldual}, one can recast the above string tension in terms of the saxionic fields $s^\alpha$. It can be shown that \eqref{AxStr_Sdual}, once coupled to \eqref{AxStr_Sstr} with \eqref{AxStr_Tstr}, delivers the cosmic string solution \eqref{CosmStr_solvarphib}, with the fundamental axion string \eqref{AxStr_Sstr} conveniently capturing the singularity at $r = 0$.

Before concluding this section, let us remark that the solution \eqref{CosmStr_solvarphib} is not a complete cosmic string solution covering the full spacetime. In fact, it breaks down when the saxions $s^\alpha = 0$. For instance, in models with only a single complex field $\varphi$, this happens at the radial distance $\bar{r} = r_0 e^{\frac{2 \pi s_0}{e}}$. In \cite{Lanza:2021udy}, in the context of $\mathcal{N}=1$ EFTs, the distance $\bar{r}$ was regarded as an energy scale $\bar{\Lambda} = \frac{1}{\bar{r}}$ at which the EFT becomes strongly coupled. Indeed, therein it was shown at the scale $\bar{\Lambda}$ the axion string tension \eqref{AxStr_Tstr} diverges. 
This behavior is typical of codimension-two objects and is ostensibly in contrast with what happens along the backreactions of objects of codimension strictly greater than two; in fact, for the latter, the backreaction becomes more and more negligible as the distance from the object increases.
However, it is worth mentioning that, as in \cite{cstring,Marchesano:2022avb}, one can `complete' the string solution \eqref{CosmStr_solvarphi} so as to encompass regions of strong coupling. In this work we will not consider such a continuation of the solution \eqref{CosmStr_solvarphi}, for \eqref{CosmStr_solvarphi}, with $r < r_0$, is already enough to explore the near boundary region of the moduli space.

\subsection{Strings as probes for polynomially and monomially tamed functions}
\label{sec:roughlylinear}

In this section we illustrate how the cosmic string solutions reviewed in the previous sections can be used as tools to probe the behavior of monomially and polynomially tamed functions. However, monomially and polynomially tamed functions are defined over a given patch of the moduli space, while the cosmic string solution \eqref{CosmStr_solvarphi} is defined in spacetime. Therefore, as a preliminary step, we need to translate the backreaction \eqref{CosmStr_solvarphib} into a path drawn by the saxionic fields $s^\alpha$ within the local patch $\mathcal{E}$ in \eqref{CosmStr_domain}. To this end, we preliminarily define the parameters
\begin{equation}
	\label{StrProbe_sigma}
	\sigma \equiv -\frac1{2\pi}\log \left(\frac{r}{r_0}\right)\,, \qquad \rho \equiv \frac{\theta}{2\pi} + \im\, \sigma\,,
\end{equation}
so that the backreaction \eqref{CosmStr_solvarphi} can be recast as
\begin{equation}
	\label{StrProbe_paths}
	\varphi^\alpha(\rho) = \varphi_{0}^\alpha + e^\alpha \rho\,.
\end{equation}
that is specified by the choice of string charge ${\bf e} = (e^\alpha)$. Then, we can map each point $\varphi^{\alpha}(r,\theta)$ along the cosmic string solution \eqref{CosmStr_solvarphi} to a point $\varphi^\alpha (\rho)$ that specifies a vacuum configuration of the effective field theory. In other words, the cosmic string solution \eqref{CosmStr_sols} maps to a linear path within the moduli space. Moreover, by further completing the above path with the non-dynamical fields $\zeta^\kappa = \zeta^\kappa_0$, paths in \eqref{StrProbe_paths} can be promoted to paths in the full domain \eqref{CosmStr_domain}. 

\begin{figure}[thb]
	\centering
	\includegraphics[width=14cm]{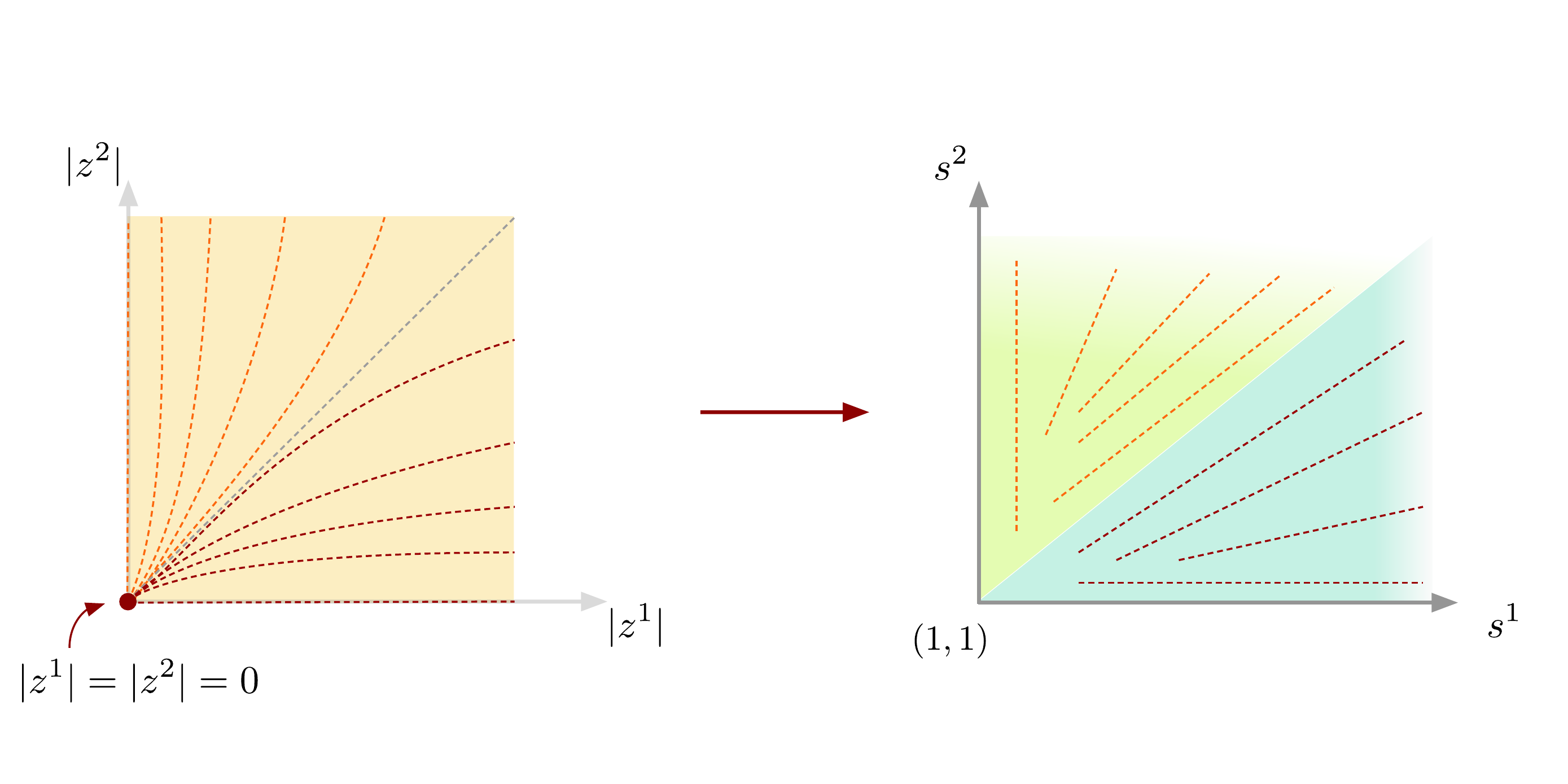}
	\caption{On the left, the subregion spanned by $|z^1|$ and $|z^2|$ within the polydisk $\{0<|z^1|<1\} \cup \{0<|z^2|<1\}$; within this region, the string backreaction \eqref{CosmStr_solvarphi} traces curves as depicted by the dotted lines; on the right, are depicted the paths within the saxionic space $(s^1,s^2)$, further partitioned via the sets \eqref{def-SigmaI}. \label{Fig:Infinite_Dist_Lin_Paths}}
\end{figure}

The paths \eqref{StrProbe_paths} are suitable to explore the near-boundary region of the moduli space. As $\rho \to \im \infty$ (corresponding to $r \to 0$ from the spacetime perspective), the saxions $s^\alpha$ are driven towards `distant' regions in the north of the upper-half plane in Figure~\ref{Fig:Disk_HalfPlane}. In Section~\ref{sec:Tame_bound} we have additionally stressed that it is convenient to cover the near-boundary region via the sets
\begin{equation} \label{def-SigmaI}
	\Sigma_I = \{ 0<a^\alpha <1 ,\ \  s^1 \geq  s^2 \geq \cdots \geq s^n \geq 1 \}\, ,
\end{equation}
where the index $I$ labels all permutations of the $s^\alpha$ in the hierarchy and we have only displayed the simplest permutation.
The paths \eqref{StrProbe_paths} may indeed be tuned in order to cover only a single among the sets $\Sigma_I$. For concreteness, let us focus on the set $\Sigma_{12\ldots n}$. Let us assume that the initial values of saxionic and axionic fields are chosen such that $s^1_0 > s^2_0 > \ldots > s^n_0 > 1$ and $0<a_0^\alpha < 1$, so that $\varphi^\alpha(0) \in \Sigma_{12\ldots n}$. Given $\varphi^\alpha(0)$, the flow of the scalar fields is fully determined by the string charges $e^\alpha$, and we need to ensure that these are chosen such that $\varphi^\alpha(\rho) \in \Sigma_{12\ldots n}$ for any $\rho$. To this end we identify the following lattice of charges\footnote{Notice that, for $\mathcal{N}=1$ EFTs and when the cosmic string are axion string solutions, the charge lattice \eqref{Strprobe_Ce} can be understood as a sublattice of the EFT charge lattice $\mathcal{C}^{\text{\tiny{EFT}}}_{\text{\tiny{S}}}$ defined in \cite{Lanza:2021udy}. Indeed, the lattice \eqref{Strprobe_Ce} can be regarded as the lattice such that the instanton corrections -- whose charges lie in `dual' lattice $\mathcal{C}_{\text{I}} = \{ {\bf m} \in \mathbb{N}^n\}$ -- are negligible in $\Sigma$.}
\begin{equation}
	\label{Strprobe_Ce}
	\mathcal{C}_{\bf e} = \{ {\bf e} \in \mathbb{N}^n \; |\;  e^1 \geq e^2 \geq \ldots \geq e^n \geq 0 \} \,.
\end{equation}
Then, provided an additional rescaling ${\rm Re} \rho \to \frac{1}{C {\rm max}\{e^\alpha\}} {\rm Re} \rho$ for some $C>0$\footnote{Such a rescaling can be equivalently understood as a rescaling of the axions $a^\alpha$ subjected to the string backreaction.}, any path $\varphi^\alpha(\rho) = \varphi_{0}^\alpha + e^\alpha \rho$ with ${\bf e} \in \mathcal{C}_{\bf e}$ is fully contained in $\Sigma_{12\ldots n}$. By choosing the string charge ${\bf e}$ and varying the initial values of the saxions $s^\alpha_0$ and the fixed values of the axions $a^\alpha_0$, one can span the full set $\Sigma_{12\ldots n}$, as pictorially depicted in Figure~\ref{Fig:Infinite_Dist_Lin_Paths}. We collect all such paths in the set
\begin{equation}
	\label{StrProbe_pathsc}
	\mathcal{P}_{\bf e} =  \{ \varphi^\alpha = \varphi_0^\alpha + e^\alpha \rho \, | \, {\bf e} \in  \mathcal{C}_{\bf e} \,, s^1_0 > \ldots >s^n_0> 1 \,, 0 < a^\alpha_0 <1 \}  \in \Sigma_{12\ldots n} \,.
\end{equation}

In the following it will be convenient to consider special subsets of paths in \eqref{StrProbe_pathsc} such that some string charges are zero. To this end, we preliminary define the charge sublattices
\begin{equation}
	\label{Strprobe_Cen}
	\mathcal{C}_{\bf e}^{(\hat n)} = \{ {\bf e} \in \mathbb{N}^n \; |\;  e^1 \geq e^2 \geq \ldots \geq e^{\hat n} > 0\,, e^{\hat n +1} = \ldots = e^n = 0 \} \subset  \mathcal{C}_{\bf e}\,,
\end{equation}
and the associated families of paths
\begin{equation}
	\label{StrProbe_pathscn}
	\mathcal{P}_{\bf e}^{(\hat n)} =  \{ \varphi^\alpha = \varphi_0^\alpha + e^\alpha \rho \, | \, {\bf e} \in  \mathcal{C}_{\bf e}^{(\hat n)} \,, s^1_0 > \ldots >s^n_0> 1  \,, 0 < a^\alpha_0 <1 \}  \in \Sigma_{12\ldots n} \,.
\end{equation}
Clearly, $\bigcup\limits_{i=1}^{n} \mathcal{P}_{\bf e}^{(i)} = \mathcal{P}_{\bf e}$. In the language of the previous section, paths belonging to $\mathcal{P}_{\bf e}^{(1)}$, with the only non-null string charge $e^1$, will be referred to as `elementary' paths, while the others, $\mathcal{P}_{\bf e}^{(\hat n)}$ with ${\hat n} \geq 2$, will be generically called `non-elementary'.

The paths \eqref{StrProbe_pathsc} play the role of \emph{test paths}, which carry important information about the behavior of the EFT couplings. Indeed, below we show how the paths \eqref{StrProbe_pathsc} can be employed to determine the behavior of monomially tamed functions and to bound the behavior of polynomially tamed functions.

\vspace{1em}

\noindent{\textbf{The behavior of monomially tamed functions}}

\noindent
Monomially tamed functions display the simple structure \eqref{g-isroughlymono}. Indeed, their leading monomial behavior can be inferred solely by how they grow on BPS cosmic string solutions. Indeed, consider a generic monomially tamed function $f \sim (s^1)^{k_1} \cdots (s^n)^{k_n}$. In order to probe the behavior of $f$, one can consider how $f$ grows along the families of paths in \eqref{StrProbe_pathscn} varying $\hat n$. Concretely, consider first the elementary paths $\mathcal{P}_{\bf e}^{(1)}$, along which only the field $\varphi^1$ is driven to boundary according to \eqref{StrProbe_paths}. Along these elementary paths $f \sim \sigma^{k_1}$. Thus, the growth of $f$ with respect to the saxion $s^1$ is solely determined by how $f$ behaves on the elementary paths $\mathcal{P}_{\bf e}^{(1)}$. Let us then consider the non-elementary paths $\mathcal{P}_{\bf e}^{(2)}$ along which only the saxions $s^1$ and $s^2$ may reach the asymptotic region. Along these paths, $f \sim \sigma^{k_1 +k_2}$. Thus, having fixed $k_1$ from the behavior on $\mathcal{P}_{\bf e}^{(1)}$, we can then fix $k_2$ from how $f$ behaves on  $\mathcal{P}_{\bf e}^{(2)}$. We can then proceed by considering all the remaining non-elementary paths, and the leading behavior of $f$ would be fully fixed by how it grows along the linear paths. However, it is worth noticing that, since $f$ is monomially tamed, fixing the exponents $k_i$ by using linear paths gives information about the leading behavior of $f$ throughout $\Sigma_I$.

We can exploit such observations in order to compare different monomially tamed functions. In fact, consider two monomially tamed functions
\begin{equation}
	f \sim (s^1)^{k_1} \cdots (s^n)^{k_n} \,, \qquad g \sim (s^1)^{m_1} \cdots (s^n)^{m_n}\,,
\end{equation}
and assume that $f \sim g$ on curves in $\Sigma_I$. In particular, this holds on the families of test paths \eqref{StrProbe_pathscn}. By reasoning as above, it is simple to show that such an ordering is preserved in $\Sigma$, i.e.
\begin{equation}
	f \sim  g \sim (s^1)^{k_1} \cdots (s^n)^{k_n} \qquad \text{on $\Sigma_I$.}
\end{equation}

Thus, the growth of monomially tamed functions can be inferred from how they grow on linear paths only. Of course, inferring the behavior of the restricted analytic functions that serve as their coefficients would require some finer arguments.

\vspace{1em}

\noindent{\textbf{The behavior of polynomially tamed functions}}

\noindent
As displayed in Section~\ref{sec:Tame_DC}, polynomially tamed functions exhibit a more involved structure than the monomially tamed functions. Indeed, the behavior of a polynomially tamed function throughout $\Sigma$ cannot be generically inferred from how it behaves on a small subset of curve: in fact, the growth of a polynomially tamed function is path dependent, and the comparison of the behavior of polynomially tamed functions would be path dependent as well. However, one can rather compare the behavior of a polynomially tamed function with a monomially tamed one. In order to do that we introduce an order relation $f \prec g$ among the polynomially or monomially tamed functions $f$ and $g$. Recall that writing $f \prec g$ means that there exists a positive constant $C$ such that $f < Cg$ on all $\Sigma_I$.
Now, one can bound a polynomially tamed function using a monomially tamed one.
This can be achieved by exploiting the following lemma formulated in \cite{BKT} and we reformulate as follows: 
\begin{importantboxtitle}{Curve-reduction lemma for polynomially tamed functions}
  Consider a polynomially tamed function $f$ and a monomially tamed function $g$. Assume that
  \begin{equation}
    |f| \prec |g|
  \end{equation}
  on all the linear paths
  \begin{equation}
    \label{Linear_Sigmalin}
    \begin{aligned}
      &\beta_1 \varphi_1 + \gamma_1 = \cdots = \beta_{n_0} \varphi_{n_0} + \gamma_{n_0}\, ,
      \\
      &\varphi_{n_0 + 1} = \delta_{n_0 + 1}, \ldots, \varphi_n = \delta_n\, ,
    \end{aligned}
  \end{equation}
  for all choices of  $1 \le n_0 \le n$, complex numbers $\delta_{n_0 + 1}, \ldots, \delta_n$ with positive imaginary parts, positive rational numbers $\beta_1, \ldots, \beta_{n_0}$, and real numbers $\gamma_1, \ldots, \gamma_{n_0}$. Then
  \begin{equation}
    |f| \prec |g|  \qquad \text{on all}\quad\Sigma_I\,.
  \end{equation}
\end{importantboxtitle}
We refer to Appendix~\ref{sec:proof_lemma4.5} for the proof of the statement above. Here, instead, we will give an idea of how the general proof works for the two moduli case. For simplicity, let us focus on a polynomially tamed function which depends on two saxions only as
\begin{equation}
	\label{StrProbe_f}
	f (a,s) = \sum\limits_{m_1, m_2} \rho_{m_1 m_2} (a^1, a^2) (s^1)^{m_1} (s^2)^{m_2} \,.
\end{equation}
Let us assume that $|f(a,s)| \prec 1$ on all the linear paths in \eqref{Linear_Sigmalin}. We want to show that $|f(a,s)| \prec 1$ in all the growth sector $\Sigma_2$. In order to achieve this, it is enough to show that
\begin{equation}
  \label{StrProbe_proofa}
  |f(a,s)| \leq \sum\limits_{m_1,m_2} | \rho_{m_1 m_2} (a^1,a^2)| (s^1)^{m_1} (s^2)^{m_2} \prec 1\,.
\end{equation}
However, since $|f(a,s)| \prec 1$ for linear paths, the powers $m_1$, $m_2$ in which the saxionic fields appear in \eqref{f-expansion} are greatly constrained. In fact, $|f(a,s)| \prec 1$ has to hold for both the path choices
\begin{equation}
  \label{StrProbe_proofb}
  \begin{aligned}
  &|f(a,s)| \prec 1 \qquad \text{on}\quad \mathcal{P}^{(1)}_{\bf e} \qquad \Rightarrow \qquad m_1 \leq 0\,,
  \\
  &|f(a,s)| \prec 1 \qquad \text{on}\quad \mathcal{P}^{(2)}_{\bf e} \qquad \Rightarrow \qquad m_1 + m_2 \leq 0\,.
  \end{aligned}	
\end{equation}
Thus, in turn, \eqref{StrProbe_proofa} holds over the full sector $\Sigma_2$.

It is then immediate to show that the linear paths \eqref{StrProbe_paths} can be identified with the test paths \eqref{StrProbe_paths}. Assume that $e^\alpha\neq 0$ for $\alpha \leq n_0$, then the paths in \eqref{StrProbe_paths} can be written as
\begin{equation}
	\label{StrProbe_pathsb}
	\begin{aligned}
		&\frac{\varphi^1 - \varphi^1_0}{e^1} = \frac{\varphi^2 - \varphi^2_0}{e^2} = \cdots = \frac{\varphi^{n_0} - \varphi^{n_0}_0}{e^{n_0}} \,,\\
		&\varphi^\alpha = \varphi^\alpha_0 \qquad \text{for}\quad  n_0 + 1 \leq  \alpha \leq n\,. 
	\end{aligned}
\end{equation}
These coincide with the test paths in \eqref{Linear_Sigmalin}, upon identifying $\beta_\alpha = 1/e^\alpha$, and choosing ${\rm Im}\, \varphi^\alpha_0 =  e^\alpha$ for $\alpha \leq n_0$. Notice that, since $e^\alpha$ can be regarded as string charges, $e^\alpha$ are quantized and we may assume that $e^\alpha \in \mathbb{Z}$. Thus, the coefficients $\beta_\alpha$ are rational numbers.

Therefore, the above statement can be rephrased as a statement for the string flows as follows:
\begin{importantboxtitle}{Curve-reduction lemma using BPS-strings}
  \label{lemma-4.5b} Consider a polynomially tamed function $f$ and a monomially tamed function $g$. Assume that
  \begin{equation}
    |f| \prec |g| \qquad \text{on} \quad  \mathcal{P}^{(i)}_{\bf e} \quad \forall\; i\,,
  \end{equation}
  then
  \begin{equation}
    |f| \prec |g|  \qquad \text{on}\quad \Sigma_I\,.
  \end{equation}
\end{importantboxtitle}

\subsection{Test strings and the Distance Conjecture}
\label{sec:DC_StrProbe}

In Section~\ref{sec:Tame_DC} we illustrated how the Distance Conjecture can be expressed in a path independent fashion by employing monomially and polynomially tamed functions. We now re-investigate the statements made in Section~\ref{sec:Tame_DC} in light of the findings of the previous sections. In fact, we will display that, in order to satisfy the Distance Conjecture in a path independent way in a wide region of the moduli space, it is enough that it satisfied on a subset of curves in that region.

Preliminarily, let us consider a simpler case than the one examined in Section~\ref{sec:Tame_DC}. Namely, let us assume that both $e^{-\lambda d(s,s_0)} $ and the masses of the candidate infinite tower of states $M_n$ are both monomially tamed on $\Sigma_I$, that is:
\begin{equation}
	\begin{array}{l}
		e^{-\lambda d(s,s_0)} \sim (s^1)^{k_1} \cdots (s^1)^{k_n} 
		\\
		M_n \sim (s^1)^{m_1} \cdots (s^1)^{m_n} 
	\end{array} \qquad \text{on} \quad \Sigma_I\,.
\end{equation}
Thus, we can apply the reasoning of the previous section: if $M_n \sim e^{-\lambda d}$ along curves, then such a relation holds throughout $\Sigma_I$.
This guarantees that the distance conjecture holds on $\Sigma_I$ and the emergence of a tower of states with masses such that $M_n \sim e^{-\lambda d(s,s_0)}$ does not depend on the path taken towards the field space boundary. Remarkably, as a byproduct of this analysis, this further shows that a single tower is enough in order to realize the Distance Conjecture.

Now let us assume the less constraining case in which $e^{-\lambda d}$ is polynomially tamed on $\Sigma_I$. We will additionally assume the existence of a finite number of tower of states, with masses $M_n^{(a)}$ that also behave as polynomially tamed functions. Analogously to the reasoning of Section~\ref{sec:Tame_DC}, one can perform an appropriate partition of the set $\Sigma_I$ into a finite number of subsets $\{\mathcal{U}_{\mathsf{A}}^{(a,k)}\}$. On each subset $\mathcal{U}_{\mathsf{A}}^{(a,k)}$, both $e^{-\lambda d}$ and the tower $M_n^{(a)}$ exhibit a leading behavior. If the towers $M_n^{(a)}$ are good candidate towers for realizing the Distance Conjecture only two cases are allowed:
\begin{itemize}
	\item on $\mathcal{U}_{\mathsf{A}}^{(a,k)}$ both $e^{-\lambda d}$ and the tower $M_n^{(a)}$ are monomially tamed. Then, in $\mathcal{U}_{\mathsf{A}}^{(a,k)}$, we can apply the same reasoning as above: if the Distance Conjecture is realized along the one-dimensional curves \eqref{StrProbe_pathsc}, it is then realized everywhere within $\mathcal{U}_{\mathsf{A}}^{(a,k)}$;
	\item on $\mathcal{U}_{\mathsf{A}}^{(a,k)}$ both $e^{-\lambda d}$ and the tower $M_n^{(a)}$ are strictly polynomially tamed, namely $e^{-\lambda d} \prec g_1$ and $M_n^{(a)} \prec g_2$ for some monomially tamed functions $g_1$ and $g_2$, but $e^{-\lambda d} \not\sim g_1$ and $M_n^{(a)} \not\sim g_2$; in order for the Distance Conjecture to hold on $\mathcal{U}_{\mathsf{A}}^{(a,k)}$, we can minimally require that $e^{-\lambda d}, M_n^{(a)} \prec g$, for some monomially tamed function such that $g \to 0$ for any path leading to the boundary. Such a requirement can be tested via the curve-reduction lemma using BPS-strings: if $e^{-\lambda d}, M_n^{(a)} \prec g$ hold for every axion-string-induced path \eqref{StrProbe_pathsc}, then $e^{-\lambda d}, M_n^{(a)} \prec g$ everywhere on $\mathcal{U}_{\mathsf{A}}^{(a,k)}$. This renders the tower $M_n^{(a)}$ a good candidate for realizing the Distance Conjecture; however, checking that $e^{-\lambda d} \sim M_n^{(a)}$ on $\mathcal{U}_{\mathsf{A}}^{(a,k)}$ goes beyond the scope of the curve-reduction lemma; as mentioned in Section~\ref{sec:Tame_DC} such a check requires to know the leading behavior \eqref{f-rphatuc} of $e^{-\lambda d}$ and $M_n^{(a)}$ on the subset $\mathcal{U}_{\mathsf{A}}^{(a,k)}$.
\end{itemize}

This novel viewpoint strengthens the Distant Axionic String Conjecture proposed in \cite{Lanza:2021udy}. The Distant Axionic String Conjecture asserts that any infinite distance point can be reached as endpoint of an axion string flow. As reviewed in Section~\ref{sec:axion_strings}, axion strings generate the backreactions \eqref{CosmStr_solvarphib} on the moduli fields; these are mapped to the families \eqref{StrProbe_pathscn} of paths in the moduli space. Moreover, as we move towards the field space boundaries along the paths \eqref{StrProbe_pathscn}, the axion string generating the flow becomes tensionless. Thus, the Distant Axionic String Conjecture delivers a bottom-up perspective on the origin of the EFT breaking at any infinite distance limit: the axion string is the object that, with its infinite tower of oscillatory modes, generates the infinite tower of states that can be a candidate to realize the Distance Conjecture. As a signal of such an EFT breaking, in \cite{Lanza:2021udy}, it was shown that, along the linear backreaction of BPS axion strings \eqref{StrProbe_pathscn}, the axion string becomes tensionless and the EFT cutoff $\Lambda_{\text{\tiny{EFT}}}$ has to consistently become smaller and smaller. In \cite{Lanza:2021udy} it was further proposed that the EFT cutoff $\Lambda_{\text{\tiny{EFT}}}$ is bounded by the tensionless axion string as
\begin{equation}
	\label{DistAx_cutoff}
	\left(\frac{\Lambda_{\text{\tiny{EFT}}}}{M_{\rm P}}\right)^2 \sim \left( \frac{\mathcal{T}_{\rm str}}{M_{\rm P}^2} \right)^{\rm w} \qquad \text{along \eqref{CosmStr_solvarphib}}\,,
\end{equation}
for some \emph{scaling weight} ${\rm w} \in \mathbb{N}_{>0}$ along the paths \eqref{StrProbe_pathscn}. We can now revisit these statements, generalizing them, in light of the discussion above.

Generically, it is too strong to assume that \eqref{DistAx_cutoff} holds along any path that leads to infinite distance, for the integrality of the scaling weight is too restrictive. However, in order for the axion string to signal the EFT breaking, it is enough that the EFT cutoff is always upper bounded by any axion string tension as $\Lambda^2_{\text{\tiny{EFT}}} \leq \mathcal{T}_{\rm str}$. Therefore, we can proceed as follows. For any $\Sigma_I$ we choose an EFT cutoff $\Lambda_{\text{\tiny{EFT}}}^{(I)}$ which dictates when the EFT is broken within the field space subregion $\Sigma_I$. The region $\Sigma_I$ can be probed via the axion string flows \eqref{StrProbe_pathscn}; let us denote with $\mathcal{T}_{\rm str}^{\rm min}$ the minimal string tension among all axion strings that generate the flow. Then, the consistency of the EFT requires that $(\Lambda_{\text{\tiny{EFT}}}^{(I)})^2 \leq \mathcal{T}_{\rm str}^{\rm min}$ everywhere in $\Sigma_I$. We rephrase this condition as
\begin{equation}
	\label{DistAx_cutoffb}
	(\Lambda_{\text{\tiny{EFT}}}^{(I)})^2 \prec \mathcal{T}_{\rm str}^{\rm min} \qquad \text{in $\Sigma_I$}\,.
\end{equation}
Such a condition guarantees that the emergence of an axion-string-induced infinite tower of states is responsible for the EFT breaking along \emph{every} path that leads to infinite distance. 

However, the more general statement \eqref{DistAx_cutoffb} follows from the scaling behavior \eqref{DistAx_cutoff} provided that the quantities appearing in \eqref{DistAx_cutoff} obey certain tameness conditions. First, we assume that the EFT cutoff scale $\Lambda_{\text{\tiny{EFT}}}$ is determined by the lightest mass of the infinite tower of states that emerge at infinite distance. Thus, as in \eqref{DC_hyp1}, we consider $\Lambda_{\text{\tiny{EFT}}}$ a polynomially tamed function of the scalar fields. The behavior of the string tension can be inferred from the very expression \eqref{AxStr_Tstr}-\eqref{AxStr_soldual}. In stringy EFTs -- as we will see in the next section for Type IIB EFTs -- $e^{K(\phi,\bar\phi)}$ is monomially tamed. Consequently, as shown in Appendix~\ref{sec:rm_rp}, the string tension \eqref{AxStr_Tstr}, being given by the derivative of the K\"ahler potential, is generically polynomially tamed. However, let us assume here the stricter hypothesis that also the string tension \eqref{AxStr_Tstr} is monomially tamed. For instance, in all the concrete EFT models considered in \cite{Lanza:2021udy}, the string tension \eqref{DistAx_cutoff} is monomially tamed in any given set $\Sigma_I$. Then, we can straightforwardly apply the conclusions of Section~\ref{sec:roughlylinear}: whenever \eqref{DistAx_cutoff} holds along any string backreaction spanning $\Sigma_I$, then \eqref{DistAx_cutoffb} follows. This result thus greatly expands and generalizes the findings of \cite{Lanza:2021udy}, and shows how the Tameness Conjecture can be used to refine pre-existing Swampland Conjectures.


\section{Tameness in Type IIB EFTs}
\label{sec:IIB}

The findings of the previous sections are general and far-reaching. The aim of this section is to give evidence to the claims made above by showing how tame couplings appear in concrete stringy EFTs. Specifically, we will focus on four-dimensional EFTs that are obtained after compactifying the ten-dimensional Type IIB string theory over a Calabi-Yau three-fold. After first reviewing some salient features of such a family of EFTs, we will introduce the central objects of our analysis, namely the \emph{Hodge inner products}. The Hodge inner products determine many of the couplings entering the EFT, such as the gauge couplings, the scalar potential, or the masses and tensions of certain BPS objects. We will then illustrate how the EFT couplings so determined are not only tame functions of the moduli, but they are either monomially tamed or polynomially tamed functions.

\subsection{Type IIB effective field theories}
\label{sec:hier_IIB}

To set the ground for the forthcoming sections, here we review some features of four-dimensional effective field theories that originate from the compactification of Type IIB string theory over Calabi-Yau three-folds or orientifolds thereof.

\subsubsection{Type IIB $\mathcal{N}=2$ effective field theories}
\label{sec:hier_N2}

We start by outlining some basic features of $\mathcal{N}=2$ four-dimensional supergravities that are obtained after compactifying the ten-dimensional Type IIB string theory over a Calabi-Yau three-fold $Y$ \cite{Louis:2002ny,Grimm:2004uq}. The resulting $\mathcal{N}=2$ four-dimensional theory is populated by the gravity multiplet, whose bosonic components are the graviton $g_{\mu\nu}$ and the graviphoton $A^0$, a set of $h^{2,1}$ vector multiplets, accommodating $h^{2,1}$ complex scalar fields $\phi^i$ and $h^{2,1}$ real vector fields $A^i$, and $h^{1,1}+1$ hypermultiplets. Throughout this section, we will disregard the hypermultiplet sector even though we expect our approach can be extended to this sector.

The scalar fields $\phi^i$, $i = 1,\ldots, h^{2,1}$ within the $h^{2,1}$ vector multiplets are associated to the deformations of complex structure of $Y$ as follows. Let us introduce a real, integral basis of three-forms $\gamma_{\mathcal{I}}$, $\mathcal{I} = 1,\ldots, 2h^{2,1}+2$ of $H^3(Y)$. The complex structure moduli appear in the expansion of the holomorphic three-form $\Omega$ as
\begin{equation}
	\label{N=2_Omega}
	\Omega = \Pi^{\mathcal{I}} (\phi) \gamma_{\mathcal{I}} = {\bm \Pi}^T (\phi) {\bm \gamma}\,, \qquad \Pi^{\mathcal{I}} (\phi) = \int_{\Gamma_{\mathcal{I}}} \Omega\ ,
\end{equation}
with the \emph{periods} $\Pi^{\mathcal{I}} (\phi)$ being holomorphic function of the fields $\phi^i$. Here we have introduced a basis $\Gamma_{\mathcal{I}}$ of three-cycles such that $\int_{\Gamma_{\mathcal{I}}}  \gamma_{\mathcal{J}} = \delta_\mathcal{J}^\mathcal{I}$.
The decomposition \eqref{N=2_Omega} is general but, in what follows, it will be useful to be more specific about the choice of basis $\gamma_{\mathcal{I}}$. In particular, let us introduce a symplectic basis of three-cycles $\Gamma^{\mathcal{I}} = (A^I, B_J)$ of $Y$, and a dual basis of three-forms $\gamma_{\mathcal{I}} = (\alpha_I, \beta^J)$, with $I, J= 1, \ldots, h^{2,1} + 1$ such that
\begin{equation}
	\label{Sympl_basis}
	\int_Y \alpha_I \wedge \beta^J = \int_{B_J} \alpha_I =  - \int_{A^I} \beta^J = \delta_I^J\,, \qquad \int_Y \alpha_I \wedge \alpha_J = \int_Y \beta^I \wedge \beta^J  = 0\,.
\end{equation}
Then, the holomorphic three-form $\Omega$ can be expanded in terms of the symplectic basis as in \eqref{N=2_Omega} with the periods
\begin{equation}
	\label{IIB_periodsymplb}
	{\bm \Pi} (\phi) =  \begin{pmatrix}
		\int_{B_I} \Omega \\  -\int_{A^I} \Omega
	\end{pmatrix}  =  \begin{pmatrix}
		X^I (\phi) \\  -\mathcal{F}_I(\phi)
	\end{pmatrix} \, ,
\end{equation}
where $X^I(\phi)$ and $\mathcal{F}_I(\phi)$ holomorphic functions of the complex structure moduli $\phi^i$.

The four-dimensional action describing the interactions among the bosonic components of the $\mathcal{N}=2$ gravity multiplet and the $h^{2,1}$ vector fields is
\begin{equation}
	\label{N2_action}
	S = \int \left( \frac{1}2 M^2_{\rm P} R *1 - M^2_{\rm P} K_{i\bar\jmath}^{\rm cs} {\rm d} \phi^i \wedge * {\rm d} \bar{\phi}^{\bar\jmath} + \frac14 {\rm Im} \mathcal{N}_{IJ} F^I \wedge * F^J + \frac14 {\rm Re} \mathcal{N}_{IJ} F^I \wedge F^J \right)\,,
\end{equation}
Here $R$ is the Ricci scalar and $K_{i\bar\jmath}^{\rm cs} = \partial_{\phi^i} \partial_{\bar{\phi}^{\bar\jmath}} K^{\rm cs}$ is the K\"ahler metric, with the K\"ahler potential $K^{\rm cs}$ specified by the periods as
\begin{equation}
	\label{IIB_Kcs}
	K^{\rm cs}(\Phi,\bar\Phi) = -\log \left(  \im \int_Y \Omega \wedge \bar\Omega \right)=- \log \im\, {\bm \Pi}^T \eta \bar{\bm \Pi} = - \log \im(\bar{X}^I \mathcal{F}_I -  X^I \bar{\mathcal{F}}_I)\,.
\end{equation}
where we have introduced the intersection matrix $\eta_{\mathcal{I} \mathcal{J}} = \int_Y \gamma_{\mathcal{I}} \wedge \gamma_{\mathcal{J}}$, computed out of the symplectic basis $(\alpha^I,\beta_J)$ employing \eqref{Sympl_basis}. 

We will assume that there exists a prepotential $\mathcal{F}(X)$: this is a homogeneous function of degree two in the projective coordinates $X^I$, using which the quantities $\mathcal{F}_I$ appearing in \eqref{IIB_periodsymplb} can be understood as derivatives of the prepotential $\mathcal{F}_I = \frac{\partial \mathcal{F}}{\partial X^I}$. The dynamics of the abelian gauge fields $A^I = (A^0, A^i)$, with field strengths $F^I = {\rm d} A^I$, is dictated by the matrix 
\begin{equation}
	\label{IIB_NIJb}
	\mathcal{N}_{IJ} = \bar{\mathcal{F}}_{IJ} + 2 \im \frac{{\rm Im} 	\mathcal{F}_{IK} X^K {\rm Im} 	\mathcal{F}_{JL} X^L}{ {\rm Im} 	\mathcal{F}_{MN} X^M X^N }\,.
\end{equation}
As is clear from \eqref{N2_action}, the matrix \eqref{IIB_NIJb} determines both the gauge couplings via its imaginary part and may deliver a $\theta$-term via its real part.

Four-dimensional Type IIB EFTs can be populated by extended objects that stem from higher-dimensional branes wrapped on some internal cycles. Here we will focus on D3-branes wrapped on internal three-cycles $\Gamma$. Thus, in the four-dimensional EFT such D3-branes appear as particles, to which we will refer as `D3-particles'. The mass $M_{\bf q}$ of a BPS D3-particle is obtained from the central charge $\mathcal{Z}_{\bf q}$ as \cite{Hull:1994ys,Ceresole:1995ca}:
\begin{equation}
	\label{IIB_D3M}
	M_{\bf q} = |\mathcal{Z}_{\bf q}| = e^{\frac{K^{\rm cs}}2} \left| \int_{\Gamma_{\bf q}} \Omega \right|  = e^{\frac{K^{\rm cs}}2} \left| \int_Y q \wedge \Omega \right| \,,
\end{equation}
with ${\bf q}$ the D3-particle \emph{elementary charges} and $q$ the three-form Poincar\'e dual to the three-cycle $\Gamma_{\bf q}$. In the following, it will be useful to expand the three-form $q$ in the symplectic basis as $q = \alpha_I p^I - q_I \beta^I$. Accordingly, the elementary charge vector ${\bf q}$ can be split
\begin{equation}
	\label{IIB_D3q}
	{\bf q} = \begin{pmatrix}
		p^I \\ -q_I
	\end{pmatrix}\,, \qquad q_I, p^I \in \mathbb{Z}\,,
\end{equation}
and \eqref{IIB_D3M} can be recast as
\begin{equation}
	\label{IIB_D3Mb}
	M_{\bf q} = e^{\frac{K^{\rm cs}}{2}} |{\bf q}^T\,\eta\, {\bm \Pi}(\phi)|\,.
\end{equation}

We will refer to $q_I$ as the D3 elementary \emph{electric} charges and $p^I$ as the D3 elementary \emph{magnetic} charges. The physical charge of a D3-particle can be obtained out of the elementary charges ${\bf q}$ as 
\begin{equation}
	\label{IIB_D3Q}
	\mathcal{Q}^2_{\bf q} = \frac12  \int_Y q \wedge \star q  = - \frac12 {\bf q}^T \mathbb{M} {\bf q}\,,
\end{equation}
with the matrix $\mathbb{M}$ that, in the symplectic basis, can be conveniently rewritten as
\begin{equation}
	\label{IIB_Mmat}
	\mathbb{M} = \begin{pmatrix}
		{\rm Im} \mathcal{N} + {\rm Re} \mathcal{N} ({\rm Im}\mathcal{N})^{-1} {\rm Re} \mathcal{N} & {\rm Re} \mathcal{N} ({\rm Im} \mathcal{N})^{-1}
		\\
		({\rm Im} \mathcal{N})^{-1} {\rm Re} \mathcal{N}  & ({\rm Im}\mathcal{N})^{-1} 
	\end{pmatrix}\,.
\end{equation}

It is worth recalling that the above definition of physical charge in \eqref{IIB_D3Q} carries information about the gauge couplings associated to the ${\rm U}(1)$ abelian gauge one-forms $A^I$. In order to exhibit this, let us restrict to electric D3-particles with sole non-null charges ${\bf q}_{\rm el} = (q_I)$. The physical charge of an electric D3-particle is
\begin{equation}
	\label{IIB_N2_Q2el}
	\mathcal{Q}^2_{{\bf q}_{\rm el}} =  - \frac12{\bf q}_{\rm el}^T\,({\rm Im} \mathcal{N})^{-1} \, {\bf q}_{\rm el} \,.
\end{equation}
As is clear from the general form of the $\mathcal{N}=2$ vector multiplet action \eqref{N2_action}, the matrix $({\rm Im} \mathcal{N})^{-1}$ delivers the gauge coupling functions $g^2_I(\phi)$. Thus, introducing a basis $\{{\bf q}_{\rm el}^{(I)}\}$ of electric elementary charges, we identify the gauge couplings
\begin{equation}
	\label{IIB_N2_g2}
	g^2_I(\phi) = - \frac12 ({\bf q}_{\rm el}^{(I)})^T ({\rm Im} \mathcal{N})^{-1} {\bf q}_{\rm el}^{(I)} = \mathcal{Q}^2_{{\bf q}_{\rm el}^{(I)}}\,.
\end{equation}

\subsubsection{Type IIB $\mathcal{N}=1$ effective field theories}
\label{sec:hier_N1}

Let us now consider the $\mathcal{N}=1$ supergravity theories obtained after compactifying Type IIB string theory over orientifolds $\hat{Y}$ of Calabi-Yau three-folds. The field content of the four-dimensional theory is thus an appropriate projection of the one characterizing the $\mathcal{N}=2$ theories examined in the previous section \cite{Grimm:2004uq}. In particular, the Calabi-Yau holomorphic three-form enjoys an expansion similar to \eqref{N=2_Omega} in terms of odd three-cycles where now ${\mathcal{I}} = 1, \ldots, 2h^{2,1}_- + 2$, and are holomorphic functions of the complex structure moduli $\phi^i$, $i = 1,\ldots, h^{2,1}_-$, embedded within $h^{2,1}_-$ chiral multiplets $\Phi^i$. Similarly, one can introduce a symplectic basis $ (\alpha_I, \beta^J)$, with $I, J= 1, \ldots, h^{2,1}_- + 1$ so that the periods can be recast as in \eqref{IIB_periodsymplb}. Furthermore, unlike the previous section, we will keep track of the axio-dilaton and the K\"ahler moduli. The former is most readily accommodated in the lowest component of a chiral multiplet $\tau = C_0 + \im e^{-\phi} $,
with $C_0$ the RR zero-form and the ten-dimensional dilaton $\phi$ related to the string coupling as $g_{\rm s} = e^{\phi}$. Instead, the K\"ahler moduli $v_\lambda$ are obtained by expanding the K\"ahler two-form $J$ over a basis of two-forms $[D^\lambda]$, $\lambda = 1, \ldots, h^{1,1}_+$, Poincar\'e dual of a basis of divisors $D^\lambda \in H_4(\hat Y, \mathbb{Z})$, as $J = v_\lambda [D^\lambda]$. For simplicity, we will restrict ourselves to the compactifications over Calabi-Yau three-folds with $h^{1,1}_- = 0$. Then, the K\"ahler moduli $v^\lambda$, alongside the $C_4$-axions $a^\lambda = \int_{D^\lambda} C_4$, are accommodated within additional $h^{1,1}_+$ chiral coordinates $u^\lambda = a^\lambda + \im s^\lambda$ with $s^\lambda = \frac12 \int_{D^\lambda} J\wedge J  = \frac12 \kappa^{\lambda\rho\sigma} v_\lambda v_\sigma$, with $\kappa^{\lambda\rho\sigma}$ intersection numbers.

Collecting all the complex scalar fields as $\varphi^\alpha = (\phi^i, u^\lambda, \tau)$, the bosonic effective action describing the interactions among them is
\begin{equation}
	\label{N1_action}
	S=\int \left(\frac{1}{2} M^2_{\rm P} R*1-M^2_{\rm P} K_{\alpha \bar \beta}\,{\rm d} \varphi^\alpha\wedge * {\rm d}\bar\varphi^{\bar \beta} -  V * 1 \right)\, .
\end{equation}
with the K\"ahler metric $K_{\alpha\bar{\beta}}\equiv \partial_\alpha\partial_{\bar \beta}K$ and $V$ the scalar potential. Under the assumption that $h^{1,1}_- = 0$, the K\"ahler potential entering \eqref{N1_action} splits as
\begin{equation}
	\label{IIB_K}
	K = K^{\rm cs} + K^{\rm ks} - \log \left[- \im (\tau -\bar \tau) \right]\, ,
\end{equation}
with 
\begin{equation}
	\label{IIB_Kcsb}
	K^{\rm cs} = -\log \left(  \im \int_{\hat Y} \Omega \wedge \bar\Omega \right)=- \log \im\, {\bm \Pi}^T \eta \bar{\bm \Pi} = - \log \im(\bar{X}^I \mathcal{F}_I -  X^I \bar{\mathcal{F}}_I)\, ,
\end{equation}
and 
\begin{equation}
	\label{IIB_Kk}
	K^{\rm ks} = -2 \log \int_{\hat Y} J \wedge J \wedge J = - 2 \log \kappa^{\lambda\rho\sigma} v_\lambda v_\rho v_\sigma\,.
\end{equation}
with the latter obeying the no-scale condition $K^{\lambda\bar\rho}_{\rm ks} K^{\rm ks}_{\lambda} K^{\rm ks}_{\bar\rho} = 3$ \cite{Grimm:2004uq}.

We will further focus on cases for which the scalar potential is generated solely by the Gukov-Vafa-Witten superpotential \cite{Gukov:1999ya}
\begin{equation}
	\label{N1_WGVW}
	W (\phi) = M_{\rm P}^3 \int_{\hat{Y}} \Omega \wedge G_3 = M_{\rm P}^3\, {\bf g}^T\, \eta\, {\bm \Pi} (\phi)\,, \qquad G_3 = F_3 - \tau H_3 =  {\bf g} \, {\bm \gamma} =({\bf f} - \tau {\bf h} )\, {\bm \gamma}  \, .
\end{equation}

Then, the scalar potential entering \eqref{N1_action} can be obtained via the usual Cremmer et al. formula \cite{Cremmer:1982en}:
\begin{equation}
	\label{IIB_Vfluxquad}
	V_{\bf g} = e^K (K^{\alpha \bar\beta} D_\alpha W \bar{D}_{\bar\beta} \bar W - 3 W \bar{W}) =  - \frac12 \, {\bf g}^T\, \eta\, \mathbb{T} \, \eta\, {\bf g}\,,
\end{equation}
where we have introduced the K\"ahler covariant derivative $D_\alpha = \partial_\alpha + K_\alpha = \partial_{\varphi^\alpha} + \frac{\partial K}{\partial \varphi^\alpha}$. Furthermore, \eqref{IIB_Vfluxquad} manifestly exhibits the quadratic dependence on the background fluxes, with positive semi-definite symmetric matrix
\begin{equation}
	\label{IIB_Tmat}
	\mathbb{T}^{\mathcal{I}\mathcal{J}} \equiv 2M^4_{\rm P}\, e^{K}{\rm Re}\left(K^{i \bar\jmath}_{\rm cs} D_i \Pi^{\mathcal{I}} \bar D_{\bar\jmath} \bar\Pi^{\mathcal{J}} + \Pi^{\mathcal{I}} \bar\Pi^{\mathcal{J}}\right)\,.
\end{equation}

Let us now consider which objects these effective field theories can be coupled to. In general, the D3-particles introduced in Section~\ref{sec:hier_N2} are not  a valid option. In fact, due to the orientifold projection, the D3-particles are here coupled to the gauge one-forms $A_1^I$, with $I = 1,\ldots, h^{2,1}_+$. Thus, in effective four-dimensional theories obtained from Calabi-Yau orientifolds characterized by $h^{2,1}_+ = 0$ the full spectrum of D3 particles is removed. 

An alternative is provided by \emph{membranes}. In four-dimensional effective theories, membranes appear as codimension-one defects, stretching in the time direction and two space directions. In the Type IIB EFTs under scrutiny, BPS membranes can be generically obtained from bound states of D5 and NS5 branes wrapped over internal, special Lagrangian odd three-cycles. In four-dimensional EFTs, membranes can be included as fundamental, semiclassical objects via the action \cite{Bandos:2010yy,Bandos:2012gz,Farakos:2017jme,Farakos:2017ocw,Bandos:2018gjp,Bandos:2019wgy,Lanza:2019nfa,Lanza:2019xxg}
\begin{equation}
	\label{IIB_Smem}
	S_{\rm mem}=-\int_{\mathcal{W}} {\rm d}^3\zeta \sqrt{-\det h}\, \mathcal{T}_{\rm mem} + {\bf q}^T\, \eta \int_{\mathcal{W}}\,  {\bf A}_3 + {\bf p}^T\, \eta \int_{\mathcal{W}}\,  \tilde{\bf A}_3\, ,
\end{equation}
where ${\bf q}, {\bf p} \in \mathbb{Z}^{b^3_-}$, with $b^3_- = 2 h^{2,1}_- + 2$ the \emph{elementary membrane charges}, $\mathcal{T}_{\rm mem} $ the moduli-dependent membrane tension and ${\bf A}_3$, $\tilde{\bf A}_3$ sets of $b^3_-$ three-forms. The three-forms ${\bf A}_3$, $\tilde{\bf A}_3$ can be obtained by reducing the ten-dimension gauge six-forms respectively dual to $C_2$ and $B_2$ \cite{Bielleman:2015ina,Carta:2016ynn,Lanza:2019nfa}. Furthermore, in the first, Nambu-Goto term in \eqref{IIB_Smem} we have introduced the coordinates $\zeta^{\hat \imath}$, ${\hat \imath} =1,2,3$ parametrizing the membrane worldvolume $\mathcal{W}$, and $h_{{\hat \imath}  {\hat \jmath}}$ is the pullback of the spacetime metric to the membrane worldvolume. Requiring that the membranes are BPS objects, maximally preserving a half of the bulk supersymmetry, fixes the tension to be
\begin{equation}
	\label{IIB_Tmemgen}
	\mathcal{T}_{\rm mem} = 2 M_{\rm P}^3 e^{\frac K2} \left| \int_{\hat Y} \Omega \wedge n \right|= 2 M_{\rm P}^3 e^{\frac K2} | ({\bf q} - \tau {\bf p} )^T\eta \, {\bf \Pi} (\phi)|\,,
\end{equation}
with $n = q - \tau p$, where $q, p \in H^3_-(\mathbb{Z})$ are Poincar\'e dual to the three-cycle wrapped by the D5-NS5-bound state. We also recall that, analogously to the D3 particles in $\mathcal{N}=2$ supergravity, we can define the \emph{physical charge} of a membrane as
\begin{equation}
	\label{IIB_Qmem}
	\mathcal{Q}_{\rm mem}^2 =  \int_{\hat Y} n \wedge \star \bar{n} = - ({\bf q} - \tau {\bf p})^T \eta\, \mathbb{T}\, \eta\, ({\bf q} - \tau {\bf p})\, ,
\end{equation}
by employing the same matrix $\mathbb{T}$ defined in \eqref{IIB_Tmat}. 

The role of membranes is to induce flux transitions across various spacetime regions. For instance, consider a single flat BPS membrane, stretching across $z=0$. Then, the membrane separates the spacetime into two regions, distinguished by the values of the background fluxes: assume that for $z<0$ the fluxes ${\bf g}_< = {\bf f}-\tau {\bf h}$; then, the membrane makes the background fluxes `jump' so that the region $z>0$ is characterized by the background fluxes ${\bf g}_> = {\bf f} + {\bf q}-\tau ({\bf h}+{\bf p})$. 

However, as noticed in \cite{Lanza:2019xxg}, it is generically not possible to include an arbitrary number of membranes within the EFT while still guaranteeing (off-shell) $\mathcal{N}=1$ supersymmetry, tadpole cancellation condition and consistency of the effective description. For Type IIB $\mathcal{N}=1$ EFTs it was shown in \cite{Lanza:2019xxg} that supersymmetry and tadpole condition impose that the maximal amount of gauge three-forms ${\bf A}_3$, $\tilde{\bf A}_3$ that the EFT may be endowed with is $b^3_-$. Consequently, the maximal number of independent elementary charges ${\bf q}$, ${\bf p}$ that may appear in the action \eqref{IIB_Smem} is also $b^3_-$. We will define ${\bf n}$ the maximal $b^3_-$ independent elementary charges, and $\mathcal{T}_{\bf n}$ the membrane tension \eqref{IIB_Tmemgen} associated to such a choice of elementary charges. Furthermore, enforcing that membranes can be treated semiclassically requires to further impose $\mathcal{T}_{\rm mem} >  M_{\rm P}^3$ and, additionally, that the jump induced by the membrane in the scalar potential is still described within the same EFT with the cutoff $\Lambda_{\text{\tiny EFT}}$ implies that the membrane charge ${\bf n}$ has to be picked in the \emph{consistent EFT flux lattice}
\begin{equation}
	\label{IIB_GammaEFT}
	\Gamma_{\text{\tiny EFT}} = \left\{ {\bf n} \in \mathbb{Z}^{b^3_-} \;|\;    \Lambda_{\text{\tiny EFT}}^3 < \mathcal{T}_{\bf n}  <\Lambda_{\text{\tiny EFT}} M_{\rm P}^2 \right\}.
\end{equation}
For instance, for  small string coupling $g_{\rm s} \ll 1$, the D5-membranes, which induce jumps of RR-fluxes, are parametrically lighter than NS5-membranes, that induce jumps of NS-NS fluxes. We may then assume that, appropriately choosing the cutoff $\Lambda$, the EFT flux lattice is at most a subset of the RR-flux lattice.

Before concluding this section, it is worth noticing that, by comparing the physical charges \eqref{IIB_Qmem} with the general expression for the scalar potential \eqref{IIB_Vfluxquad}, it becomes clear that the physical charges of membranes \eqref{IIB_Qmem} can also be thought of as the scalar potential generated by a flux that is equal to the membrane charge. Namely, given ${\bf n} \in \Gamma_{\text{\tiny EFT}}$
\begin{equation}
	\label{IIB_Q2n}
	\mathcal{Q}_{\bf n}^2 = - {\bf n}^T \eta\, \mathbb{T}\, \eta\, {\bf n}= 2 V_{\bf n}\,.
\end{equation}
Alternatively, \eqref{IIB_Q2n} can be understood as the potential generated by a BPS membrane that interpolates between an EFT with null scalar potential ${\bf f} = {\bf h} = 0$, and one with a scalar potential as in \eqref{IIB_Vfluxquad}, with ${\bf g} = {\bf n}$. This identification will be useful in order to infer properties of the scalar potential from the properties of the generating membranes.

\subsection{Type IIB complex structure sector and Hodge theory}
\label{sec:hodgereview}

The Type IIB EFTs reviewed in the previous section can be neatly and generically described by using Hodge theory. In this section we review some basic facts about Hodge theory so as to setup the notation, and highlight the main results that we will employ in the following section. Here we will be brief and we refer to \cite{MR840721,Grimm:2018ohb,Grimm:2018cpv} for further details on the subject.

Let us focus on the complex structure moduli space of a Calabi-Yau three-fold $Y$. Denote $H^3(Y, \bbZ)$ the middle cohomology of the Calabi-Yau with integer coefficients (similarly for rational and complex coefficients). Let $Q$ be the (anti-symmetric) intersection pairing: For three-forms $u$ and $v$, 
\begin{equation}
	Q(u, v) = \int_{Y} u \wedge v\, .
\end{equation}
The complex cohomology $H^3(Y, \bbC)$ carries a Hodge decomposition
\begin{equation}
	H^3(Y, \bbC) = \bigoplus_{p + q = k} H^{p, q}\, ,
\end{equation}
such that $H^{p, q} = \conj{H^{q, p}}$. Equivalently one can define a filtration $F^p = \bigoplus_{r \ge p} H^{r, s}$ on $H^3(Y, \bbC)$ satisfying $F^{p} \oplus \conj{F}^{4 - p} \cong H^3(Y, \bbC)$ for all $p$. To go back to the Hodge decomposition, set $H^{p, q} = F^p \cap \conj{F}^q$. We will use the Hodge filtration and the Hodge decomposition interchangeably in the following to denote a Hodge structure and we suppress the superscript in $F^p$ to denote the entire Hodge filtration. With a Hodge structure $F$, we have the Weil operator $C_F$ acting on $H^3(Y, \bbC)$:
\begin{equation}
	C_F u = \im^{p - q} u\, ,\quad\textrm{for } u \in H^{p, q}\, .
\end{equation}
Note that the pairing $Q$ is almost an hermitian inner product. To make it into a genuine hermitian inner product, we use the Weil operator $C_F$. The Hodge inner product $h$ is defined as follows. For $u, v$ complex three-forms,
\begin{equation}
	\label{Hodge_hq}
	h_F(u, v) = Q(u, C_F\conj{v})\, .
\end{equation}
And the \emph{Hodge norm} of a three-form given by
\begin{equation}
	\label{Hodge_hn}
	\norm{u}^2_F := h_F(u, u)\, .
\end{equation}

On the middle cohomology of Calabi-Yau threefolds, the Weil operator coincides with the Hodge star operator, so the Hodge inner product can also be written in the usual way
\begin{equation}
	\label{Hodge_h}
	h_F(u, v) = \int_{Y}  u \wedge \star \conj{v}\, .
\end{equation}

Indeed, some of the quantities that characterize the Type IIB EFTs reviewed in the previous section can be elegantly recast as Hodge inner products or Hodge norms. Consider first the $\mathcal{N}=2$ Type IIB EFTs reviewed in Section~\ref{sec:hier_N2}. By recalling that $\star \bar \Omega = \im \bar\Omega$, it is immediate to see that the complex structure K\"ahler potential \eqref{IIB_Kcs} can be written as
\begin{equation}
	\label{IIB_N2_Kcsb}
	e^{-K^{\rm cs}} = \norm{ {\bm \Pi} }_F^2\,.
\end{equation}
By comparing with \eqref{Hodge_h} and employing the just found \eqref{IIB_N2_Kcsb}, it can be shown that the mass of a BPS D3-particle \eqref{IIB_D3M} can be written as follows
\begin{equation}
	\label{IIB_N2_D3Mh}
	M_{\bf q}^2 = M_{\rm P}^2 \frac{| \langle {\bf q}, {\bf \Pi} \rangle |^2}{\norm{ {\bf \Pi} }^2}\,.
\end{equation}
On the other hand, the physical charge of a D3-particle \eqref{IIB_D3Q} can be most readily recast as a Hodge norm by using \eqref{Hodge_h} as
\begin{equation}
	\label{IIB_N2_D3Qh}
	\mathcal{Q}_{\bf q}^2 = \norm{ {\bf q}}_F^2\,.
\end{equation}

A similar analysis can be carried out for the quantities that distinguish the $\mathcal{N}=1$ Type IIB EFTs reviewed in Section~\ref{sec:hier_N1}. The complex structure K\"ahler potential \eqref{IIB_Kcsb} can be written as in \eqref{IIB_N2_Kcsb} in terms of the $h^{2,1}_-$ complex structure moduli. Moreover, the tension of BPS membranes \eqref{IIB_Tmemgen} can be recast as
\begin{equation}
	\label{IIB_N1_T2b}
	\mathcal{T}^2_{\bf n} =  2 M_{\rm P}^6 \frac{e^{K^{\rm ks}}}{g_{\rm s}} \frac{| \langle {\bf n}, {\bf \Pi} \rangle |^2}{\norm{ {\bf \Pi} }^2} \,,
\end{equation}
while its physical charge \eqref{IIB_Qmem} can be expressed as a Hodge norm
\begin{equation}
	\label{IIB_N1_Q2b}
	\mathcal{Q}^2_{\bf n} = 2 \norm{ {\bf n} }^2\, .
\end{equation}

Now let us vary the complex structure moduli of $Y$. In a local patch of the singularity in the moduli space, we use $\varphi^i$ to denote the complex structure moduli. We adopt the same convention as in Section \ref{sec:axion_strings}, so the singularity will be at $\varphi^i \to \im \infty$. Locally in the moduli space, the cohomology $H^3(Y, \bbZ)$ can be regarded as fixed, and by varying the complex structure of the Calabi-Yau, we get a family of Hodge structures labelled by $\varphi^i$. We denote $F_\varphi$ the Hodge structure on the middle cohomology $H^3(Y, \bbC)$ when the complex structure moduli take value $\varphi^i$. And the variation of the Weil operator, hence the Hodge inner product, is denoted by adding a subscript $h_\varphi(u, v) = Q(u, C_{F_\varphi} \conj{v})$ (also $\norm{u}^2_\varphi$). In this paper, we are mostly interested in the behavior of the inner product $h_\varphi$ as $\varphi$ approaches some singularities in the moduli space. For later convenience, we also recall the nilpotent orbit theorem: Near the singularity $\varphi^i \to \im\infty$, the Hodge structure $F_\varphi$ has the following asymptotic form
\begin{equation} \label{eqn:nilpotent_orbit}
	F_\varphi = e^{\varphi^i N_i} F_{\textrm{nil}} + \cO(e^{2\pi\im\varphi^i})\, ,
\end{equation}
where, for each $i$, the nilpotent matrix $N_i$ is the logarithm of the monodromy matrix as one loops around $z^i \to e^{2\pi\im} z^i$, and $F_{\textrm{nil}}$ is the so called nilpotent orbit. Equation \eqref{eqn:nilpotent_orbit} clearly distinguishes between the contributions that are polynomial in $\varphi^i$ and those that are exponential in $\varphi^i$.

To study the asymptotic behaviors of the Hodge norm, one can invoke the well-known growth theorem \cite{MR0382272,MR840721,MR817170}. Namely, there is a decomposition of $H^3(Y, \bbQ)$ into rational subspaces
\begin{equation}\label{eqn:J-splitting}
	H^3(Y, \bbQ) = \bigoplus_{\ell_1, \ldots, \ell_n} V_{\ell_1, \ldots, \ell_n}\, ,
\end{equation}
and such decomposition naturally extends to the complex cohomology. Each subspace $V_{\ell_1, \ldots, \ell_n}$ is characterized by the property that for every three-form $u \in V_{\ell_1, \ldots, \ell_n}$, we have
\begin{equation}
	\label{Hodge_normest}
	\norm{u}^2_\varphi \sim \left( \frac{s^1}{s^2} \right)^{\ell_1} \cdots \left( \frac{s^{n - 1}}{s^n} \right)^{\ell_{n - 1}} (s^n)^{\ell_n}\, ,
\end{equation}
where we recall that $s^i = \Im \varphi^i$. 

For instance, \eqref{Hodge_normest} can be exploited in order to estimate the growth of the complex structure K\"ahler potential \eqref{IIB_N2_Kcsb}, or the physical charges of D3-particles \eqref{IIB_N2_D3Qh} or of D5-NS5-bound state membranes \eqref{IIB_N1_Q2b}. In the upcoming sections we will also deliver estimates for the BPS masses \eqref{IIB_N2_D3Mh} and tensions \eqref{IIB_N1_T2b}.

\subsection{The polynomially tamed behavior of the Hodge inner product}
\label{sec:hodgemetric}

In Section~\ref{sec:Tame_bound} we have introduced two families of functions: the monomially tamed and the polynomially tamed functions, the former with definite, path independent growth in the set \eqref{def-SigmaI}, and the latter with a generic path dependent growth. Indeed, in Section~\ref{sec:Tame_bound} we claimed that such special classes of tame functions are enough to study the behavior of most of the couplings entering any EFT that ought to be consistent with quantum gravity. In this section we prove that this is the case for the Type IIB EFTs that we introduced in Section~\ref{sec:hier_IIB}.

In fact, the Hodge inner product \eqref{Hodge_hq} and consequently the Hodge norm \eqref{Hodge_hn} are quantities for which we can predict their polynomial behavior close to any singularity in the complex structure moduli space. In \cite{BKT} the following statements, that are crucial for the following analysis, have been delivered:
\begin{importantboxtitle}{Hodge inner product growth}
	Let $u \in V_{\ell_1, \ldots, \ell_n}$ and $v \in V_{\ell_1', \ldots, \ell_n'}$, then
	\begin{enumerate}
		\item $\norm{u}^2_\varphi$ is monomially tamed; \label{lemma-4.7:Part1}
		\item $h_\varphi(u, v)$ is polynomially tamed. \label{lemma-4.7:Part3}
	\end{enumerate}
\end{importantboxtitle}

We will refer to these two statements as \emph{Hodge inner product growth}.  Statement (\ref{lemma-4.7:Part1})  implies that \emph{any} Hodge norm displays a definite growth in the set \eqref{def-SigmaI}. The growth of the Hodge norm in (\ref{lemma-4.7:Part1}) has been already extensively used in literature: in \cite{Grimm:2018ohb,Grimm:2018cpv,Grimm:2019bey,Bastian:2020egp} it was employed in relation to estimate the growth, or in \cite{Grimm:2019ixq,Bastian:2021hpc,Grimm:2021ckh} to explore the structure of EFTs' vacua. The statement (\ref{lemma-4.7:Part1}) is related to \eqref{Hodge_normest}, and the monomially tamed behavior of $\norm{u}^2_\varphi$ is fixed by the location of $u$ within certain ${\rm sl}_2$-eigenspaces $V_{\ell_1,\ldots,\ell_n}$. However, it is worth stressing that the statement (\ref{lemma-4.7:Part1}) carries more information, for it explains how the growth of the actual norm may differ from the estimate in \eqref{Hodge_normest}. Indeed, recalling the meaning of the symbol $\sim$ from Section~\ref{sec:roughlylinear}, the statement (\ref{lemma-4.7:Part1}) can be recast as
\begin{equation}
	\label{Hodge_normestb}
	\left( \frac{s^1}{s^2} \right)^{\ell_1} \cdots \left( \frac{s^{n - 1}}{s^n} \right)^{\ell_{n - 1}} (s^n)^{\ell_n} \prec \norm{u}^2_\varphi \prec \left( \frac{s^1}{s^2} \right)^{\ell_1} \cdots \left( \frac{s^{n - 1}}{s^n} \right)^{\ell_{n - 1}} (s^n)^{\ell_n}\, .
\end{equation}
Namely, the norm $\norm{u}^2_\varphi$ is upper and lower bounded by the \emph{same} monomial, whose coefficient is a real restricted analytic function $\rho = \rho(a^1, \ldots, a^n, s^1, \ldots, s^n)$:
\begin{equation}
	\label{Hodge_normestc}
	C_1 \rho \left( \frac{s^1}{s^2} \right)^{\ell_1} \cdots \left( \frac{s^{n - 1}}{s^n} \right)^{\ell_{n - 1}} (s^n)^{\ell_n} \leq \norm{u}^2_\varphi \leq C_2 \rho \left( \frac{s^1}{s^2} \right)^{\ell_1} \cdots \left( \frac{s^{n - 1}}{s^n} \right)^{\ell_{n - 1}} (s^n)^{\ell_n}\, ,
\end{equation}
where $C_1$ and $C_2$ are positive numbers.

However, the Hodge inner product growth does not only allow for recovering the Hodge norm estimates of \cite{Grimm:2018ohb,Grimm:2018cpv,Grimm:2019bey,Bastian:2020egp}, but it additionally provides information about the growth of the off-diagonal terms of the Hodge inner product in its part \ref{lemma-4.7:Part3}. Indeed, since $h(u,v)$ is polynomially tamed, by employing the results of Section~\ref{sec:roughlylinear}, one can further bound $h(u,v)$ with a monomially tamed function in $\Sigma$, provided that such a bound holds for curves \eqref{Linear_Sigmalin}:
\begin{equation}
	|h(u,v)| \prec  \left( \frac{s^1}{s^2} \right)^{\ell'_1} \cdots \left( \frac{s^{n - 1}}{s^n} \right)^{\ell'_{n - 1}} (s^n)^{\ell'_n}\
\end{equation}
for some $\ell'_i$.

\subsection{Tame EFT couplings and the Distance Conjecture in Type IIB EFTs}
\label{sec:tame_IIB}

Here we are going to apply the results of the previous section to the class of Type IIB EFTs introduced in Section~\ref{sec:hier_IIB}. We will illustrate that all the couplings that enter either the $\mathcal{N}=2$ and the $\mathcal{N}=1$ Type IIBs EFT reviewed therein are tame, and we will estimate the polynomially tamed and monomial growth for all of them. We will then illustrate the implications for the Distance Conjecture.

\subsubsection{Tameness in $\mathcal{N}=2$ Type IIB EFTs}
\label{sec:tame_IIB_N2}

We start by revisiting the four-dimensional $\mathcal{N}=2$ Type IIB EFTs reviewed in Section~\ref{sec:hier_N2}. The field space metric $K_{i\bar\jmath}^{\rm cs}$ appearing in the effective action \eqref{N2_action} is determined by the K\"ahler potential \eqref{IIB_Kcs}.  However, the K\"ahler potential \eqref{IIB_Kcs} can be written as an Hodge norm as in \eqref{IIB_N2_Kcsb}. As such, the statement (\ref{lemma-4.7:Part1}) of the Hodge inner product growth implies that $e^{-K^{\rm cs}}$ is a monomially tamed function; its behavior close to the singularity $s^i \to \infty$ is then 
\begin{equation}
	\label{IIB_N2_Kcsc}
	e^{-K^{\rm cs}} \sim \left( \frac{s^1}{s^2} \right)^{d_1} \cdots \left( \frac{s^{n - 1}}{s^n} \right)^{d_{n - 1}} (s^n)^{d_n}\, ,
\end{equation}
for some $d_i \in \mathbb{Z}$, $i = 1, \ldots, h^{2,1}$. In turn, the K\"ahler metric can be computed as
\begin{equation}
	\label{IIB_N2_Kcscij}
	K_{i\bar\jmath}^{\rm cs} = e^{2 K^{\rm cs}} \left(  \partial_i e^{- K^{\rm cs}} \partial_{\bar\jmath} e^{- K^{\rm cs}} -  e^{- K^{\rm cs}} \partial_i \partial_{\bar\jmath} e^{- K^{\rm cs}}   \right)\,.
\end{equation}
As proved in Appendix~\ref{sec:rm_rp_prop}, the derivative of a monomially tamed function is a polynomially tamed function. Thus, the K\"ahler metric $K_{i\bar\jmath}^{\rm cs}$ is tame, and specifically polynomially tamed.

The K\"ahler potential is not the sole quantity introduced in Section~\ref{sec:hier_N2} that stems from the Hodge inner product, for also the couplings of the gauge sector are related to Hodge inner products. Indeed, the matrix $\mathbb{M}$ that appears in \eqref{IIB_D3Q} can be regarded as a Hodge inner product. According to the second Hodge inner product growth statement (\ref{lemma-4.7:Part3}), then the elements of the matrix $\mathbb{M}$ are tame, and exhibit a polynomially tamed behavior. Therefore, also the elements of the matrices ${\rm Im}\mathcal{N}_{IJ}$ and ${\rm Re}\mathcal{N}_{IJ}$ may be generically assumed to behave polynomially tamed. This proves that also the dynamics of the gauge fields $A^I$ is regulated by polynomially tamed couplings. 

These features in turn influences the physical properties of the D3-particles. In fact, as reviewed in Section~\ref{sec:hodgereview}, the physical charge \eqref{IIB_D3Q} can be most readily recast as a Hodge norm as in \eqref{IIB_N2_D3Qh}. Therefore, the physical charge exhibits a monomially tamed behavior
\begin{equation}
	\label{IIB_N2_D3Qm}
	\mathcal{Q}_{\bf q}^2 \sim \left( \frac{s^1}{s^2} \right)^{\ell_1} \cdots \left( \frac{s^{n - 1}}{s^n} \right)^{\ell_{n - 1}} (s^n)^{\ell_n}\, ,
\end{equation}
for some $\ell_i \in \mathbb{Z}$ that depends on the elementary charges ${\bf q}$. 
The behavior of the mass of BPS D3-particles \eqref{IIB_D3M} is slightly more subtle. In the language introduced in Section~\ref{sec:hodgereview}, the mass \eqref{IIB_D3M} of such BPS D3-particles can be written as in \eqref{IIB_N2_D3Mh}. Thus, stemming from a general inner product, the mass \eqref{IIB_D3M} is generically polynomially tamed, as predicted by the statement (\ref{lemma-4.7:Part3}).

In turn, by recalling that the behaviors of the gauge couplings can be inferred from the growth of the physical charges of electric D3-particles as in \eqref{IIB_N2_Q2el}, we conclude that gauge couplings share a similar monomially tamed behavior:
\begin{equation}
	\label{IIB_N2_D3Q}
	g^2_I = \norm{ {\bf q}_{\rm el}^{(I)} }^2 \sim \left( \frac{s^1}{s^2} \right)^{\ell_1} \cdots \left( \frac{s^{n - 1}}{s^n} \right)^{\ell_{n - 1}} (s^n)^{\ell_n}\,.
\end{equation}
The monomial tameness of the gauge couplings has profound phenomenological implications. 
In fact, within any given set $\Sigma_I$ as defined in \eqref{def-SigmaI}, it is always possible to single out a set of gauge couplings that falls off faster than any other and in a path-independent way. Thus, as the field space boundary is approached, such a set of gauge couplings may ungauge some of the zero-form gauge symmetries associated to the gauge fields $A^{(I)}$, signalling the appearance of zero-form global symmetries in these limits. However, in any consistent theory of quantum gravity, such corners of the moduli space in which global symmetries emerge ought to be obstructed. Indeed,  limits of vanishing gauge couplings are related to the emerge of an infinite tower of states. 

In fact, in \cite{Grimm:2018ohb,Grimm:2018cpv,Gendler:2020dfp,Palti:2021ubp} it was proposed that in the $\mathcal{N}=2$ EFTs Type IIB under examination the Distance Conjecture is realized by infinite towers of BPS D3-particles. We here revisit and expand these results. Let us preliminarily recall some basic features of the construction of the infinite towers proposed in \cite{Grimm:2018ohb,Grimm:2018cpv}, and we refer to the original works for a detailed discussion. Consider an infinite tower of D3-particles specified by elementary charges ${\bf q}^{(k)}$. In order for this tower to be a candidate for realizing the Distance Conjecture, its constituting BPS D3-particles have to be exhibit the following features:
\begin{description}
	\item[Weak Coupling] The D3-particles constituting the infinite tower are weakly coupled. Namely, $\mathcal{Q}^2_{{\bf q}^{(k)}} \to 0 $ as the singularity $s^i \to \infty$ is reached;
	\item[Stability] The tower is stable under decays.
\end{description}
The weak coupling condition is enough to guarantee that the D3-particles in the tower become massless as the singularity is reached. In fact, by exploiting the Cauchy-Schwarz inequality, the D3-particle masses \eqref{IIB_D3M} can be generically bounded by their physical charge as
\begin{equation}
	\label{IIB_N2_D3MQ_CS}
	M_{{\bf q}^{(k)}}^2  = M_{\rm P}^2 \frac{| \langle {{\bf q}^{(k)}} , {\bf \Pi} \rangle |^2}{\norm{ {\bf \Pi} }^2} \leq M_{\rm P}^2  \mathcal{Q}^2_{{\bf q}^{(k)}} \, .
\end{equation}
Thus, requiring that $\mathcal{Q}^2_{{\bf q}^{(k)}} \to 0$ asymptotically, it is enough to guarantee that also the masses $M_{{\bf q}^{(k)}}$ fall down asymptotically, rendering the tower massless towards the boundary. As in \cite{Grimm:2018ohb,Grimm:2018cpv,Gendler:2020dfp,Palti:2021ubp}, one can show that the infinite towers are constituted by electric particles. Moreover, their stability can be guaranteed as follows. Consider a \emph{seed charge} ${\bf q}_{\rm s}$. We assume such a seed charge to be of electric type such that its physical charge $\mathcal{Q}^2_{{\bf q}_{\rm s}} \to 0$ towards the field space boundary. Then, a tower of states can be built out of the seed charge by exploiting the infinite-order monodromy matrix $T=e^{N_{(n)}}$ as
\begin{equation}
	\label{IIB_N2_Qtower}
	{\bf q}^{(0)} = {\bf q}_{\rm s} \,, \quad  {\bf q}^{(1)} = T {\bf q}_{\rm s}\,, \quad\ldots\, , \quad {\bf q}^{(k)} = T^k {\bf q}_{\rm s}\,, \quad \ldots
\end{equation}
An infinite tower so constructed is stable against decays into constituents, for no walls of marginal stability is crossed.

But does the tower so built remain relevant by approaching the singularity along any path in  any set $\Sigma_I$ defined in \eqref{def-SigmaI}? This question can be addressed by exploiting the same arguments introduced in Section~\ref{sec:Tame_DC}. In fact, since the physical charges $\mathcal{Q}^2_{{\bf q}^{(k)}}$ of the D3-particles constituting the infinite tower of states are monomially tamed in the saxions, then their behavior is path independent in any given set $\Sigma_I$. Thus, the tower of D3-particles so built remains weakly coupled, with $\mathcal{Q}^2_{{\bf q}^{(k)}} \to 0$, along any path in $\Sigma_I$. Then, using the inequality \eqref{IIB_N2_D3MQ_CS}, we infer that the states become massless along any path in $\Sigma_I$ that approaches the infinite distance boundaries.

\subsubsection{Tameness in $\mathcal{N}=1$ Type IIB EFTs}
\label{sec:tame_IIB_N1}

Let us show how tameness reflects on the couplings that characterize the $\mathcal{N}=1$ Type IIB EFT action \eqref{N1_action}. Due to the similarities to the $\mathcal{N}=2$ Type IIB investigated in the previous section, here we will much briefer. Indeed, since the K\"ahler potential has the same structure as in the $\mathcal{N}=2$ EFTs for the $h^{2,1}_-$ complex structure moduli, also in $\mathcal{N}=1$ Type IIB EFTs $e^{-K^{\rm cs}}$ has a monomially tamed behavior as in \eqref{IIB_N2_Kcsc}. Therefore, the field space metric for the complex structure moduli, which can be computed as in \eqref{IIB_N2_Kcscij}, has tame, polynomially tamed behavior. Moreover, under the assumptions made in Section~\ref{sec:hier_N1}, it is simple to show that the other moduli sectors display terms with tame behavior. In fact, the field space metric for the axio-dilaton $\tau$ is trivially monomially tamed. Moreover, given \eqref{IIB_Kk}, also $e^{-K^{\rm ks}}$ is monomially tamed; thus, the metric for the K\"ahler moduli is tame and polynomially tamed.

Moreover, also the charge of D5-NS5-membranes \eqref{IIB_Qmem} is an Hodge norm and thus exhibits a monomially tamed behavior as for the D3-charges \eqref{IIB_N2_D3Qm}:
\begin{equation}
	\label{IIB_N1_D5NS5Qm}
	\mathcal{Q}_{\bf n}^2 \sim \left( \frac{s^1}{s^2} \right)^{\ell_1} \cdots \left( \frac{s^{n - 1}}{s^n} \right)^{\ell_{n - 1}} (s^n)^{\ell_n}\, ,
\end{equation}
for some $\ell_i \in \mathbb{Z}$ related to the choice of the elementary charges ${\bf n} = {\bf q}-\tau {\bf p}$. On the other hand, the tension of BPS membranes \eqref{IIB_Tmemgen} can be recast as in \eqref{IIB_N1_T2b}, which rather exhibits a polynomially tamed behavior.

The Distance Conjecture can be here realized by considering, for instance, infinite towers of membranes as in \cite{Font:2019cxq,Lanza:2020qmt}.  For simplicity, we will assume that the EFT is defined within regions of weak string coupling, so that the spectrum of EFT membranes determined by \eqref{IIB_GammaEFT} is composed by D5-membranes only, with membrane tension and physical charges given by \eqref{IIB_N1_T2b} and \eqref{IIB_N1_Q2b} with ${\bf n} = {\bf q}$. First, introduce a basis $\{{\bf q}^{(I)}\}$, $I = 1, \ldots , b^3_-$ for the elementary membrane charges within the maximal EFT lattice, namely the RR-fluxes. Then, one can construct towers of stable, weakly coupled BPS D5-membranes as for D3-particles in \eqref{IIB_N2_Qtower}. As for D3-particles, the tameness of the EFT couplings guarantees that such infinite towers of membranes remains relevant along any path in the set $\Sigma_I$.

However, the membrane picture allows one to infer crucial information about the $\mathcal{N}=1$ F-term scalar potential. In fact, recall that the physical charge $\mathcal{Q}^2_{\bf n}$ of a membrane with elementary charge ${\bf n} \in \Gamma_{\text{\tiny EFT}}$ is related to the scalar potential generated by the ${\bf n}$ background fluxes as in \eqref{IIB_Q2n}. Therefore the scalar potential must obey Statement~\ref{lemma-4.7:Part1} of the Hodge inner product growth: 
\begin{equation}
	\label{IIB_N1_VQm}
	V_{\bf n}^2 \sim \left( \frac{s^1}{s^2} \right)^{\ell_1} \cdots \left( \frac{s^{n - 1}}{s^n} \right)^{\ell_{n - 1}} (s^n)^{\ell_n}\, .
\end{equation}
This implies that within any F-term scalar potential of an EFT consistent with quantum gravity it is always possible to single out a term that grows or fall-off faster than any other in a given set \eqref{def-SigmaI}.


\section{Conclusions}

In this work we have expanded on the Tameness Conjecture recently proposed in \cite{Grimm:2021vpn}. In its strong version, it asserts that any EFT coupling, field space, and parameter space ought to be definable in the o-minimal structure $\mathbb{R}_{\rm an,exp}$. Here we have taken a step forward, by refining the focus of the Tameness Conjecture for studying stringy EFTs. Indeed, we observed that the o-minimal structure $\mathbb{R}_{\rm an,exp}$ may be too vast for specifying the couplings in most of the known string theory-originated effective field theory: it is enough to concentrate on a subset of functions definable in $\mathbb{R}_{\rm an,exp}$ that are asymptotically bounded by polynomials. We discussed such functions in 
detail in Section~\ref{sec:Tame_bound} and termed them monomially and polynomially tamed functions. In prominent cases of stringy EFTs, one can indeed \emph{prove} that the EFT couplings do belong to such families. As an important example, in Section~\ref{sec:IIB}, we have shown that the couplings involving the complex structure sectors of the four-dimensional EFTs obtained compactifying Type IIB over a Calabi-Yau three-fold are polynomially or monomially tamed following from a mathematical result of \cite{BKT}.  Albeit in these EFTs the tameness of the field spaces and couplings are a consequence of the underlying Calabi-Yau geometry, the Tameness Conjecture is more general and does neither rely on holomorphicity properties  encountered in these settings 
nor the fact that Calabi-Yau moduli spaces admit a complex structure. Indeed, the polynomial tameness of the couplings can be proved in other contexts and we plan to investigate more general settings in the future.

The proposed refinement of the Tameness Conjecture offered us a novel possibility on how to \emph{test} the behavior of the couplings near any field space boundary. In fact, one can probe the leading behavior of any polynomially tamed function by focusing on a smaller set of paths leading to the boundary. As illustrated in Section~\ref{sec:roughlylinear}, assuming that the field space boundary is reached as the saxionic fields $s^i \to \infty$, the leading behavior of any monomially tamed function is fully determined by how these functions behave on linear paths that the saxions draw towards the boundary; polynomially tamed functions are instead bounded by monomially tamed functions provided that they are bounded by said functions on linear paths spanning the field space region of interest.

The tameness of the EFT couplings is crucial to fully comprehend the physics that emerge in the near-boundary region of the moduli space. In particular, by knowing the generic behavior of the couplings towards the field space boundary one can grasp pathologies that the EFT might exhibit, such as those predicted by the Distance Conjecture. For instance, as explained in Section~\ref{sec:Tame_DC}, by assuming that the EFT couplings are sufficiently tame -- i.e.~they are either monomially tamed or polynomially tamed -- we were able to illustrate how the Distance Conjecture can be realized in a path independent fashion. Namely, if along a given set of paths that leads toward the boundary an infinite tower of states become massless, the tameness properties of the masses of the states constituting the tower guarantees that such a tower becomes massless along any other path reaching the boundary, at least, within a certain sector. Indeed, without assuming that the EFT couplings are tamed it would have been very hard to deliver such path independent statements. Moreover, the tameness of the EFT couplings can be employed to make additional statements on how the Distance Conjecture is realized. Indeed we shown that, within a definable EFT, each boundary region can be partitioned into only finitely many sectors and hence that only a finite number of different towers is needed in order to realize the Distance Conjecture. In turn, following \cite{Lee:2018urn,Lee:2018spm,Lee:2019tst,Lee:2019xtm,Lee:2019wij,Klaewer:2020lfg}, such a statement can be rephrased by asserting that only a finite number of dual theories is required to fully grasp the physics emerging towards any infinite field distance boundary.

Let us note that, to our current understanding, tameness alone is not enough 
in order to guarantee that the Distance Conjecture holds. First, knowledge of the UV completion is required in order to show the existence of the infinite tower of states that should invalidate the EFT at infinite field distance. Additionally, albeit tameness is helpful to identify the subsets where the relation \eqref{Mass+ineq} can be enforced path-independently, it is hard to generically single out the behavior of $e^{-\lambda d}$ in a given asymptotic regime. In fact, it is not clear how the tameness of the field space metric is reflected onto the shape of the geodesic paths and, consequently, on the functional form of geodesic distance. In turn, the knowledge of the specific functional form of the geodesic distance is indeed crucial in order to compute the parameter $\lambda$ in \eqref{Mass+ineq} that appears in the Distance Conjecture. Nevertheless, it is worth remarking that the predictions of the Tameness Conjecture for the realization of the Distance Conjecture can be ameliorated if one renounces the feature that the fall-off of the masses is dictated by the geodesic distance as in \eqref{Mass+ineq}, rather replacing it with some simpler notion of field distance. We leave such an investigation for future work.

Furthermore, the picture we delivered ties in nicely with the Distant Axionic String Conjecture proposed in \cite{Lanza:2020qmt,Lanza:2021udy}. In fact, the linear paths that serve as test paths for the behavior of monomially and polynomially tamed functions may be regarded as induced by the backreaction of axion strings. On the one hand, our findings deliver a mathematical motivation of why the axion strings proposed in \cite{Lanza:2020qmt,Lanza:2021udy} are good candidates to study the near-boundary physics. On the other hand, we have been able to vastly generalize the implications of the Distant Axionic String Conjecture. In fact, in \cite{Lanza:2021udy} it was shown that infinite towers of state emerge along the linear backreaction of axion strings; the tameness of the EFT couplings guarantees that such infinite towers remain relevant for any arbitrary path that leads to the field space boundary.

This work has revolved around the interconnection between the Distance Conjecture and tameness, with the latter helping inquiring how the EFT breaks down towards infinite distance limits  in full generality. However, the implications of the Tameness Conjecture are not limited to the study of the near-boundary physics. The tameness of the EFT couplings can deliver important information about other phenomenological properties that any EFT consistent with quantum gravity is endowed with. For instance, the Tameness Conjecture can be useful to better and more generally address other Swampland Conjectures, and we leave such this exploration for future work.

\vspace{3em}

\noindent{\textbf{Acknowledgments}}

\noindent We would like to thank Fernando Marchesano, Miguel Montero, Mick van Vliet for insightful discussions. 
This research is partly supported by the Dutch Research Council (NWO) via a Start-Up grant and a Vici grant.

\appendix


\section{A primer on monomially and polynomially tamed functions}
\label{sec:rm_rp}

In this appendix, we examine some useful properties of monomially and polynomially tamed functions in detail.
We will first elaborate on restricted analytic functions defined over a polydisk $\Delta^n$, and then move into the definition of monomially and polynomially tamed functions defined over the sets defined in \eqref{def-SigmaI}.
In order to lighten the presentation we will show typical examples and non-examples of the corresponding types of functions.
Then we derive some properties of monomially and polynomially tamed functions that are useful in their application in physics.

As this appendix is very general, we would like to adopt a set of notations that is slightly different from the main text, but more suitable for a mathematical discussion.
We denote a disk in $\bbR^2$ by
\begin{equation}
	\Delta = \{z = u + \im v \mid |z| < 1\}\, ,
\end{equation}
and a polydisk is $\Delta^n$ with coordinates $(z^1, \ldots, z^n)$.
A punctured disk is defined via
\begin{equation}
	\Delta^* = \{ 0 < |z| < 1 \}\, ,
\end{equation}
and a punctured polydisk is denoted by $(\Delta^*)^n$.
The punctured polydisk $(\Delta^*)^n$ is not simply connected. Its universal covering space is the $n$-dimensional upper half plane
\begin{equation}
	\cH^n = \{\varphi^k = x^k + \im y^k \mid y^k > 0 \}\, ,
\end{equation}
and the covering map $p: \cH^n \to (\Delta^*)^n$ is given by
\begin{equation}\label{eqn:covering_map}
	p(\varphi^k) = e^{2 \pi \im \varphi^k}\, .
\end{equation}

\subsection{Restricted analytic functions}\label{app:restricted_analytic_functions}

In order to analyze the monomially and polynomially tamed functions, let us first clarify the definition of restricted analytic functions.
Let us recall that an analytic function defined on a domain is a function that coincides with its own Taylor series on that domain.
Analytic functions are necessarily smooth, but the converse is not true.\footnote{Let $\rho$ be a smooth function defined on an open set $U \subs \bbR^n$.
	Then $\rho$ is analytic on $U$ if for every $x \in U$, there is an open ball $V$ satisfying $x \in V \subs U$, and positive constants $C, R$, such that, over the entire $V$,
	\begin{equation*}
		\left|\frac{\partial_{\mu_1} \ldots \partial_{\mu_n} \rho}{\mu_1!\cdots\mu_n!}\right| \le \frac{C}{R^{\mu_1 + \cdots + \mu_n}}\, . 
	\end{equation*}
	For more information, see \cite{MR1916029}.}
A \emph{restricted analytic function} $f: B(R) \to \bbR$ is a real analytic function defined on an open ball of radius $R$ inside some $\bbR^n$ that can be extended to an analytic function on a \emph{strictly larger} $B(R')$ with $R' > R$.

An example of a restricted analytic function is the sine function restricted to the interval $(-1, 1)$.
The corresponding non-example would be the sine function defined on the whole $\bbR$.
Another non-example is given by the series
\begin{equation}
	f(x) = \sum_{n = 0}^\infty x^n\, ,
\end{equation}
which converges on $(-1, 1)$ to $\frac{1}{1 - x}$.
When regarded as an analytic function defined on $(-1, 1)$, the function $f$ is \emph{not} restricted analytic, since there is no strictly larger domain in $\bbR$ over which the series converges.
However, its restriction $f|_{(-a, a)}$ is a restricted analytic function whenever $0 < a < 1$.

It turns out that the precise shape of the domain of convergence $B(R)$ is not really important in defining the $\Ranexp$-structure.
The crucial point is that the functions are required to be `over-convergent' in the sense that they converge in open sets that are strictly larger than their defining domain.
This intuition is implicitly assumed in the following discussions.
The readers will find that $B(R)$ is replaced by the multi-cube $[0, 1]^n$ in much of the literature on tame geometry.
Their definition using $[0, 1]^n$ and ours using $B(R)$, following \cite{BKT}, all generate the same class of $\Ranexp$-definable subsets.

Let us also comment on restricted analytic functions with defining domains contained in $\bbC^n$.
In fact, in our applications, the domain of a restricted analytic function is always in $\bbC^n$.
However, it is in general non-trivial to directly work with notions like `$\bbC_{\mathrm{an, exp}}$' \cite{zbMATH01787782} and the way to bypass this issue is to identify $\bbC^n$ with $\bbR^{2n}$ by the usual decomposition into real and imaginary parts, when we talk about the $\Ranexp$-structure on $\bbC^n$.
Then, for any $z \in \bbC^n$, we decompose
\begin{equation}
	z^k = u^k + \im v^k \in \bbC\, ,
\end{equation}
and a real restricted analytic function $f$ is defined as a power series over some open ball $B(R)$ that converges on a strictly larger ball $B(R')$ with $R' > R$
\begin{equation}
	f(u, v) = \sum_{\ui, \uj \in \bbN^n} a_{\ui\uj} (u^1)^{i_1}\cdots(u^n)^{i_n} (v^1)^{j_1}\cdots(v^n)^{j_n}\, ,
\end{equation}
where $\bbN$ is the set of non-negative integers, and we have used the multi-index notation $\ui = (i_1, \ldots, i_n)$.

To be more concrete, we use the notion of real restricted analytic functions over the punctured polydisk $(\Delta^*)^n \subs \bbR^{2n}$ in the following discussion. According to the above discussion, these are the functions that are real analytic on $(\Delta^*)^n$ and are actually also analytic on some larger domain containing $(\Delta^*)^n$ inside $\bbR^{2n}$. What is especially important is that such functions have good behavior at the puncture $z = 0$ as they come from functions that are analytic at $z = 0$.
As an example of an analytic but not restricted analytic function on $\Delta^*$, take $n = 1$, and consider $f(z) = 1/z$ over the punctured disk $\Delta^*$.
This function is analytic over $\Delta^*$, but its singularity at $z = 0$ forbids it being restricted analytic over $\Delta^*$.

Later we will frequently use the coordinates on the covering space $\cH^n$ as arguments in a restricted analytic function $f$ defined over $\Delta^n$.
Let us be clear about what we actually mean using $n = 1$ as an example.
Decomposing the covering map $z = e^{2\pi\im\varphi}$, for $\varphi \in \cH$, into real and imaginary parts, we have
\begin{equation}\label{eqn:real_covering_map}
	u = e^{-2 \pi y}\cos(2 \pi x)\, ,\quad v = e^{-2 \pi y}\sin(2 \pi x)\, .
\end{equation}
For a restricted analytic function $f(u, v)$ defined over $\Delta$, we write $f(x, y)$ for the function
\begin{equation}
	f(e^{-2 \pi y}\cos(2 \pi x), e^{-2 \pi y}\sin(2 \pi x))\, ,
\end{equation}
and such a function is sometimes also written as $f(z, \zbar)$ or $f(\varphi, \bar\varphi)$ to stress that $f$ is real-analytic instead of holomorphic.

\subsection{Generalities of monomially and polynomially tamed functions}

Monomially and polynomially tamed functions have been recently introduced in \cite{BKT}, where they were called roughly monomial and roughly polynomial functions. As stressed throughout this work, these special kinds of functions are ubiquitous in effective field theories emerging from string theory. Indeed, couplings and physical quantities -- that are $\mathbb{R}_{\rm an,exp}$-definable -- typically belong to these special families of functions.
Thus, due to their importance, we here systematically discuss the monomially and polynomially tamed functions, and we collect some of their properties.

Our aim is inquiring the growth of physical quantities within the region $\mathcal{E}$ in \eqref{CosmStr_domain} close to the boundary $\varphi^\alpha = \im \infty$. Within $\mathcal{E}$ we identify the subregion, in terms of the $\varphi^\alpha$-coordinates,
\begin{equation}
	\label{Linear_Sigma}
	\Sigma_n = \{ 0 < x^k < 1,\, y^1 \ge y^2 \ge \cdots \ge y^n > 1 \}\, ,
\end{equation}
to which we will oftentimes refer as \emph{growth sector} -- see Figure~\ref{Fig:Infinite_Dist_Sectors} for  a pictorial representation. The region \eqref{Linear_Sigma} singles out a specific ordering for the $y^k$ and dictates the allowed hierarchies among their values. However, we can consider analogous regions with different orderings just by reshuffling the indices in \eqref{Linear_Sigma}.

Given a general, real-analytic function $f$ defined over the growth sector \eqref{Linear_Sigma}. we would like to classify such functions according to their growth or fall-off within the region $\Sigma_n$. For the sake of clarity, we will start with some simple examples. Let us first focus on the case for which the boundary is a codimension-one locus $z = 0$, so that the region in \eqref{Linear_Sigma} is (real) two-dimensional. Let us then consider the following polynomial function:
\begin{equation}
	\label{Linear_1d_f}
	f (x, y) = \sum\limits_{m = m_0}^{\hat m} \rho_{m} (x) y^{m} \,,
\end{equation}
with $m_0, \hat{m}$ being integers and $ \rho_{m} (x)$ real-analytic functions of $x$. Since $x$ parametrizes only the open unit interval, we can safely assume that $|\rho_{m}|$ is upper bounded for any $m$. Consequently, exploiting the fact that $y>1$ in $\Sigma_1$, also $|f|$ can be minimally upper bounded as
\begin{equation}
	\label{Linear_1d_fb}
	| f |  \leq  C y^{\hat m} \quad {\rm in} \quad \Sigma_1\,,
\end{equation}
for some positive number $C$. It is worth stressing that \eqref{Linear_1d_fb} is true for any value of $y$, and thus also true for any path $y(\sigma)$ within $\Sigma_1$. Furthermore, defining an upper bound as in \eqref{Linear_1d_fb} is possible because of the simple structure of \eqref{Linear_1d_f}, which unequivocally allows to single out a monomial with maximal growth within $\Sigma_1$.

However, let us now consider a codimension-two singularity, the region around which is described by a four-dimensional growth sector $\Sigma_2$. In analogy to \eqref{Linear_1d_f}, let us investigate the possible behaviors of the following class of functions
\begin{equation}
	\label{Linear_2d_f}
	f (x,y) = \sum\limits_{m_1, m_2} \rho_{m_1 m_2} (x^1, x^2) (y^1)^{m_1} (y^1)^{m_2} \,,
\end{equation}
where we understand that the sum runs over finitely many integers $m_1$ and $m_2$. Albeit the definition of $\Sigma_2$ constrains $y^1$ and $y^2$ to be mutually bounded as $y^1 \ge y^2>1$, this is \emph{not enough} to single out a leading monomial in \eqref{Linear_2d_f}. In order to better illustrate the issue, consider the function: 
\begin{equation}
	\label{Linear_2d_exf}
	f(y^1, y^2) = \frac{1}{y^1}  + \left( \frac{y^2}{y^1}\right)^2\,,
\end{equation}
and let us investigate the behavior of $|f(y^1,y^2)|$ along the following paths:
\begin{equation}\label{Linear_2d_paths}
	\begin{IEEEeqnarraybox}[][c]{rCl"l}
		\mathcal{P}^1(\sigma) & = & ( y^1 = C \sigma \, , y^2 = y^2_0)\,,         & C\, , y^2_0 >1\, ,\\
		\mathcal{P}^2(\sigma) & = & ( y^1 = C_1 \sigma \, , y^2 = C_2 \sigma )\,, & C_1 > C_2 >1\, ,
	\end{IEEEeqnarraybox}
\end{equation}
specified by $\sigma > 1$ and in which the parameters have been chosen in compatibility with the definition of $\Sigma_2$ in \eqref{Linear_Sigma}. Along these paths, the function \eqref{Linear_2d_exf} can be differently bounded as
\begin{equation}
	\begin{aligned}
		|f(y^1, y^2)| &\leq  \frac{C}{y^1} &{\rm along}\quad \mathcal{P}^1(\sigma) \,,
		\\
		|f(y^1, y^2)| &\leq  C' \left(\frac{y^2}{y^1}\right)^2 &{\rm along}\quad \mathcal{P}^2(\sigma) \,.
	\end{aligned}
\end{equation}
with positive numbers $C$ and $C'$. In other words, the identification of the leading monomial in \eqref{Linear_2d_exf} is \emph{path-dependent}. Similar obstructions in identifying a leading monomial also appear for generic functions defined on a multi-dimensional $\Sigma_n$.

The above example illustrates that it is in general not possible to identify a leading term that determines the growth or the fall-off of even simple functions throughout the full growth sector $\Sigma_n$. It might then seem that minimal bounds such as those above are path-dependent statements. However, we will now show that, under certain conditions, bounds can indeed be formulated throughout $\Sigma_n$, and we will provide a recipe to identify when this is attainable. 

However, we first need to be more specific about the family of functions on which our investigation will be focused. For instance, in the one- and two-moduli cases, the function \eqref{Linear_1d_f} and \eqref{Linear_2d_f} are definitely not general: on the one hand, the non-singular $\rho$-functions appearing in both \eqref{Linear_1d_f} and \eqref{Linear_2d_f} are only $x$-dependent; secondly, if we allow such $\rho$-functions to acquire a $y$-dependence, then we need to be sure that this inclusion does not deliver new singularities spoiling the polynomial growth. The appropriate generalization of the $\rho$-functions in \eqref{Linear_1d_f}, \eqref{Linear_2d_f} and in general multi-moduli cases is given by \emph{restricted analytic functions}. Furthermore, let us note that a main inspiration for our constructions arise from the study of the  growth of physical quantities determined in terms of the Hodge inner product. As argued in Section~\ref{sec:hodgemetric} and Appendix \ref{sec:proof_lemma4.7}, elements of the Hodge inner product are special types of Laurent polynomials with restricted analytic functions as coefficients.

We denote $\cO$ the space of real restricted analytic functions on $(\Delta^*)^n$ expressed in the $\varphi^k$-coordinates:
\begin{equation} \label{eqn:defn_of_cO}
	\cO = \left\{ \rho(\varphi^k, \bar\varphi^k) \, , \textrm{ with } \rho(z^k, \zbar^k)\textrm{ real restricted analytic over } (\Delta^*)^n \right\}\, ,
\end{equation}
where we have used equation \eqref{eqn:covering_map} to transform $z^k = e^{2\pi\im \varphi^k}$. In other words, the functions in $\cO$ are obtained in two steps: take all restricted analytic functions $\rho(z^k, \zbar^k)$ defined on $(\Delta^*)^n$, which are functions of $u^k = \Re z^k$ and $v^k = \Im z^k$, and then transform back to the variables $\varphi^k$. The point is that, over suitable domains, functions in $\cO$ can be expanded in Taylor series in terms of $u^k$ and $v^k$. Such functions encode exponentially corrected quantities in $\varphi^k$.

We further denote the space of polynomials $\cO[x, y, y^{-1}]$ with coefficients in $\cO$ and indeterminate $x^k, y^k, (y^k)^{-1}$. A typical element of this space looks like a finite sum
\begin{equation} \label{eqn:example_cOpolynomial}
	g(x, y) = \sum_{\uk, \um}\rho_{\uk, \um}(u^1, v^1; \cdots; u^n, v^n) (a^1)^{k_1} \cdots (a^n)^{k_n} (s^1)^{m_1} \cdots (s^n)^{m_n}\, ,
\end{equation}
where $\rho_{\uk, \um} \in \cO$ are functions defined in \eqref{eqn:defn_of_cO}, $\uk = (k_1, \ldots, k_n)$ are non-negative integers, and $\um = (m_1, \ldots, m_n)$ are integers. We remind the reader that $u^k$ and $v^k$ are related to $x^k$ and $y^k$ via \eqref{eqn:real_covering_map}. For simplicity we will omit the long list of arguments appearing in \eqref{eqn:example_cOpolynomial} in the following discussion, and whenever we write $g \in \cO[x, y, y^{-1}]$, the function $g$ is assumed to be of the form displayed in \eqref{eqn:example_cOpolynomial}.

We will be mostly interested in ratios between polynomials of the form \eqref{eqn:example_cOpolynomial}, so we define a space $\cO(x, y)$ containing all fractions
\begin{equation}\label{eqn:defn_fraction_field}
	f = \frac{g}{h}\, ,\qquad\textrm{ with } g, h \in \cO[x, y, y^{-1}] \textrm{ and } h \ne 0\, .
\end{equation}
The growth of the functions in $\cO(x, y)$, as one approaches the singularity $\varphi^\alpha \to \im \infty$, can be compared, and it is convenient to recollect the definition of the order relation used in the main text: For any $f, g \in \cO(x, y)$, we write $f \prec g$ if there is a positive constant $C$ such that $f < Cg$ over the entire $\Sigma_n$. We write $f \sim g$, if $f \prec g$ and $g \prec f$.

We are now in the position to introduce the functions defined over $\Sigma_n$ with which we will work in the remainder of the paper. A function $f \in \cO(x, y)$ is \emph{monomially tamed} if
\begin{equation}
	\label{Linear_Romon}
	f  \sim (y^1)^{m_1} \cdots (y^n)^{m_n}  \quad \text{over}\quad \Sigma_n\,,
\end{equation}
for some integers $m_\alpha$. In other words, a monomially tamed function is a function in which we can single out a definite leading monomial throughout the region \eqref{Linear_Sigma}. 

A function $f \in \cO(x, y)$ is \emph{polynomially tamed} if it can be written as a ratio
\begin{equation}
	\label{Linear_Ropol}
	f = \frac{g}{h}\, ,
\end{equation}
where $h \in \cO(x, y)$ is a monomially tamed function, and $f \in \cO[x, y, y^{-1}]$. Intuitively speaking, in contrast with the monomially tamed function, a function being polynomially tamed indicates that there could be several competing leading terms. Namely, a polynomially tamed function is such that 
\begin{equation}
	\label{Linear_Ropolb}
	f  \sim \sum\limits_{i} (y^1)^{m_1^{(i)}} \cdots (y^n)^{m_n^{(i)}}  \quad \text{over}\quad \Sigma_n\,,
\end{equation}
for some sets of integers $m_\alpha^{(i)}$.

\subsection{Characterization of monomially and polynomially tamed functions}\label{app:characterization_m_and_p_tamed_functions}

From the previous discussion we see that the form of monomially and polynomially tamed functions are rather constrained.
The constraints can be utilized to write down these functions more explicitly.
This is the goal of this section.
The organizing principle is to distinguish the units and non-units in $\cO[x, y, y^{-1}]$.
Recall that a unit\footnote{We avoid using the term `invertible element' to distinguish multiplicatively invertible elements from invertible mappings.} in a ring is an element with a multiplicative inverse.

Firstly let us examine the coefficient ring $\cO$.
From the general theory of power series, it can be shown that an element $a \in \cO$, regarded as an analytic function over a strictly larger domain containing $\Delta^n$ in $\bbR^{2n}$, is a unit if and only if $a(\uz = 0) \ne 0$, i.e. $a$ has non-vanishing constant term.
The units in $\cO$ have nice growth property over $\Sigma_n$, namely
\begin{equation}
	a \prec 1\, ,
\end{equation}
for any invertible $a$.
To see this, note that for $n = 1$, in coordinates $(x, y) \in \Sigma_1$,
\begin{equation}
	a(x, y) = a_0 + \sum_{i + j > 0} a_{ij} e^{-2 \pi (i + j)y} \cos^i(2 \pi x) \sin^j(2 \pi x) \prec 1 \quad \textrm{ over } \Sigma_1\, ,
\end{equation}
where $a_0 \ne 0$ and $a_{ij}$ are complex coefficients.
The cases for $n > 1$ follow inductively.

Note that a general unit $a$ in $\cO$ is not necessarily asymptotic to a constant, because its absolute value may not be bounded by any positive constant from below.
For a simple example of such phenomena, take $n = 1$, and $a(u, v) = 1 - 2v$.
This function is clearly restricted analytic on the disk $u^2 + v^2 < 1$.
And it is a unit in $\cO$ because $a(0, 0) = 1$ is non-zero.
Over the disk, the function satisfies $|a| < 3$, so $a \prec 1$ as expected.
However, we have $a(u, \frac{1}{2}) = 0$ for all $u$.
Hence the function $a$ is not bounded by any positive constant from below, and we cannot say that $a \sim 1$.

Now we focus on the polynomial ring $\cO[x, y, y^{-1}]$.
From the general theory of Laurent polynomials, it can be shown that any unit of $\cO[x, y, y^{-1}]$ must be of the form
\begin{equation}
	a(y^1)^{j_1}\cdots(y^n)^{j_n}\, ,
\end{equation}
where $a$ is a unit in $\cO$ and $\uj = (j_1, \ldots, j_n) \in \bbZ^n$.
Thus, units in $\cO[x, y, y^{-1}]$ are almost monomially tamed functions in the sense that $a(y^1)^{j_1}\cdots(y^n)^{j_n} \prec \uy^\uj$.
The converse is obviously not true.

With the above preparation, we can classify each term in a function in $\cO[x, y, y^{-1}]$ into three classes.
More precisely, the general form of each term looks like
\begin{equation}
	a_{\ui\uj} (x^1)^{i_1} \cdots (x^n)^{i_n} (y^1)^{j_1} \cdots (y^n)^{j_n}\, ,
\end{equation}
where $a_{\ui\uj} \in \cO$.
Then we distinguish each term according to the behavior of $a_{\ui\uj}$ and the exponents $\ui = (i_1, \ldots, i_n)$.
We write $\ui = 0$ to denote $i_1 = \cdots = i_n = 0$, and $\ui \ne 0$ means that one of the $i_k \ne 0$.
The asymptotics of each term can be divided into the following three classes
\begin{equation}\label{eqn:classification-of-polynomial-terms}
	a_{\ui\uj} \ux^\ui \uy^\uj \prec
	\begin{cases}
		\textrm{or} \sim \uy^\uj\, , & \textrm{if } a_{\ui\uj} \textrm{ is unit, and } \ui = 0\, ,\\
		\uy^\uj\, ,                  & \textrm{if } a_{\ui\uj} \textrm{ is unit, and } \ui \ne 0\, ,\\
		e^{-2 \pi y^n} \uy^\uj\, ,   & \textrm{if } a_{\ui\uj} \textrm{ is non-unit}.
	\end{cases}
\end{equation}
From Equation \eqref{eqn:classification-of-polynomial-terms}, we see that if a polynomial $f \in \cO[x, y, y^{-1}]$ has a definite leading term, then this term comes from the units, i.e. terms that look like $a\uy^\uj$ for $a \in \cO$ unit.

We will see later that analyzing monomially tamed functions as fractions in $\cO(x, y)$ can be reduced to studying monomially tamed functions in $\cO[x, y, y^{-1}]$.
So let us elaborate on the forms of monomially tamed functions in $\cO[x, y, y^{-1}]$.
Using Equation \eqref{eqn:classification-of-polynomial-terms}, we can fix the forms of such functions rather explicitly.
Namely, the most general polynomial $f \in \cO[x, y, y^{-1}]$ can be split into three parts according to \eqref{eqn:classification-of-polynomial-terms}
\begin{equation}
	f = f_1 + f_2 + f_3\, ,
\end{equation}
where $f_1$ consists of the terms that are of the form $a \uy^\uj$, $a(0) \ne 0$, $f_2$ consists of $a \ux^\ui \uy^\uj$ where $\ui \ne 0$, and $f_3$ consists of $b \ux^\ui \uy^\uj$ where $b(0) = 0$.
Of these three parts, only the $f_1$ can impose a non-vanishing lower bound on $|f|$, while $f_2$ and $f_3$ fail to do so.
Indeed, $|f_2|$ and $|f_3|$ are bounded by $\uy^\uj$ from above, but there is no positive lower bound on these terms.
So they do not restrict $|f|$ from below.
In order to have $f \sim \uy^\us$ for some $\us \in \bbZ^n$, part $f_1$ must be present.
Moreover, one of the terms in $f_1$ has to be asymptotic to $\uy^\us$.

We can spell out the general form of a function $f \in \cO[x, y, y^{-1}]$ satisfying $f \sim \uy^\us$ using the above reasoning.
Write $f = f_1 + f_2 + f_3$, with sums over finitely many terms
\begin{equation}\label{eqn:decomposing_f_into_three_parts}
	f_1 = a_{\us} y^{\us} + \sum_{\uj_1} a_{\uj_1} \uy^{\uj_1}\, ,\qquad f_2 = \sum_{\ui_2, \uj_2} a_{\ui_2, \uj_2} \ux^{\ui_2} \uy^{\uj_2}\, ,\qquad f_3 = \sum_{\ui_3, \uj_3} b_{\ui_3, \uj_3} \ux^{\ui_3} \uy^{\uj_3}\, ,
\end{equation}
where $a_{\uj_1}$ and $a_{\ui_2, \uj_2}$ are units in $\cO$, and $b_{\ui_3, \uj_3}$ are non-units.
Moreover, $a_{\us} \sim 1$ because we assume $f \sim \uy^\us$, implying that there must be a term in $f_1$ providing this leading behavior.

Let us restrict further the sums in $f_1, f_2$ and $f_3$.
Before doing so, we need to present a simple fact.
Namely, assume that $\uy^\us \prec 1$ over $\Sigma_n$, we ask what possible values that the exponents $\us = (s_1, \ldots, s_n)$ can take.
Since we can rewrite
\begin{equation}\label{eqn:ys_in_fractions}
	\uy^\us = \left( \frac{y^1}{y^2} \right)^{-m_1} \left( \frac{y^2}{y^3} \right)^{-m_2} \cdots \left( \frac{y^{n - 1}}{y^n} \right)^{-m_{n - 1}} (y^n)^{-m_n}\, ,
\end{equation}
where
\begin{equation}
	\begin{IEEEeqnarraybox}[][c]{rCl}
		-m_1 & =      & s_1\, ,\\
		-m_2 & =      & s_1 + s_2\, ,\\
		& \vdots &                    \\
		-m_n & =      & s_1 + \cdots + s_n\, .
	\end{IEEEeqnarraybox}
\end{equation}
We deduce that the expression $\uy^\us$ must take the form in \eqref{eqn:ys_in_fractions} with exponents 
\begin{equation}\label{eqn:boundedness_condition_exponents}
	m_1, \ldots, m_n \ge 0\, ,
\end{equation}
so that $\uy^\us \prec 1$.

Focusing on $f_1$, and applying the above fact, we require that $a_{\uj_1}\uy^{\uj_1} \prec \uy^\us$ for all $\uj_1$.
Since $a_{\uj_1} \prec 1$, this translates to a condition on $\uj_1 = (j_{1, 1}, \ldots, j_{1, n})$ that $\uy^{\uj_1 - \us} \prec 1$, we can factor out globally $a_{\us}\uy^\us$ in $f_1$, and then $f_1$ takes the following form
\begin{equation}\label{eqn:cond_f1}
	f_1 = a_{\us} y^{\us} \left\{ 1 + \sum_{\um \ge 0} a_{\um} \left( \frac{y^1}{y^2} \right)^{-m_1} \cdots \left( \frac{y^{n - 1}}{y^n} \right)^{-m_{n - 1}} (y^n)^{-m_n} \right\}\, ,
\end{equation}
where $a_\um$ are units in $\cO$.

Similarly, we factor out $a_\us \uy^\us$ in part $f_2$, and it needs to be of the following form
\begin{equation}\label{eqn:cond_f2}
	f_2 = a_{\us} y^{\us} \sum_{\uk, \um \ge 0} a_{\uk, \um} \ux^\uk \left( \frac{y^1}{y^2} \right)^{-m_1} \cdots \left( \frac{y^{n - 1}}{y^n} \right)^{-m_{n - 1}} (y^n)^{-m_n}\, ,
\end{equation}
where $a_{\uk, \um}$ are units in $\cO$.

There is a minor difference in $f_3$.
Note that $b_{\ui_3, \uj_3}$ are non-units, meaning that $b_{\ui_3, \uj_3}(0) = 0$.
For $n = 1$, such a function looks like
\begin{equation}
	b = \sum_{i + j > 0} b_{ij} e^{-2\pi (i + j) y} \cos^i(2\pi x) \sin^j(2\pi x)\, ,
\end{equation}
and we can pull out an overall factor of $e^{-2\pi y}$ from a non-unit $b$.
A similar conclusion holds for $n > 1$, so we can factor out a $e^{-2 \pi y^n}$ in $b_{\ui_3, \uj_3}$.
Part $f_3$ then takes the form
\begin{equation}
	f_3 = \sum_{\ui_3, \uj_3} \tilde{b}_{\ui_3, \uj_3} e^{-2 \pi y^n} \ux^{\ui_3} \uy^{\uj_3}\, ,
\end{equation}
for some $\tilde{b}_{\ui_3, \uj_3}$.
Then requiring $f_3 \prec \uy^\us$ amounts to requiring that
\begin{equation}
	(y^1)^{j_{3, 1} - s_1} \cdots (y^n)^{j_{3, n - 1} - s_{n - 1}} \prec 1\, .
\end{equation}
So we only need to apply \eqref{eqn:ys_in_fractions} up to index $n - 1$.
With this in mind, pulling out a factor of $a_\us \uy^\us$, part $f_3$ can be written as
\begin{equation}\label{eqn:cond_f3}
	f_3 = a_{\us} y^{\us} \sum_{\substack{m_1, \ldots, m_{n - 1} \ge 0\\m_n \in \bbZ\, , \uk \ge 0}} b_{\uk, \um} \ux^\uk \left( \frac{y^1}{y^2} \right)^{-m_1} \cdots \left( \frac{y^{n - 1}}{y^n} \right)^{-m_{n - 1}} (y^n)^{-m_n}\, ,
\end{equation}
where $b_{\uk, \um}$ are non-units in $\cO$.
Note that in $f_3$ there is no restriction on $m_n$ because of the overall factor $e^{-2 \pi y^n}$ suppressing all powers of $y^n$.

In summary, if $f \in \cO[x, y, y^{-1}]$ is monomially tamed with $f \sim \uy^\us$ on $\Sigma_n$, then $f$ has the following form
\begin{equation}
	f = \rho_0\uy^\us \left\{ 1 + \sum_{\uk, \um \ge 0}\rho_{\uk\um} \ux^\uk \left( \frac{y^1}{y^2} \right)^{-m_1} \cdots \left( \frac{y^{n - 1}}{y^n} \right)^{-m_{n - 1}} \left[(y^n)^{-m_n} + b_{\uk\um} (y^n)^{m_n}\right]\right\}\, ,
\end{equation}
where $\rho_0, \rho_{\uk\um}$ are units in $\cO$, with $\rho_0 \sim 1$, and $b_{\uk\um}$ are non-units in $\cO$ satisfying $b_{\uk\um}(0) = 0$.
The sum contains finitely many non-zero terms.
Note that in the above expression we have combined the conditions \eqref{eqn:cond_f1}, \eqref{eqn:cond_f2}, and \eqref{eqn:cond_f3} into a summation over non-negative multi-indices $\uk$ and $\um$.
We have also used the fact that the sum of a unit and a non-unit is again a unit in $\cO$, as can be seen by evaluating the sum at $\uz = 0$.

With the above preparation on the form of monomially tamed functions in $\cO[x, y, y^{-1}]$, we are now ready to provide a concrete characterization of monomially tamed functions.
Recall that a monomially tamed function $f \in \cO(x, y)$ can be written as a ratio
\begin{equation}
	f = \frac{g}{h}\, ,
\end{equation}
where $g, h \in \cO[x, y, y^{-1}]$, $h \ne 0$, and
\begin{equation}
	f \sim (y^1)^{s_1} \cdots (y^n)^{s_n}\, ,
\end{equation}
for some $(s_1, \ldots, s_n) \in \bbZ^n$.
The observation is that, when the polynomials $g$ and $h$ have no common factor, they must be separately monomially tamed, namely
\begin{equation}
	h \sim 1\, ,\quad\textrm{and}\quad g \sim (y^1)^{s_1} \cdots (y^n)^{s_n}\, .
\end{equation}
To see this, note that any function $f \in \cO[x, y, y^{-1}]$ can be decomposed into three parts as in \eqref{eqn:decomposing_f_into_three_parts} (set $a_\us = 0$ for generality).
If in $g$ and $h$, the corresponding parts $g_1$ and $h_1$ are vanishing, then the fraction $f = g / h$ cannot be monomially tamed; the coefficients are unbounded.
Hence, both $g$ and $h$ must contain parts $g_1$ and $h_1$.
Moreover, there must be unit coefficients $a$ in $g_1$ and $h_1$ that satisfy $a \sim 1$, otherwise the fraction $f$ is still not monomially tamed.
It then follows that $g$ and $h$ must be separately monomially tamed over $\Sigma_n$.
If this is not true, then we can partition $\Sigma_n$ into subsectors, over each of which the functions $g$ and $h$ asymptote to different monomials.
Since we assume that $g$ and $h$ have no common factors, this means that over these partitions the fraction $f$ also asymptotes to different monomials, contradicting the monomially tamed condition of $f$.

\subsection{Properties of monomially and polynomially tamed functions}
\label{sec:rm_rp_prop}

We conclude this section with two basic properties of monomially and polynomially tamed functions.
The first one is that the derivative of a monomially tamed function is polynomially tamed.
To see this, suppose $f = g / h$ is a monomially tamed function, where $g, h \in \cO[x, y, y^{-1}]$ and $g, h$ have no common factors.
Then we can assume that $g$ and $h$ are separately monomially tamed in $\cO[x, y, y^{-1}]$.
Any derivative $f'$ of $f$ then has the following form
\begin{equation}
	f' = \frac{g' h - g h'}{h^2}\, .
\end{equation}
Since both $g'$ and $h'$ are functions in $\cO[x, y, y^{-1}]$, and $h$ is monomially tamed, by definition, we conclude that $f'$ is a polynomially tamed function.

The second one is that the sum of two polynomially tamed functions is again polynomially tamed.
Let $f_1 = g_1 / h_1$ and $f_2 = g_2 / h_2$ be two polynomially tamed functions, where $h_1, h_2 \in \cO(x, y)$ are monomially tamed and $g_1, g_2 \in \cO[x, y, y^{-1}]$.
We further write $h_1 = p_1 / q_1$ and $h_2 = p_2 / q_2$, with the assumption that $p_1, q_1$ have no common factors and are separately monomially tamed in $\cO[x, y, y^{-1}]$.
The same applies to $p_2, q_2$.
Then
\begin{equation}
	f_1 + f_2 = \frac{g_1 q_1 p_2 + g_2 q_2 p_1}{p_1 p_2}
\end{equation}
is polynomially tamed, as the numerator is in $\cO[x, y, y^{-1}]$ and the denominator is monomially tamed.
The last conclusion implies that the set of polynomially tamed functions form a ring.

\section{Monomial bounds for polynomially tamed functions}
\label{sec:proof_lemma4.5}

In Section~\ref{sec:roughlylinear} we stated that polynomially tamed functions can be bounded by a monomial in a wide region of the moduli space, provided that they are bounded only on a given set of curves by said monomial. In this section we deliver a proof for this statement, following closely \cite{BKT}. Since the statement is very general, we will use slightly different notation than in the main text. We denote $\varphi^\alpha \in \Sigma$ a point in the set $\Sigma$, and decompose it as
\begin{equation}
  \varphi^\alpha = x^\alpha + \im y^\alpha\, ,
\end{equation}
so that
\begin{equation}
  \Sigma = \left\{ 0 < x^\alpha < 1, y^1 \ge y^2 \ge \cdots \ge y^n > 1 \right\}\, .
\end{equation}
Moreover, we write $z^\alpha = e^{2\pi\im \phi^\alpha} \in \Delta^*$ as usual. Since we are going to deal with polynomials that depend on many variables, we use the following abbreviations
\begin{equation*}
  f(\ux) := f(x^1, \ldots, x^n)\, , \qquad\textrm{and}\qquad \ux^{\ui} := (x^1)^{i_1} \cdots (x^n)^{i_n}\, ,
\end{equation*}
for $\ui = (i_1, \ldots, i_n)$ integer powers. With $\ui > 0$, we indicate that the inequality holds component-wise.

For convenience, let us first repeat here the statement with the mathematical language of \cite{BKT}:

\begin{lemma} \label{lemma-4.5app}
  Let $f, g \in \cO(x, y)$ with $f$ polynomially tamed and $g$ monomially tamed. Assume that $|f| \prec
  |g|$ when restricted to any set of the form
  \begin{equation}\label{eqn:app:test-paths}
  \Sigma \cap \{\alpha^1 \varphi^1 + \beta^1 = \cdots = \alpha^{n_0} \varphi^{n_0} + \beta^{n_0}, \varphi^{n_0 + 1} = \zeta^{n_0 + 1}, \ldots, \varphi^n = \zeta^n \}
  \end{equation}
  for some $1 \le n_0 \le n, \zeta^{n_0 + 1}, \ldots, \zeta^n \in \cH, \alpha^1, \ldots, \alpha^{n_0} \in \bbQ_+$, and $\beta^1, \ldots, \beta^{n_0} \in \bbR$. Then $|f| \prec |g|$ on  all of $\Sigma$.
\end{lemma}

We would like to stress that the significance of Lemma \ref{lemma-4.5app} is that it allows to establish a \emph{uniform} bound of a polynomially tamed function by a monomially tamed one. Reading this lemma without caution could lead to confusion, as setting $n_0 = 1$ in the test path \eqref{eqn:app:test-paths} seems enough to conclude (incorrectly) that $|f| \prec |g|$ over the entire $\Sigma$. However, one should be careful, as setting $n_0 = 1$ really gives a point-wise condition: Unrolling the definition of the `$\prec$' notation, we see that condition \eqref{eqn:app:test-paths} with $n_0 = 1$ is equivalent to that, for every $\underline{\varphi} = (\varphi^1, \ldots, \varphi^n) \in \Sigma$, one has
\begin{equation}
  |f(\underline{\varphi})| < C(\underline{\varphi}) |g(\underline{\varphi})|\, ,
\end{equation}
where $C(\underline{\varphi}) > 0$ is positive and depends on $\underline{\varphi}$. Thus, condition \eqref{eqn:app:test-paths} does not imply $|f| \prec |g|$ over $\Sigma$, as the latter requires that the prefactor $C$ does not depend on $\underline{\varphi}$.

We now go into its proof. The first step is to notice that it suffices to prove Lemma \ref{lemma-4.5app} for $g = 1$. Indeed, since $g$ is monomially tamed, so is $|g|$. By the definition of polynomially tamed functions, $f$ being polynomially tamed implies that $f/g$ is polynomially tamed. The condition $|f| \prec |g|$ is then equivalent to the condition $|f|/|g| \prec 1$. Lemma \ref{lemma-4.5app} can then be proved by induction on $n$.

The initial case is $n = 1$. Lemma \ref{lemma-4.5app} holds trivially in this case since for $n = 1$ the condition in Lemma \ref{lemma-4.5app} does not restrict $y^1$, and the statement is vacuous. We now assume that Lemma \ref{lemma-4.5app} is true up to $n - 1$ and deduce Lemma \ref{lemma-4.5app} for $n$.

Since $f$ is polynomially tamed, the idea is to analyze each term in the function $f$. Let us first examine the form of $f$. Recall that, as a polynomially tamed function, $f$ lives in $\cO(x, y)$. Rolling out the definition, we have
\begin{equation}
f(x, y) = \sum_{(\ui, \uj) \in \bbZ^{2n}} f_{\ui, \uj}(\ut) \ux^{\ui} \uy^{\uj}\, ,
\end{equation}
where each $f_{\ui, \uj}$ is a restricted analytic function on $(\Delta^*)^n$. Note that the sum is finite as $f$ is polynomially tamed. In particular, this means that $f_{\ui, \uj}$ has a power series expansion in $(\Delta^*)^n$
\begin{align}
f_{\ui, \uj}(\ut) & = \sum_{\uk \ge 0} f_{\ui, \uj; \uk}\ut^{\uk}\nonumber\\
& = \sum_{k_2, \ldots, k_n \ge 0} (f_{\ui, \uj; 0, k_2, \ldots, k_n} + f_{\ui, \uj; 1, k_2, \ldots, k_n} z^1 + \cdots)(z^2)^{k_2}\cdots(z^n)^{k_n}\nonumber\\
& = \sum_{k_2, \ldots, k_n \ge 0} (f_{\ui, \uj; 0, k_2, \ldots, k_n} + f_{\ui, \uj; 1, k_2, \ldots, k_n}e^{-2\pi y^1}e^{2\pi\im x^1} + \cdots)(z^2)^{k_2}\cdots(z^n)^{k_n} \, , \label{eqn:fij}
\end{align}
where each coefficient $f_{\ui, \uj; \uk}$ is real, and we have written the sum over $k_1$ explicitly for later use. From this discussion we see that the form of $f$ is already strongly constrained. This makes the proof viable.

To proceed, we organize $f$ around $x^1$ and $y^1$:
\begin{equation*} 
f(x, y) = \sum_{(i_1, j_1) \in \bbZ^2} a_{i_1, j_1} (x^1)^{i_1} (y^1)^{j_1}\, .
\end{equation*}
Again, the sum contains finitely many terms. The coefficients $a_{i_1, j_1}$ contain the coefficients $f_{\ui, \uj; \uk}$ in \eqref{eqn:fij}, as well as the (Laurent) polynomial-dependency on all other $x$'s and $y$'s. The following two observations will help simplifying the analysis further. First, from \eqref{eqn:fij}, we safely assume that $a_{i_1, j_1}$ does not depend on $z^1$, as those terms depending on $z^1$ will fall-off quicker than $e^{-2\pi y^1}$, thus will not interfere with our estimates. Second, we also assume that the power of $x^1$ is non-negative $i_1 \ge 0$. Indeed, if the function $f$ contains negative powers of $x^1$, we can just multiply the entire $f$ with sufficiently many $x^1$'s to eliminate all negative powers of $x^1$. Note that within $\Sigma$, one has $0 < x^1 < 1$, so multiplying $f$ by $x^1$ does not alter the `$\prec$' relation, either. In summary, the coefficients $a_{i_i, j_1}$ depend on $(x^2, y^2, z^2; \cdots; x^n, y^n, z^n)$, so we have
\begin{equation} \label{eqn:fx1y1}
  f(x, y) = \sum_{i_1 \ge 0, j_1} a_{i_1, j_1}(x^2, y^2, z^2; \cdots; x^n, y^n, z^n) (x^1)^{i_1} (y^1)^{j_1}\, .
\end{equation}

Now we examine the consequence of the condition $|f| \prec 1$. Immediately, we see that $j_1$ cannot be positive, otherwise it violates the condition $|f| \prec 1$ on the linear path where $z^2, \ldots, z^n$ are fixed. Since $0 < x^1 < 1$ is bounded, showing that $|f| \prec 1$ on $\Sigma$ amounts to examining the coefficients $a_{i_1, j_1}$ closer. This motivates the following claim, which implies Lemma \ref{lemma-4.5app}.

\begin{claim} \label{the-claim}
  For every $(i_1, j_1)$, one has $|a_{i_1, j_1}|(y^1)^{j_1} \prec 1$ on $\Sigma$.
\end{claim}

The remaining task is to prove Claim \ref{the-claim}. First, we have a crucial observation. Since $y^1 \ge y^2$ and $j_1 \le 0$, we have
\begin{equation}
|a_{i_1, j_1}| (y^1)^{j_1} \le |a_{i_1, j_1}| (y^2)^{j_1}\, .
\end{equation}
Hence, if one can show that
\begin{equation} \label{eqn:reduction_to_linear_at_y1y2}
|a_{i_1, j_1}| (y^2)^{j_1} \prec 1\, ,
\end{equation}
for all $(i_1, j_1)$ then the claim is proven. Condition \eqref{eqn:reduction_to_linear_at_y1y2} can be rephrased in a nicer form. Spelling out the definition of the symbol `$\prec$', there is a positive constant $C'$ such that
\begin{equation}
|a_{i_1, j_1}| (y^2)^{j_1} < C'\, .
\end{equation}
Let $C$ be any positive constant, then the above condition is further equivalent to
\begin{equation}
|a_{i_1, j_1}| (C y^2)^{j_1} < C' C^{j_1} \quad\Longleftrightarrow\quad |a_{i_1, j_1}| (C y^2)^{j_1} \prec 1\, .
\end{equation}
And this condition is exactly Claim \ref{the-claim} realized for the linear path $y^1 = C y^2$. So, to prove Claim \ref{the-claim}, it suffices to prove \eqref{eqn:reduction_to_linear_at_y1y2}, which is equivalent to proving Claim \ref{the-claim} along a linear path $y^1 = C y^2$ with $C > 0$. Note that the reasoning also shows that if Claim \ref{the-claim} is true for a linear path with a particular choice of $C$, then it is true for all linear paths.

Next we would like to show \eqref{eqn:reduction_to_linear_at_y1y2} by induction. Since $|a_{i_1, j_1}| (y^2)^{j_1}$ is a polynomially tamed function in $z^2, \ldots, z^n$, it would be nice to apply the induction hypothesis on it. So we will check if $|a_{i_1, j_1}| (y^2)^{j_1}$ satisfies the condition in Lemma \ref{lemma-4.5app}. The idea is to extract the term $|a_{i_1, j_1}| (y^2)^{j_1}$ from $f$ by substitution of a series of properly chosen $z^1$'s. Taking a linear combination of these will yield the term $|a_{i_1, j_1}| (y^2)^{j_1}$. In this process, the hypothesis of Lemma \ref{lemma-4.5app} is never violated, and the number of variables is reduced by one. Hence by induction, Claim \ref{the-claim} will be proven for $n$.

To this end, define
\begin{equation}
f_{m, c}(z^2, \ldots, z^n) := f(z^1 = m z^2 + c, z^2, \ldots, z^n)\, ,
\end{equation}
where $m$ and $c$ are integers. Such particular assumptions on $m$ and $c$ will be used later. Since $f$ satisfies the condition in Lemma \ref{lemma-4.5app}, $f_{m, c}$ also satisfies the condition for $n - 1$ variables. It then follows by induction that
\begin{equation} \label{eqn:f_mc}
|f_{m, c}| = \left| \sum_{i_1, j_1} a_{i_1, j_1} (m x^2 + c)^{i_1} (m y^2)^{j_1} \right| \prec 1\, ,\quad\textrm{on the entire } \Sigma_{n - 1}\, .
\end{equation}

To proceed, we need a small technical result.
\begin{fact} \label{lemma:Stirling2}
  For natural numbers $0 \le k \le n$,
  \begin{equation}
  \sum_{j = 0}^{n} (-1)^j \binom{n}{j} j^{k} =
  \begin{cases}
  0\, ,          & \textrm{ for } 0 \le k < n\, ,\\
  (-1)^n\,n!\, , & \textrm{ for } k = n\, .
  \end{cases}
  \end{equation}
\end{fact}
There is a generalization to the cases where $k > n$, but these are irrelevant to our application. This fact can be computed by induction on $k$ and $n$.

Now we focus on the factor $(m x^2 + c)^{i_1}$ in \eqref{eqn:f_mc}. By Fact \ref{lemma:Stirling2}, a straightforward computation shows that
\begin{equation}
\sum_{k = 0}^{\hi} (-1)^k \binom{\hi}{k} (m x^2 + k)^{i_1} =
\begin{cases}
0\, ,                & \textrm{ for } i_1 < \hi\, ,\\
(-1)^{\hi}\,\hi!\, , & \textrm{ for } i_1 = \hi\, .
\end{cases}
\end{equation}
This computation gives a clue to proceed: We start with the highest $i_1$ power, and form a linear combination of the above form. This kills all terms with a lower $i_1$, while keeping all terms with the same highest $i_1$. These terms are further accompanied with $(m y^2)^{j_1}$ with different $j_1$, and by plugging in different values of $m$, each single term of the form $a_{i_1, j_1} y^{j_1}$ can be obtained.

More precisely, let $(\hi_1, \hj_1)$ be the lexicographically maximal $(i_1, j_1)$ that is present in \eqref{eqn:fx1y1}, with $a_{\hi_1, \hj_1} \ne 0$. Define
\begin{align}
F_m & := \frac{(-1)^{\hi_1}}{\hi_1!}\sum_{i = 0}^{\hi_1} (-1)^{i} \binom{\hi_1}{i} f_{m, i}\\
& = \sum_{j_1} a_{\hi_1, j_1} (m y^2)^{j_1}\, ,\nonumber
\end{align}
where the second equality follows from Fact \ref{lemma:Stirling2}. Note that, by induction, each $|f_{m, i}| \prec 1$, implying that $|F_m| \prec 1$. Finally, by taking a linear combination of $F_m$'s with different $m$'s, we can solve for $a_{\hi_1, j_1} (y^2)^{\hj_1}$ . This implies that $|a_{\hi_1, \hj_1}| (y^2)^{\hj_1} \prec 1$. Subtracting this term multiplied by $(mx^2 + c)^{\hi_1} m^{\hj_1}$ from $f_{m, c}$ and continue inductively, we have thus shown \eqref{eqn:reduction_to_linear_at_y1y2} for all $(i_1, j_1)$, hence Claim \ref{the-claim}. This completes the proof of Lemma \ref{lemma-4.5app}.

\section{The Hodge inner product growth}
\label{sec:proof_lemma4.7}

In this appendix, we discuss the proof of the Hodge inner product growth stated in Section~\ref{sec:hodgemetric}. We will again follow \cite{BKT}, and display the proof of a broader theorem as follows. Let us recall here the statements in \cite{BKT} that determine the growth of the Hodge inner products:

\begin{theorem} \label{app:lemma-4.7}
  Let $u \in I^{p, q_1, \ldots, q_n}$ and $v \in I^{p', q_1', \ldots, q_n'}$.
  \begin{enumerate}
    \item $\norm{u}^2$ is monomially tamed; \label{app:lemma-4.7:Part1}
    \item $\norm{\gamma(z)u}^2$ is monomially tamed; \label{app:lemma-4.7:Part2}
    \item $h(u, v)$ is polynomially tamed. \label{app:lemma-4.7:Part3}
  \end{enumerate}
\end{theorem}

The particular subspaces $I^{p, q_1, \ldots, q_n}$ will be defined later. For the moment, the reader can regard $I^{p, q_1, \ldots, q_n}$ as a subspace of the $V_{q_1, \ldots, q_n}$ space discussed in \eqref{eqn:J-splitting}.

The statement and the proof of Theorem \ref{app:lemma-4.7} requires a deeper understanding of asymptotic Hodge theory. Let us first review the necessary ingredients of asymptotic Hodge theory. We will use the same notation as in Appendix \ref{sec:proof_lemma4.5}, with the modification that we denote the indices in the moduli space by $i$ instead of $\alpha$ for prettier presentation. So a singular point is at $z^i = 0$, which is equivalent to $\varphi^i = x^i + \im y^i \to \im \infty$. We also denote the cohomology vector space as $V_\bbQ, V_\bbR, V_\bbC$, where the subscripts distinguishes the fields of coefficients.

\subsection{Lightning review of asymptotic Hodge theory}

The first theorem that we need is the nilpotent orbit theorem \cite{MR0382272,MR840721}. Around each $z^i = 0$, there is a monodromy operator, and its logarithm is denoted by $N_i$. Then the nilpotent orbit theorem says that there is a normal form of the period mapping around the singularity
\begin{equation}
  \Phi(\underline{\varphi}) = e^{\varphi^i N_i} e^{\Gamma(\uz)} F\, ,
\end{equation}
where $F$ is called the limiting Hodge filtration\footnote{This filtration is not necessarily Hodge, i.e. it is not necessarily $k$-opposed.}, and $\Gamma$ is a holomorphic function such that $\Gamma(\uz = 0) = 1$. We will not use the function $\Gamma$ so we refer interested reader to \cite{MR840721,MR1042802,Grimm:2020ouv} for more information. In the following, we write
\begin{equation} \label{app:eqn:defn-gamma}
  \gamma(\underline{\varphi}) = e^{\varphi^i N_i} e^{\Gamma(\uz)}\, ,
\end{equation}
so that
\begin{equation} \label{app:eqn:map-F-to-Phi}
  \Phi(\underline{\varphi}) = \gamma(\underline{\varphi}) F\, .
\end{equation}

To define the limiting mixed Hodge structure, we need the monodromy weight filtration. For every nilpotent operator $N$, there is a unique increasing filtration $0 \subs W(N)_0 \subs \cdots \subs W(N)_{2k} = V_\bbC$ such that
\begin{equation}
  N(W(N)_p) \subs W(N)_{p - 2}\, ,\quad\textrm{and}\quad N^s: \Gr{W(N)}{k + p} \xrightarrow{\sim} \Gr{W(N)}{k - p}\, ,
\end{equation}
where $\Gr{W(N)}{p} := W(N)_{p} / W(N)_{p - 1}$.

From the operators $N_i$, we define $n$ different monodromy weight filtrations. Let
\begin{equation}
  N_{(j)} := \sum_{i = 1}^j N_i\, ,
\end{equation}
and we define
\begin{equation}
  W^{(j)} := W(N_{(j)})\, .
\end{equation}
The $\slt$-orbit theorem \cite{MR0382272,MR840721} then implies that $(F, W^{(n)})$ is a mixed Hodge structure.

For a mixed Hodge structure, say $(F, W^{(n)})$, we have the well-known Deligne splitting
\begin{equation}
  V_\bbC = \bigoplus_{p, q} I^{p, q}\, ,
\end{equation}
such that
\begin{equation}
  F^s = \bigoplus_{p \ge s} I^{p, q}\, ,\quad\textrm{and}\quad W^{(n)}_s = \bigoplus_{p + q \le s} I^{p, q}\, ,
\end{equation}
and a conjugation condition \cite{MR840721} that is not important for our discussion. In the above expressions, the omitted indices are implicitly summed over their possible ranges. What we need in the following is a generalization of the Deligne splitting for not only a single monodromy weight filtration $W^{(n)}$, but all of them. Such a splitting is given in \cite{MR817170} and let us now review its definition.

According to Lemma 2.4.1 and Corollary 1.8.3 in \cite{MR817170}, the family of filtrations
\begin{equation}
  (F, W^{(1)}, \ldots, W^{(n)})
\end{equation}
admits a common splitting\footnote{Our convention on the indices aligns with the convention in \cite{BKT}, which is different from the original \cite{MR817170}. Denote the splitting in \cite{MR817170} by $H^{p, q_1, \ldots, q_n}$, then our $I^{p, q_1, \ldots, q_n} = H^{p, p + q_1, \ldots, p + q_n}$.}
\begin{equation}
  V_\bbC = \bigoplus_{p, q_1, \ldots, q_n} I^{p, q_1, \ldots, q_n}\, ,
\end{equation}
such that
\begin{equation} \label{app:eqn:defn_Ipq}
  F^s = \bigoplus_{p \ge s} I^{p, q_1, \ldots, q_n}\, , \quad\textrm{and}\quad W^{(j)}_s = \bigoplus_{p + q_j \le s} I^{p, q_1, \ldots, q_n}\, ,
\end{equation}
where again the omitted indices are implicitly summed over their possible ranges. This decomposition generalizes the Deligne splitting of a mixed Hodge structure: If we put just $W^{(n)}$ in, then one immediately reads out the properties of Deligne splitting.

On the other hand, recall that there is also a rational splitting
\begin{equation}
  V_\bbQ = \bigoplus_{\ell_1, \ldots, \ell_n} V_{\ell_1, \ldots, \ell_n}\, ,
\end{equation}
satisfying
\begin{equation} \label{app:eqn:defn-Js}
  W^{(j)}_s = \bigoplus_{\ell_j \le \ell} V_{\ell_1, \ldots, \ell_n}\, .
\end{equation}
This splitting is characterized by its relation to the growth of the Hodge norm \cite{MR840721,MR817170}, namely, for every $u \in V_{\ell_1, \ldots, \ell_n}$, one has
\begin{equation}
  \norm{ u}^2 \sim \left( \frac{s^1}{s^2} \right)^{\ell_1} \cdots \left( \frac{s^{n - 1}}{s^n} \right)^{\ell_{n - 1}} (s^n)^{\ell_n}\, .
\end{equation}
We will revisit this property in Theorem \ref{app:lemma-growth-Hodge-norm} in the following section.

Using properties \eqref{app:eqn:defn_Ipq} and \eqref{app:eqn:defn-Js}, we see that
\begin{equation} \label{app:eqn:iso-I-J}
  V_{\ell_1, \ldots, \ell_n} \quad\cong\quad \Gr{W^{(1)}}{\ell_1} \cdots \Gr{W^{(n)}}{\ell_n}(V_\bbC) \quad\cong\quad \bigoplus_{p} I^{p, \ell_1 - p, \ldots, \ell_n - p}\, ,
\end{equation}
meaning that each element in ${\ell_1, \ldots, \ell_n}$ can be decomposed into finitely many components living in different $I$-subspaces. Technically speaking, if we use the $\slt$-orbit theorem in \cite{MR0382272,MR840721}, we can actually obtain a nice expression characterizing and relating the $I$- and $V$-splittings. For simplicity, we assume that the nilpotent orbit $(F, N_1, \ldots, N_n)$ is $\bbR$-split. By the multi-variable $\slt$-orbit theorem in \cite{MR840721}, there exists a series of $\bbR$-split $\slt$-orbits
\begin{equation}
  (F_{(1)}, W^{(1)})\, , \ldots\, , (F_{(n)}, W^{(n)})\, ,
\end{equation}
constructed out of the original nilpotent orbit $(F, N_1, \ldots, N_n)$. Let $I^{p_1, q_1}_{(1)}, \ldots, I^{p_n, q_n}_{(n)}$ be their corresponding Deligne splittings. Then we can define
\begin{equation}
  V_{\ell_1, \ldots, \ell_n} \quad = \quad \bigcap_{i = 1}^n \bigoplus_{p_i + q_i = \ell_i} I^{p_i, q_i}_{(i)}\, .
\end{equation}
Moreover, we define
\begin{equation}
  I^{p, \ell_1, \ldots, \ell_n} \quad = \quad I^{p, \ell_n}_{(n)} \cap \bigcap_{i = 1}^{n - 1} \bigoplus_{p_i + q_i = \ell_i} I^{p_i, q_i}_{(i)}\, ,
\end{equation}
so that we have
\begin{equation} \label{app:eqn:equal-I-J}
  V_{\ell_1, \ldots, \ell_n} \quad = \quad \bigoplus_{p} I^{p, \ell_1 - p, \ldots, \ell_n - p}\, ,
\end{equation}
a genuine equality realizing the isomorphism \eqref{app:eqn:iso-I-J} between the two splittings $I$ and $V$.

\subsection{Proof of Theorem \ref{app:lemma-4.7}}

For later convenience, let us repeat here the well-known theorem on the growth of Hodge norm \cite{MR0382272,MR840721,MR817170}.
\begin{theorem} \label{app:lemma-growth-Hodge-norm}
  Let $u \in V_{s_1, \ldots, s_n}$. Then on $\Sigma$ we have
  \begin{enumerate}
    \item $\norm{u}^2 \sim \left( \frac{y^1}{y^2} \right)^{s_1} \cdots \left( \frac{y^{n - 1}}{y^n} \right)^{s_{n - 1}} (y^n)^{s_n}\, ;$ \label{app:lemma-growth-Hodge-norm:Part1}
    \item $\norm{e^{\sum z^i N_i} u}^2 \sim \left( \frac{y^1}{y^2} \right)^{s_1} \cdots \left( \frac{y^{n - 1}}{y^n} \right)^{s_{n - 1}} (y^n)^{s_n}\, ;$ \label{app:lemma-growth-Hodge-norm:Part2}
    \item $\norm{\gamma(z) u}^2 \sim \left( \frac{y^1}{y^2} \right)^{s_1} \cdots \left( \frac{y^{n - 1}}{y^n} \right)^{s_{n - 1}} (y^n)^{s_n}\, .$ \label{app:lemma-growth-Hodge-norm:Part3}
  \end{enumerate}
\end{theorem}

Before we dive into the proof of Theorem \ref{app:lemma-4.7}, let us note that the conclusion of Theorem \ref{app:lemma-4.7} also holds for vectors living in $V_{s_1, \ldots, s_n}$. To show this, we use the relation \eqref{app:eqn:equal-I-J}. For any $u \in V_{s_1, \ldots, s_n}$, there is a decomposition
\begin{equation}
  u = \sum_{p} u^p\, ,\qquad u^p \in I^{p, s_1 - p, \ldots, s_n - p}\, ,
\end{equation}
where the sum is finite. For each $u^p$, Theorem \ref{app:lemma-4.7} holds. Particularly, $\norm{u^p}^2 \in \cO(x, y)$. This implies that $\norm{u}^2 \in \cO(x, y)$. Using the growth Theorem \ref{app:lemma-growth-Hodge-norm}, we see that $\norm{u}^2$ is monomially tamed. The same reasoning applies to $\norm{e^{\sum z^i N_i} u}^2$ and $\norm{\gamma(z) u}^2$.

Now we would like to address the proof of Theorem \ref{app:lemma-4.7}. It turns out that to show Theorem \ref{app:lemma-4.7} for all weights $k$, one needs to separate the cases between even $k = 2m$ and odd $k = 2m - 1$ weights. This is mainly because, later in the proof, we will crucially use Lemma \ref{app:lemma-Gram-Schmidt}, which is only applicable to the case of even weights, whose polarization form is symmetric. Fortunately, one can transform any odd-weight VHS to an even one, preserving the Hodge inner product, without too much effort. So let us first describe how to reduce the proof for odd weights to even weights.

The idea \cite{Bakker:2021uqw} is to define a good auxiliary Hodge structure $\hH$ of weight $1$. Then by tensoring our odd-weight VHS with this auxiliary Hodge structure, the weight is raised by one, and the problem is neatly transformed into an even weight problem where the original and new Hodge inner products are related by a constant factor. The auxiliary Hodge structure is given by the Hodge structure on the middle cohomology of a special elliptic curve. We will not bother with the geometry and only discuss the algebraic data. Its underlying integral module is
\begin{equation}
  \hH_\bbZ = \bbZ \oplus \bbZ\, .
\end{equation}
We pick the canonical integral basis of $\hH_\bbZ$ and denote any element in $\hH_\bbZ$ by a pair of integers $(\ha, \hb)$. This choice of integral basis extends to the complexification $\hH_\bbC = \hH_\bbZ \otimes_\bbZ \bbC$, so we also denote $(\ha, \hb) \in \hH_\bbC$, where for the complex case $\ha, \hb$ are complex numbers. Then the Hodge structure is defined as
\begin{equation}
  \hH_\bbC   = \hH^{1, 0} \oplus \hH^{0, 1}\, , \quad\textrm{with}\quad \hH^{1, 0} := \bbC(1, \im)\, , \quad\textrm{and}\quad \hH^{0, 1} := \bbC(1, -\im)\, .
\end{equation}
Obviously, $\hH^{0, 1} = \conj{H}^{1, 0}$. It is straightforward to verify that the associated Weil operator $\hC: \hH_\bbC \to \hH_\bbC$ acts as
\begin{equation}
  \hC(\ha, \hb) = (\hb, -\ha)\, .
\end{equation}
Finally, the Hodge structure $\hH$ is polarized by the anti-symmetric bilinear form $\hQ: \hH \times \hH \to \bbZ$ defined as
\begin{equation}
  \hQ((\ha, \hb), (\hc, \hd)) = \ha\hd - \hb\hc\, ,
\end{equation}
which also extends to a bilinear form on $\hH_\bbC$. Combine these together, we have the Hodge inner product $\hh$ on $\hH$
\begin{equation}
  \hh((\ha, \hb), (\hc, \hd)) = \hQ(C(\ha, \hb), (\conj{\hc}, \conj{\hd})) = \ha\conj{\hc} + \hb\conj{\hd}\, .
\end{equation}

Turning back to the odd-weight case, let $H$ be any pure Hodge structure with odd weight $k = 2m - 1$. The tensor product $\tH = H \otimes \hH$ is a pure Hodge structure of even weight $k + 1 = 2m$. Its underlying integral module is given by
\begin{equation}
  \tH_\bbZ = H_\bbZ \otimes_\bbZ \hH_\bbZ \cong H_\bbZ \oplus H_\bbZ\, ,
\end{equation}
and we denote $(a, b) \in \tH_\bbZ$ an element, with $a, b \in H_\bbZ$. The Hodge decomposition is given by\footnote{In general, $\tH^{p, q} = \bigoplus_{\substack{r + t = p\\s + u = q}} H^{r, s} \otimes \hH^{t, u}$. Recall that $\hH^{1, 0} \cong \hH^{0, 1} \cong \bbC$ as complex vector spaces, and $V \otimes \bbC \cong V_\bbC$ for any complex vector space $V_\bbC$.}
\begin{eqnarray}
  \tH = \bigoplus_{p + q = k + 1} \tH^{p, q}\, ,\quad\textrm{with}\quad \tH^{p, q} := H^{p - 1, q} \oplus H^{p, q - 1}\, .
\end{eqnarray}
It is straightforward to check that the Weil operator $\tC: \tH_\bbC \to \tH_\bbC$ acts as $\tC = C \otimes \hC$, where $C$ is the Weil operator of $H$. More explicitly, we have
\begin{equation}
  \tC(a, b) = (Cb, -Ca)\, .
\end{equation}
And similarly, the polarization form is now $\tQ = -Q \otimes \hQ$, with $Q$ the polarization form of $H$. The minus sign accompanying $Q$ is to make sure that the positivity in the polarization condition is satisfied. We have
\begin{equation}
  \tQ((a, b), (c, d)) = - Q(a, d) + Q(b, c)\, .
\end{equation}
Assemble everything together, we have the new Hodge inner product $\tilh$ on $\tH$ given by
\begin{equation}
  \tilh((a, b), (c, d)) = \tQ(\tC(a, b), (\conj{c}, \conj{d})) = - Q(Ca, \conj{c}) - Q(Cb, \conj{d}) = h(a, c) + h(b, d)\, .
\end{equation}
In particular,
choosing $(a, b) = (u, u)$ and $(c, d) = (v, v)$ with $u, v \in H_\bbC$ yields
\begin{equation} \label{app:eqn:even-odd-Hodge-inner-product}
  \tilh((u, u), (v, v)) = 2 h(u, v)\, .
\end{equation}
This relates the Hodge inner product in the even-weight structure $\tH$ to the one in the original odd-weight structure $H$.

Starting with an odd-weight VHS, we tensor it with the constant VHS with Hodge structure $\hH$ and arrive at an even-weight VHS. Moreover, their Hodge inner products are related by \eqref{app:eqn:even-odd-Hodge-inner-product}. So if Theorem \ref{app:lemma-4.7} is proven for even weights, then it is also true for odd weights.

Now let us turn to the proof of Theorem \ref{app:lemma-4.7} for even weights $k = 2m$. We first show part \ref{app:lemma-4.7:Part1}. By part \ref{app:lemma-growth-Hodge-norm:Part1} of Theorem \ref{app:lemma-growth-Hodge-norm}, it suffices to show that $\norm{u}^2 \in \cO(x, y)$. Namely, we are going to show that $\norm{u}^2$ can be written as a ratio between two functions in $\cO[x, y, y^{-1}]$, polynomials with restricted analytic functions as coefficients.

The idea is to compute the norm $\norm{u}^2$ by decomposing $u$ with respect to a nice basis. This basis is constructed as follows. Firstly, we choose a basis that is adapted to the limiting Hodge filtration $F$. Namely, we choose a basis $w_i$ of $V_\bbC$ such that each $w_i \in I^{p_i, q_1^i, \ldots, q_n^i}$, and we order them such that $p_i$ is non-increasing. For any $\varphi \in \Sigma$, recall from equation \eqref{app:eqn:map-F-to-Phi} that we have $\Phi(\varphi) = \gamma(\varphi) F$. We define
\begin{equation} \label{app:eqn:defn-wiz}
  w_i(\varphi) := \gamma(\varphi)w_i \in \Phi^i(\varphi)\, .
\end{equation} Then, because of the ordering of $p_i$ and taking property \eqref{app:eqn:defn_Ipq} into account, we have
\begin{equation} \label{app:eqn:Weil-on-wi}
  w_i(\varphi) \in H^{p_i, k - p_i}_\varphi \qquad\Longrightarrow\qquad C_\varphi (w_i(\varphi)) = \im^{2 p_i - k} w_i(\varphi)\, .
\end{equation}

Because of the above property \eqref{app:eqn:Weil-on-wi}, the evaluation of the Hodge inner product $h$ on the basis $w_i$ is eventually reduced to a constant multiple of $Q$. To simplify notation, we define $B(u, v) := Q(u, \conj{v})$ and $B^2(u) := B(u, u)$. The next step is to construct a $B$-orthogonal basis out of $w_i(\varphi) \in \Phi^i(\varphi)$. This is done by the Gram-Schmid process. We actually need an extended version of it, so let us present the process in the following technical Lemma \ref{app:lemma-Gram-Schmidt}.

\begin{lemma} \label{app:lemma-Gram-Schmidt}
  \emph{(Gram-Schmidt)} Let $V_\bbC$ be a complex finite dimensional vector space equipped with an hermitian inner product $B$. Let $\{v_i\}$ be a basis of $V_\bbC$. Define inductively
  \begin{equation} \label{app:eqn:defn-Gram-Schmidt}
    \vtil_i :=
    \begin{cases}
      v_1\, ,                                                                            & \textrm{for } i = 1\, ,\\
      v_i - \sum_{j < i} \frac{B(v_i, \vtil_j)}{B(\vtil_j, \vtil_j)} \vtil_j\, , & \textrm{for } i \ge 2\, ,
    \end{cases}
  \end{equation}
  then $\{\vtil_i\}$ is also a basis of $V_\bbC$ satisfying, for any $v \in V_\bbC$ and all $i$,
  \begin{equation} \label{app:eqn:Gram-Schmidt-inner-product}
    B(v, \vtil_i) = \frac{B(v_1 \wedge \cdots \wedge v_{i - 1} \wedge v, v_1 \wedge \cdots \wedge v_i)}{B^2(v_1 \wedge \cdots \wedge v_{i - 1})}\, ,
  \end{equation}
  where \footnote{This is the extension of the inner product $B$ to the $n$-th tensor power of $V_\bbC$ for any $n$.}
  \begin{equation} \label{app:eqn:defn-B-extended-to-wedge}
    B(u_1 \wedge \cdots \wedge u_n, w_1 \wedge \cdots \wedge w_n) := \det(B(u_i, w_j))\, ,\quad\textrm{for all } n, \textrm{and } u_i, w_j \in V_\bbC\, .
  \end{equation}

  Moreover, property \eqref{app:eqn:Gram-Schmidt-inner-product} implies
  \begin{equation} \label{app:eqn:Gram-Schmidt-norm}
    B(\vtil_i, \vtil_j) =
    \begin{cases}
      0\, ,                                                                   & \textrm{if } i \ne j\, ,\\
      \frac{B^2(v_1 \wedge \cdots \wedge v_i)}{B^2(v_1 \wedge \cdots \wedge v_{i - 1})}\, , & \textrm{if } i = j\, .
    \end{cases}
  \end{equation}
  So $\{ \vtil_i \}$ is an orthogonal basis with respect to $B$.
\end{lemma}
This lemma can be shown by induction, with the use of determinant identities relating a matrix, its minors, and its cofactor.

Now we turn back to the proof of part \ref{app:lemma-4.7:Part1} of Theorem \ref{app:lemma-4.7}. Apply the Gram-Schmidt process to the basis $w_i(\varphi)$, we obtain a new basis $\wtil_i(\varphi)$. It is easy to check by induction that,
\begin{equation}
  h_z(\wtil_i(\varphi), \wtil_j(\varphi)) = B(C_\varphi \wtil_i(\varphi), \wtil_j(\varphi)) =
  \begin{cases}
    \im^{2p_i - k} B(\wtil_i(\varphi), \wtil_j(\varphi))\, ,  & \textrm{ if } i \le j\, ,\\
    \im^{-2p_j + k} B(\wtil_i(\varphi), \wtil_j(\varphi))\, , & \textrm{ if } i > j\, .
  \end{cases}
\end{equation}
In particular, this implies that the Hodge norm of $\wtil_i(\varphi)$ satisfies, for all $i$,
\begin{equation} \label{app:eqn:Hodge-norm-wtil}
  \norm{\wtil_i(\varphi)}^2 = \im^{2p_i - k} B^2(\wtil_i(\varphi))\, .
\end{equation}
So although, in general, $\wtil_i(\varphi)$ does not belong to a single $I^{p_i, q_1^i, \ldots, q_n^i}$, when computing the Hodge norm $\norm{\wtil_i(\varphi)}$, the Weil operator $C_\varphi$ still factorizes out of the bilinear form. Note that to show this fact, one must use the orthogonality relation \eqref{app:eqn:Gram-Schmidt-norm} of the Gram-Schmidt basis.

Next, expanding our $u \in I^{p, q_i, \ldots, q_n}$ with respect to the basis $\wtil_i(\varphi)$, we get
\begin{equation}
  u = \sum_i \util_i(\varphi)\, ,
\end{equation}
where each $\util_i(\varphi)$ is a multiple of $\wtil_i(\varphi)$. We can actually compute the multiplication factor, due to property \eqref{app:eqn:Gram-Schmidt-norm} of the basis $\wtil_i(\varphi)$. We have
\begin{equation}
  \util_i(\varphi) = \frac{B(u, \wtil_i(\varphi))}{B^2(\wtil_i(\varphi))} \wtil_i(\varphi)\, .
\end{equation}
We do the same for any $v \in I^{p', q_1', \ldots, q_n'}$ to get $\vtil_i(\varphi)$. And the Hodge inner product between $\util_i(\varphi)$ and $\vtil_j(\varphi)$ can be computed (for $i \le j$)
\begin{align}
  h(\util_i(\varphi), \vtil_i(\varphi)) & = Q(C_\varphi \util_i(\varphi), \conj{\vtil_i(\varphi)})\nonumber\\
                            & = \im^{2p_i - k} B(\util_i(\varphi), \vtil_i(\varphi))\nonumber\\
                            & = \im^{2p_i - k} \frac{B(u, \wtil_i(\varphi))B(\wtil_i(\varphi), v)}{B^2(\wtil_i(\varphi))}\, , \label{app:eqn:h-utili-vtilj}
\end{align}
where in the second equality we have used \eqref{app:eqn:Hodge-norm-wtil}.

On the other hand, by Lemma \ref{app:lemma-Gram-Schmidt}, we have
\begin{align}
  B(u, \wtil_i(\varphi)) & = \frac{B(w_1(\varphi) \wedge \cdots \wedge w_{i - 1}(\varphi) \wedge u, w_1(\varphi) \wedge \cdots \wedge w_i(\varphi))}{B^2(w_1(\varphi) \wedge \cdots \wedge w_{i - 1}(\varphi))}\, ; \label{app:eqn:B-u-wtili}\\
  B^2(\wtil_i(\varphi))  & = \frac{B^2(w_1(\varphi) \wedge \cdots \wedge w_i(\varphi))}{B^2(w_1(\varphi) \wedge \cdots \wedge w_{i - 1}(\varphi))}\, . \label{app:eqn:B2-wtili}
\end{align}

Now from the expression \eqref{app:eqn:defn-gamma} for $\gamma(\varphi)$ and the definition \eqref{app:eqn:defn-wiz} of $w_i(\varphi)$, we conclude that both the numerator and the denominator in \eqref{app:eqn:B-u-wtili} and \eqref{app:eqn:B2-wtili} are in $\cO[x, y, y^{-1}]$. Hence we conclude that the Hodge norm of $u$
\begin{equation}
  \norm{u}^2_\varphi = \sum_{i, j} h_\varphi(\util_i(\varphi), \util_j(\varphi)) \in \cO(x, y)\, .
\end{equation}
Combine with part \ref{app:lemma-growth-Hodge-norm:Part1} of Theorem \ref{app:lemma-growth-Hodge-norm}, part \ref{app:lemma-4.7:Part1} of Theorem \ref{app:lemma-4.7} is proven. The proof of part \ref{app:lemma-4.7:Part2} is similar.

To prove part \ref{app:lemma-4.7:Part3}, it remains to show that the denominator of \eqref{app:eqn:h-utili-vtilj} is a monomially tamed function. This follows from part \ref{app:lemma-4.7:Part2} of Theorem \ref{app:lemma-4.7}. Let us consider $B^2(w_1(\varphi) \wedge \cdots \wedge w_n(\varphi))$. Using the definition \eqref{app:eqn:defn-B-extended-to-wedge}, we have
\begin{equation} \label{app:eqn:B^2-w1-wn}
  B^2(w_1(\varphi) \wedge \cdots \wedge w_n(\varphi)) = \det(B(w_i(\varphi), w_j(\varphi)))\, .
\end{equation}
Let the angle between $w_i(\varphi)$ and $w_j(\varphi)$ be $\theta_{ij}$, and it satisfies the usual relation to the inner product
\begin{equation} \label{app:eqn:defn-theta-ij}
  B(w_i(\varphi), w_j(\varphi))^2 = B^2(w_i(\varphi))B^2(w_j(\varphi))\cos^2\theta_{ij}\, .
\end{equation}
Note that $0 \le \cos^2\theta_{ij} < 1$ is always true as $w_i(\varphi)$ form a basis of $V_\bbC$. Next we expand the determinant in \eqref{app:eqn:B^2-w1-wn} and plug \eqref{app:eqn:defn-theta-ij} into it. We get
\begin{align}
  \det(B(w_i(\varphi), w_j(\varphi))) & = \sum_{\sigma} (-1)^\sigma B(w_1(\varphi), w_{\sigma(1)}(\varphi)) \cdots B(w_i(\varphi), w_{\sigma(i)}(\varphi))\\
                          & = (B^2(w_1(\varphi)) \cdots B^2(w_i(\varphi)))(1 + \cdots)\, ,
\end{align}
where $\sigma$ runs over all permutations of $(1, 2, \ldots, i)$, and the omitted part in the second line consists of a summation of various $|\cos\theta_{ij}|$, and is always bounded by a positive constant. The second equality is based on the observation that each index appears exactly twice in each term. In summary, we have shown that, using \eqref{app:eqn:Hodge-norm-wtil} and \eqref{app:eqn:defn-wiz},
\begin{equation}
  B^2(w_1(\varphi) \wedge \cdots \wedge w_n(\varphi)) \propto B^2(w_1(\varphi)) \cdots B^2(w_i(\varphi)) \propto \norm{\gamma(\varphi) w_1}^2_z \cdots \norm{\gamma(\varphi) w_i}^2_z\, .
\end{equation}

And using part \ref{app:lemma-4.7:Part2} of Theorem \ref{app:lemma-4.7}, this quantity is monomially tamed. Finally, by \eqref{app:eqn:h-utili-vtilj}, part \ref{app:lemma-4.7:Part3} of Theorem \ref{app:lemma-4.7} is proven.

\providecommand{\href}[2]{#2}\begingroup\raggedright\endgroup

\end{document}